\documentclass[useAMS,usenatbib]{mn2e}
\usepackage{color,graphicx}
\def\aj{AJ}             	
\def\araa{ARA\&A}       	
\def\apj{ApJ}           	
\def\apjl{ApJ}          	
\def\aap{A\&A}          	
\def\mnras{MNRAS}       	
\def\nat{Nature}        	

\def\gsim{\hspace{0.3em}\raisebox{0.4ex}{$>$}\hspace{-0.75em}\raisebox{-.7ex}{$\sim$}\hspace{0.3em}}
\def\lsim{\hspace{0.3em}\raisebox{0.4ex}{$<$}\hspace{-0.75em}\raisebox{-.7ex}{$\sim$}\hspace{0.3em}}

\title[Spiral-arm instability]{Spiral-arm instability: giant clump formation via fragmentation of a galactic spiral arm}
\author[S. Inoue \& N. Yoshida]
{\parbox[t]{\textwidth} 
{Shigeki Inoue\thanks{E-mail: shigeki.inoue@ipmu.jp} \& Naoki Yoshida}
\\ \\
Kavli Institute for the Physics and Mathematics of the Universe (WPI), UTIAS, The University of Tokyo, Chiba 277-8583, Japan\\
Department of Physics, School of Science, The University of Tokyo, Bunkyo, Tokyo 113-0033, Japan
}

\begin{document}

\pagerange{\pageref{firstpage}--\pageref{lastpage}} \pubyear{2014}

\maketitle

\label{firstpage}

\begin{abstract}
Fragmentation of a spiral arm is thought to drive the formation of giant clumps in galaxies. Using linear perturbation analysis for self-gravitating spiral arms, we derive an instability parameter and define the conditions for clump formation. We extend our analysis to multi-component systems that consist of gas and stars in an external potential. We then perform numerical simulations of isolated disc galaxies with isothermal gas, and compare the results with the prediction of our analytic model. Our model describes accurately the evolution of the spiral arms in our simulations, even when spiral arms dynamically interact with one another. We show that most of the giant clumps formed in the simulated disc galaxies satisfy the instability condition. The clump masses predicted by our model are in agreement with the simulation results, but the growth time-scale of unstable perturbations is overestimated by a factor of a few. We also apply our instability analysis to derive scaling relations of clump properties. The expected scaling relation between the clump size, velocity dispersion, and circular velocity is slightly different from that given by the Toomre instability analysis, but neither is inconsistent with currently available observations. We argue that the spiral-arm instability is a viable formation mechanism of giant clumps in gas-rich disc galaxies.

\end{abstract}

\begin{keywords}
instabilities -- methods: numerical -- methods: analytical -- galaxies: spiral -- galaxies: kinematics and dynamics.

\end{keywords}

\section{Introduction}
\label{Intro}
Disc galaxies with spiral arms, including the Milky Way, occupy a large fraction of galaxies in the current Universe although the fraction depends on mass and colour \citep[e.g.][and references therein]{bst:88}. Such spiral galaxies generally have relatively smooth stellar distribution, and their star clusters in the discs are small: $M_{\rm cl}\lsim10^3~{\rm M_\odot}$ \citep[e.g.][]{ll:03}. Their spiral arms are considered to drive the formation of such star clusters.

In the high-redshift Universe, disc galaxies have been observed to have clumpy morphologies, rather than smooth density distribution, in which several star-forming `giant clumps' of $M_{\rm cl}\lsim10^8~{\rm M_\odot}$ are hosted in their disc regions. Although some fraction of such clumpy galaxies are ongoing mergers \citep[e.g.,][]{w:06,fgb:09,p:10,r:17}, a number fraction of clumpy galaxies generally increases with redshift, depending on galaxy mass, up to $\sim50$ per cent at a redshift of $z\simeq2$--$3$ \citep{tkt:13II,mkt:14,gfb:14,sok:16,bmo:17}.\footnote{Clumpy fractions at further high redshifts are poorly known. \citet{sok:16} argued that clumpy fractions decrease after a peak at a redshift $z\sim2$ until $z\sim8$, whereas \citet{r:17} claimed that the fractions keep increasing at least until $z\sim5$.} Clumpy disc galaxies are also observed at low redshifts, but their abundance is lower by far than at high redshifts \citep[e.g.][]{ees:13,bgf:14,fgb:14,f:17,gpm:15}. Although the high-redshift disc galaxies that are currently observable seem to have too high mass-concentrations of baryon to be the progenitors of low-redshift spiral galaxies \citep{gsu:17}, the clumpy discs have also been argued to evolve into spiral galaxies while their clumps are disrupted \citep[e.g.][]{g:12,hqm:12,okh:17} or merge into their central bulges \citep[e.g.][]{n:98,n:99,is:11,is:12,cdt:14}.

Spiral arms and giant clumps in low- and high-redshift discs are clearly different structures; nevertheless, the formation mechanisms of both structures are often attributed to Toomre instability (e.g. \citealt{n:98,n:99,g:08,gnj:11,dsc:09,rck:17} for clump formation, or at least relevant to low values of Toomre's instability parameters $Q$, possibly via swing amplification for spiral-arm formation, e.g. \citealt{h:71,t:81,bt:08,fbs:11,hs:15,bmm:16}). It is still unclear, however, why these high-redshift galaxies form giant clumps whereas the low-redshift ones do not. Recently, \citet{idm:16} have shown, by utilizing cosmological simulations, that a significant fraction of giant clumps start forming with high values of $Q\gsim2$ \citep[see also][]{okh:17}, which are indicative of gravitationally stable states for axisymmetric linear perturbations \citep{s:60,t:64} in their simulated clumpy galaxies.\footnote{However, recent observations estimated $Q$ in clumpy galaxies to be typically below unity in the extended disc regions \citep{gnj:11,gfl:14,fga:17}. Thus, still there is tension between the observations and the simulations of high-redshift clumpy galaxies.} Hence, the formation of giant clumps in high-redshift may not be driven by Toomre instability. In addition, the high values of $Q\gsim2$ in their simulated clumpy galaxies are close to those determined in nearby spiral galaxies and the solar neighbourhood in the Milky Way \citep[e.g.][]{r:01,lwb:08,flw:14,wab:14,ttt:17}, which implies a similarity between clumpy and spiral galaxies in terms of the stability against axisymmetric perturbations. Thus, it is still an open question what physical processes are responsible for formation of spiral arms and giant clumps. In this sense, the current status of our understanding of the evolution of disc galaxies from high to low redshifts is unsatisfactory.

Our purpose in this study is to discuss a physical mechanism of disc instability --- other than Toomre instability --- that can form giant clumps and possibly explain the morphological difference between spiral and clumpy galaxies. Recently, \citet[][hereafter TTI]{tti:16} have performed two-dimensional hydrodynamic simulations for proto-planetary discs and found that spiral arms in their discs fragment and form knotty structures (proto-planets) when Toomre $Q$ parameters measured on the arms are lower than $0.6$. Although they empirically obtained their criterion for the spiral-arm fragmentation from a large number of isolated simulations, they also performed their linear perturbation analysis for a spiral arm and found the TTI criterion, $Q\lsim0.6$, to be consistent with their their analysis (see also Appendix \ref{TTIana}). Hence, TTI instability --- which means spiral-arm fragmentation occurring when $Q\lsim0.6$ on an arm --- can be explained as gravitational instability of the gas arm. The analysis for TTI instability can also be applied to galactic spiral arms, and the spiral-arm instability (hereafter, SAI) might explain the presence (absence) of giant clumps as the instability (stability) against fragmentation of spiral arms in high- (low-) redshift galaxies. Because TTI focused on planet formation in a Keplarian gas disc, however, their instability criterion may not be applicable to a galactic spiral arm. Therefore, we need to obtain a general formalisation of the criterion for the SAI. Moreover, since galaxies are generally multi-component system such as gas and stars, we have to extend the TTI analysis to a multi-component model. 

This paper is organised as follows. In Section \ref{basiceq}, we present our linear perturbation analysis, based on that proposed by TTI, for spiral-arm fragmentation to derive an instability parameter and its criterion. Next ,we extend our linear perturbation theory to a two-component model. In Section \ref{sims}, we perform $N$-body/hydrodynamic simulations with isolated disc galaxy models to test our theory. In Section \ref{result}, we adopt our simulation data to our analysis, such as instability parameters, wavelengths and growth time-scales of unstable perturbations. In Section \ref{discussion}, we discuss whether SAI can be a possible mechanism of giant clump formation in gas-rich disc galaxies. Using our analysis, we obtain scaling relations for properties of clumps forming from SAI and compare the scaling relations obtained from our SAI models and Toomre analysis with observations of clumpy galaxies. In Section \ref{conclusions}, we present our conclusions and summary of this study.

\section{Linear perturbation analysis}
\label{basiceq}
Our analysis basically follows the same manner as TTI until we derive a single-component dispersion relation. We perform local linear perturbation analysis in which a pitch angle of a spiral arm is assumed to be negligibly small (the tight-winding approximation), and the arm is rotating around the galactic centre on a disc plane with an angular velocity $\Omega$. In this case, the spiral arm can be locally considered as a structure resembling a ring. If the spiral arm is self-gravitating, it can be assumed to have a rigid rotation, and the Oort's constant $B=-\Omega$ in the arm. In the polar coordinates $(R,\phi)$, we consider azimuthal perturbations propagating along the arm, which are assumed to be proportional to $\exp[i(ky-\omega t)]$, where $y\equiv\phi R$. For the perturbations, if their wavelengths are small enough compared with the radius of the arm, i.e. $kR\gg1$, then the curvature of the spiral arm is negligible. 

We adopt a Gaussian distribution to a radial surface density profile of the spiral arm, $\Sigma(R)=\Sigma_0\exp(-\xi^2/2w^2)$, where $\xi\equiv R-R_0$, $R_0$ is the radius of the density peak in the arm, and $\Sigma_0$ is the surface density at $R_0$. As is done in TTI, we define the edges of the spiral arm to be the inner and outer radii where $\Sigma(R)=0.3\Sigma_0$. In this case, the half width $W\simeq1.55w$, and $2W$ corresponds to  the full width of the spiral arm. The line-mass\footnote{This quantity is defined as the mass per unit length. Although some previous papers termed it `line-density', we follow the terminology in TTI.} of the arm is given as 
\begin{equation}
\Upsilon\equiv2\int^W_0\Sigma(R)~\textrm{d}\xi=AW\Sigma_0,
\label{linemass}
\end{equation}
where $A\simeq1.4$ for a Gaussian density distribution.

\subsection{A single-component model}
\label{singleanalysis}
\begin{figure}
	\includegraphics[bb=0 0 409 281,width=\hsize]{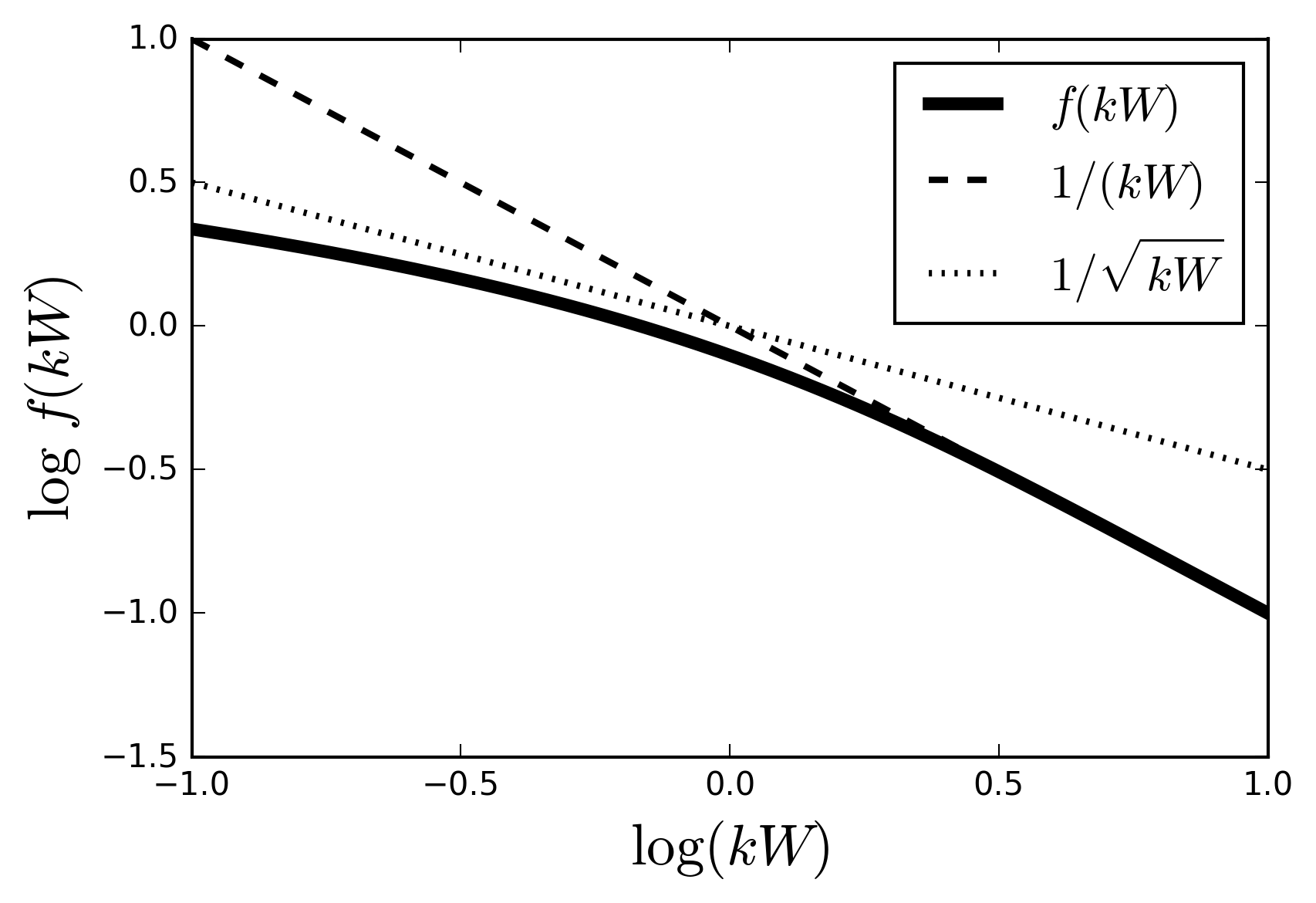}
	\caption{The function of $f(kW)$ indicated with the solid line. The dashed and dotted lines show power-low relations of $1/(kW)$ and $1/\sqrt{kW}$ as reference.}
	\label{fkw}
\end{figure}

First, we consider the linear perturbation analysis for a gas component in a spiral arm. With the aforementioned assumptions and settings, the linearised equations of continuity, $R$- and $\phi$-momenta for the azimuthal perturbations are obtained as
\begin{equation}
\omega\delta \Upsilon = k\Upsilon\delta v_\phi,
\label{linearlized1}
\end{equation}
\begin{equation}
-i\omega\delta v_R = 2\Omega\delta v_\phi,
\label{linearlized2}
\end{equation}
\begin{equation}
-i\omega\delta v_\phi = -2\Omega\delta v_R - ik\frac{\sigma^2}{\Upsilon}\delta \Upsilon - ik\delta\Phi
\label{phimom}
\end{equation}
(see TTI), where the prefix $\delta$ means the perturbation of a physical value following it. In equation (\ref{phimom}), $\Phi$ is gravitational potential, and $\sigma^2\equiv\sigma_\phi^2+c^2$, where $\sigma_\phi$ and $c$ are azimuthal dispersion of turbulent velocities and sound velocity of gas. Combining the above equations, one can obtain the dispersion relation, 
\begin{equation}
\label{DR}
\omega^2 = \left(\sigma^2+\frac{\Upsilon}{\delta \Upsilon}\delta\Phi\right)k^2 + 4\Omega^2.
\end{equation}
The Poisson equation for the perturbed potential of a razor-thin ring with a Gaussian density distribution is given as
\begin{equation}
\label{poisson}
  \delta\Phi =-\pi G\delta\Upsilon\left[K_0(kW)L_{-1}(kW) + K_1(kW)L_0(kW)\right],
\end{equation}
where $K_i$ and $L_i$ are modified Bessel and Struve functions of order $i$ (see TTI for the derivation of equation \ref{poisson}). By substituting the above into equation (\ref{DR}), the dispersion relation is described as,
\begin{equation}
\omega^2 = \left[\sigma^2-\pi G\Upsilon f(kW)\right]k^2 + 4\Omega^2,
\label{DL1}
\end{equation}
where $f(kW)\equiv[K_0(kW)L_{-1}(kW) + K_1(kW)L_0(kW)]$, and Fig. \ref{fkw} illustrates how $f(kW)$ varies.

Although TTI have presented the same dispersion relation in their paper, they did not derive an instability parameter like Toomre's $Q$. To define the instability parameter of SAI, we follow a manner similar to previous studies of Toomre instability analysis \citep{js:84_2,js:84,j:96,r:01}. Equation (\ref{DL1}) is transformed as
\begin{equation}
\frac{\sigma^2k^2 + 4\Omega^2 - \omega^2}{\pi Gf(kW)\Upsilon k^2}=1.
\label{DL11}
\end{equation}
This equation is equivalent to the dispersion relation, always satisfied for the linear perturbations. In the boundary case of $\omega^2=0$, the above is
\begin{equation}
S\equiv\frac{\sigma^2k^2 + 4\Omega^2}{\pi Gf(kW)\Upsilon k^2}=1.
\label{Crit1C}
\end{equation}
By comparing with equation (\ref{DL11}), one can see that $S<1$ in the unstable cases of $\omega^2<0$, where perturbations can grow exponentially with time. Conversely, $S>1$ in the stable cases of $\omega^2>0$, where perturbations oscillate with time. Thus, one can consider $S$ as the instability parameter of SAI, with the boundary of $S=1$ like Toomre's instability parameter $Q$. However, unlike $Q$, our instability parameter $S$ is a function of $k$, therefore the instability condition is satisfied when $\min[S(k)]<1$. In Appendix \ref{TTIana}, we show that our instability criterion $\min[S(k)]<1$ is almost consistent with the TTI criterion $Q<0.6$.

\subsection{A two-component model}
Next, we extend the above instability analysis to a multi-component model. In this study, we approximate a galaxy to be composed of gas and stellar discs in a background potential. Because the different components generally have their own physical properties, we consider distinct values of $\Upsilon$, $\sigma$, $\Omega$ and $W$ for gas and stars.\footnote{Gas and stellar discs can have similar values of $\Omega$. Especially, approximately $\Omega_{\rm g}\simeq\Omega_{\rm s}$ in kinematically cold discs \citep{bt:08}.} Hereafter, let suffixes `g' and `s' denote gas and stellar values (except growth time-scale $t_{\rm g}$). We adopt the fluid approximation to the stellar component, in which stars are assumed to have the same form of dispersion relation of gas. It should be noted, however, that this approximation may not be appropriate because of the collisionless nature of stars \citep{t:64,ls:66,bt:08,r:01,e:11}. We examine whether the fluid approximation holds in our $N$-body simulations in Section \ref{Nbodyruns}.

The dispersion relation is obtained for each component from equation (\ref{DR}): 
\begin{equation}
\omega^2 = \left(\sigma_{\rm g}^2+\frac{\Upsilon_{\rm g}}{\delta \Upsilon_{\rm g}}\delta\Phi\right)k^2 + 4\Omega_{\rm g}^2,
\label{DRgas_prev}
\end{equation}
\begin{equation}
\delta \Upsilon_{\rm g} = k^2\frac{\Upsilon_{\rm g}}{\omega^2-4\Omega_{\rm g}^2-\sigma_{\rm g}^2k^2}\delta\Phi,
\label{DRgas}
\end{equation}
for gas, and 
\begin{equation}
\omega^2 = \left(\sigma_{\rm s}^2+\frac{\Upsilon_{\rm s}}{\delta \Upsilon_{\rm s}}\delta\Phi\right)k^2 + 4\Omega_{\rm s}^2,
\label{DRstar_prev}
\end{equation}
\begin{equation}
\delta \Upsilon_{\rm s} = k^2\frac{\Upsilon_{\rm s}}{\omega^2-4\Omega_{\rm s}^2-\sigma_{\rm s}^2k^2}\delta\Phi,
\label{DRstar}
\end{equation}
for stars, where $\sigma_{\rm s}$ is azimuthal component of stellar velocity dispersion. Because gas and stars interact through gravity, they share the same perturbed potential that is described as the superposition of perturbations on gas and stellar potentials, i.e. $\delta\Phi=\delta\Phi_{\rm g}+\delta\Phi_{\rm s}$ \citep{js:84_2,js:84,r:92,j:96,r:01}. The Poisson equation (\ref{poisson}) connects the two component as
\begin{equation}
\label{twopoisson}
  \delta\Phi =-\pi G\left[\delta \Upsilon_{\rm g}f(kW_{\rm g}) + \delta \Upsilon_{\rm s}f(kW_{\rm s})\right].
\end{equation}
Substituting equations (\ref{DRgas}) and (\ref{DRstar}) into equation (\ref{twopoisson}), the two-component dispersion relation is obtained as 
\begin{equation}
\pi Gk^2\left[\frac{\Upsilon_{\rm g}f(kW_{\rm g})}{\sigma_{\rm g}^2k^2+4\Omega_{\rm g}^2-\omega^2} + \frac{\Upsilon_{\rm s}f(kW_{\rm s})}{\sigma_{\rm s}^2k^2+4\Omega_{\rm s}^2-\omega^2}\right] = 1.
\label{DR2C}
\end{equation}
Eventually, as in the case of the single-component model (equation \ref{Crit1C}), the two-component instability parameter can be defined as 
\begin{equation}
S_{\rm 2}\equiv\frac{1}{\pi Gk^2}\left[\frac{\Upsilon_{\rm g}f(kW_{\rm g})}{\sigma_{\rm g}^2k^2+4\Omega_{\rm g}^2} + \frac{\Upsilon_{\rm s}f(kW_{\rm s})}{\sigma_{\rm s}^2k^2+4\Omega_{\rm s}^2}\right]^{-1}.
\label{Crit2C}
\end{equation}
Again, the instability condition is $\min[S_{\rm 2}(k)]<1$.

By solving equation (\ref{Crit2C}) for all $k$, one can obtain the most unstable mode $k_{\rm MU}$ that gives the lowest $S_{\rm 2}$. The perturbation with $k_{\rm MU}$ is expected to grow first if $S_{\rm 2}(k_{\rm MU})<1$, and the most unstable wavelength can be estimated to be $\lambda_{\rm MU}=2\pi/k_{\rm MU}$. Furthermore, by substituting $k=k_{\rm MU}$ into equation (\ref{DR2C}), the frequency $\omega_{\rm MU}$ at $k_{\rm MU}$ can be numerically computed, and the growth time-scale of the most unstable perturbation is estimated to be $t_{\rm g}=2\pi/\sqrt{|\omega_{\rm MU}^2|}$. If $S_{\rm 2}(k_{\rm MU})>1$, $\lambda_{\rm MU}$ and $t_{\rm g}$ mean the wavelength and the time-scale of the perturbative oscillation with the lowest frequency $\omega_{\rm MU}$.

\section{Simulations}
\label{sims}
To test our theory proposed in Section \ref{basiceq}, we perform numerical simulations using models of disc galaxies in isolation and utilize the simulation data in Section \ref{result}. We use the moving-mesh hydrodynamics code {\sc Arepo} \citep{arepo} to perform simulations of self-gravitating gas and stellar discs with an isothermal equation of state. Haloes and bulges are represented with rigid potentials of Navarro-Frenk-White and Hernquist models, respectively \citep{nfw:97,h:90}. Our simulations do not take into account gas cooling, star formation, stellar feedback or magnetic fields. 

\subsection{Initial Conditions}
\label{ICs}
\begin{table}
  \caption{The initial conditions of our simulations. The values of $f_{\rm g}$ are gas fractions uniform in the discs: $f_{\rm g}\equiv M_{\rm g,d}/(M_{\rm g,d}+M_{\rm s,d})$, where $M_{\rm g,d}$ and $M_{\rm s,d}$ are masses of gas and stellar discs. The coldness parameter $Q_{\rm min}$ gives normalization of radial velocity dispersion profiles in the models. The mass distribution models in the rightmost column are listed in Table \ref{modellist}.}
  \label{paramlist}
  $$ 
  \begin{tabular}{ccccccccccc}
    \hline
     name & $f_{\rm g}$ & $Q_{\rm min}$ & mass model & \\
    \hline
    Df25Q15 & $0.25$ & $1.5$ & Disc-dominant & \\
    Df20Q13 &   $0.2$ & $1.3$ & Disc-dominant & \\
    Df20Q15 &   $0.2$ & $1.5$ & Disc-dominant & \\
    Df00Q10 &      $0$ & $1.0$ & Disc-dominant & \\
    Df00Q13 &      $0$ & $1.3$ & Disc-dominant & \\
    Bf70Q13 &   $0.7$ & $1.3$ & Bulge-dominant & \\
    Bf60Q13 &   $0.6$ & $1.3$ & Bulge-dominant & \\
    \hline
  \end{tabular}
  $$ 
\end{table}
Our initial conditions of the galactic discs are generated with the method proposed by \citet{h:93}, in which radial profiles of surface densities and radial velocity dispersions are assumed to be exponential functions with the same scale radius $R_{\rm d}$, and kinematic coldness is parameterised by an arbitrary Toomre parameter $Q_{\rm min}$ at $R\simeq2.5R_{\rm d}$. The mean and a dispersion of azimuthal velocities are computed from the Jeans equation and the epicyclic approximation \citep{bt:08}. Vertical structures of the discs are constructed with a density function of ${\rm sech}^2[z/(2z_{\rm d})]$ and a velocity distribution determined from vertical Jeans equilibrium. The gas and stellar discs have the same scale height $z_{\rm d}=50~{\rm pc}$ independent of radius. Because our simulations are aimed at testing our analysis assuming a razor-thin spiral arm, we set $z_{\rm d}$ to this small value although stellar discs in observed spiral galaxies are typically thicker than $z_{\rm d}=50~{\rm pc}$ (see also Section \ref{real}). The gas and stars share the same initial density and velocity distributions. The whole halo region of $200^3~{\rm kpc^3}$ is filled with diffuse gas the density of which is $n_{\rm H}=10^{-6}~{\rm cm^{-3}}$; however the halo gas hardly affects our simulation results. Our simulations with the isothermal equation of state keeps the gas temperature at $10^4~{\rm K}$ independent of density. In all runs, we use the same numbers of gas and stellar elements in the discs: the numbers of stellar particles $n_{\rm s}=5\times10^6$ and gas cells $n_{\rm g}=1\times10^6$. The simulation code operates mesh regulations such as motions of gas cells, refinement and derefinement so that each gas cell keeps its initial mass within a factor of $2$. A gravitational softening length of a stellar particle is set to $\epsilon_{\rm s}=50~{\rm pc}$, and that of a gas cell varies with its cell volume $V_{\rm cell}$ as $\epsilon_{\rm g}=2.5(3V_{\rm cell}/4\pi)^{1/3}$ with the lower limit of $\epsilon_{\rm g,min}=50~{\rm pc}$; therefore gas contraction is limited to the scale $\sim\epsilon_{\rm g,min}$. The initial kinematic coldness $Q_{\rm min}$ and gas fractions $f_{\rm g}$ of the discs are arbitrary parameters in our runs. All of our initial conditions are tabulated in Table \ref{paramlist}.

\begin{table*}
  \caption{Mass-distribution models in the initial states of our simulations. Values of $M_{\rm d}\equiv M_{\rm g,d}+M_{\rm s,d}$, $M_{\rm 200}$ and $M_{\rm b}$ are the total masses of the disc, halo and bulge in each model, and $R_{\rm d}$, $r_{\rm h}$ and $r_{\rm b}$ are their scale radii. The value of $M_{\rm 200}$ is defined to be the halo mass enclosed within the galactocentric radius $r_{\rm 200}$ at which $3M_{\rm 200}/(4\pi r_{\rm 200}^3)=2.9\times10^{4}{\rm M_{\odot}~kpc^{-3}}$.}
  \label{modellist}
  $$ 
  \begin{tabular}{ccccccccccc}
    \hline
     mass model & $M_{\rm d}$ [${\rm M_\odot}$] & $R_{\rm d}$ [${\rm kpc}$] & $M_{\rm 200}$ [${\rm M_\odot}$] & $r_{\rm 200}$ [${\rm kpc}$] & $r_{\rm h}$ [${\rm kpc}$] & $M_{\rm b}$ [${\rm M_\odot}$] & $r_{\rm b}$ [${\rm kpc}$] & \\
    \hline
    Disc-dominant & $4.1\times10^{10}$ & $3.0$ & $1.1\times10^{12}$  & $206$ & $20.6$  & $4.3\times10^{9}$  & $0.3$ & \\
    Bulge-dominant & $2.1\times10^{10}$ & $3.0$ & $1.8\times10^{11}$  & $114$ & $15$  & $4.3\times10^{10}$  & $0.45$ & \\
    \hline
  \end{tabular}
  $$ 
\end{table*}

\begin{figure}
	\includegraphics[bb=0 0 1626 1970,width=\hsize]{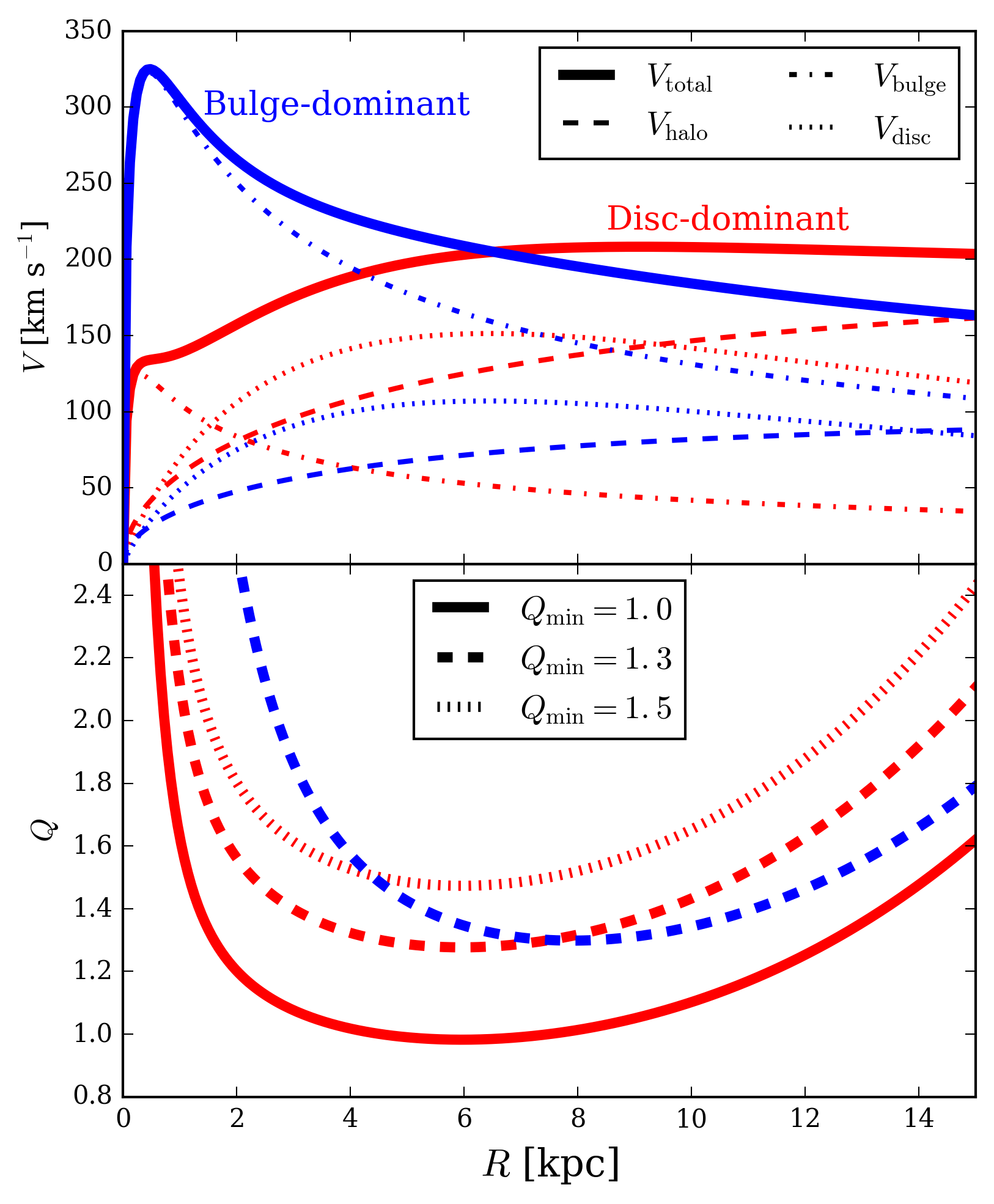}
	\caption{\textit{Top panel}: circular-velocity curves in the initial conditions of the disc-dominant (red lines) and the bulge-dominant (blues lines) models. The thick solid lines indicate circular velocity curves of the total mass distributions. The dashed, dot-dashed and dotted lines correspond to contributions from haloes, bulges and discs, respectively. \textit{Bottom panel}: Toomre's instability parameters $Q$ as functions of $R$ in the initial conditions with different $Q_{\rm min}$. The line colours are the same as in the top panel. The value of $Q$ is defined as $Q\equiv\sigma_R\kappa/(\pi G\Sigma_{\rm d})$, where $\sigma_R$ is radial component of velocity dispersion, $\kappa$ is epicyclic frequency measured from circular velocity curves, and $\Sigma_{\rm d}$ is the total surface density of the disc.}
	\label{VcircQ}
\end{figure}
We perform our simulations with two mass-distribution models whose structural parameters are listed in Table \ref{modellist}. Fig. \ref{VcircQ} illustrates circular-velocity curves in our galaxy models. As our fiducial model, the first one labelled as a `disc-dominant' model is approximately similar to the mass distributions of the Milky Way; this model has a relatively small bulge, and the circular velocity curve gently increases in inner radii of $R\lsim6~{\rm kpc}$ and becomes flat in outer radii. The disc component dominates the circular velocities in the range from $R\simeq2$ to $9~{\rm kpc}$. The second model labelled as `bulge-dominant' has, as the name suggests, a quite massive bulge, and the masses of the disc and the bulge are smaller than those of the fiducial model. The circular velocity curve has a strong peak at $R\simeq r_{\rm b}$ and continuously decreases with radius because of the massive bulge that dominates the circular velocities in the whole disc regions. 

\begin{figure*}
	\begin{minipage}{\hsize}
		\includegraphics[bb=0 0 952 443,width=\hsize]{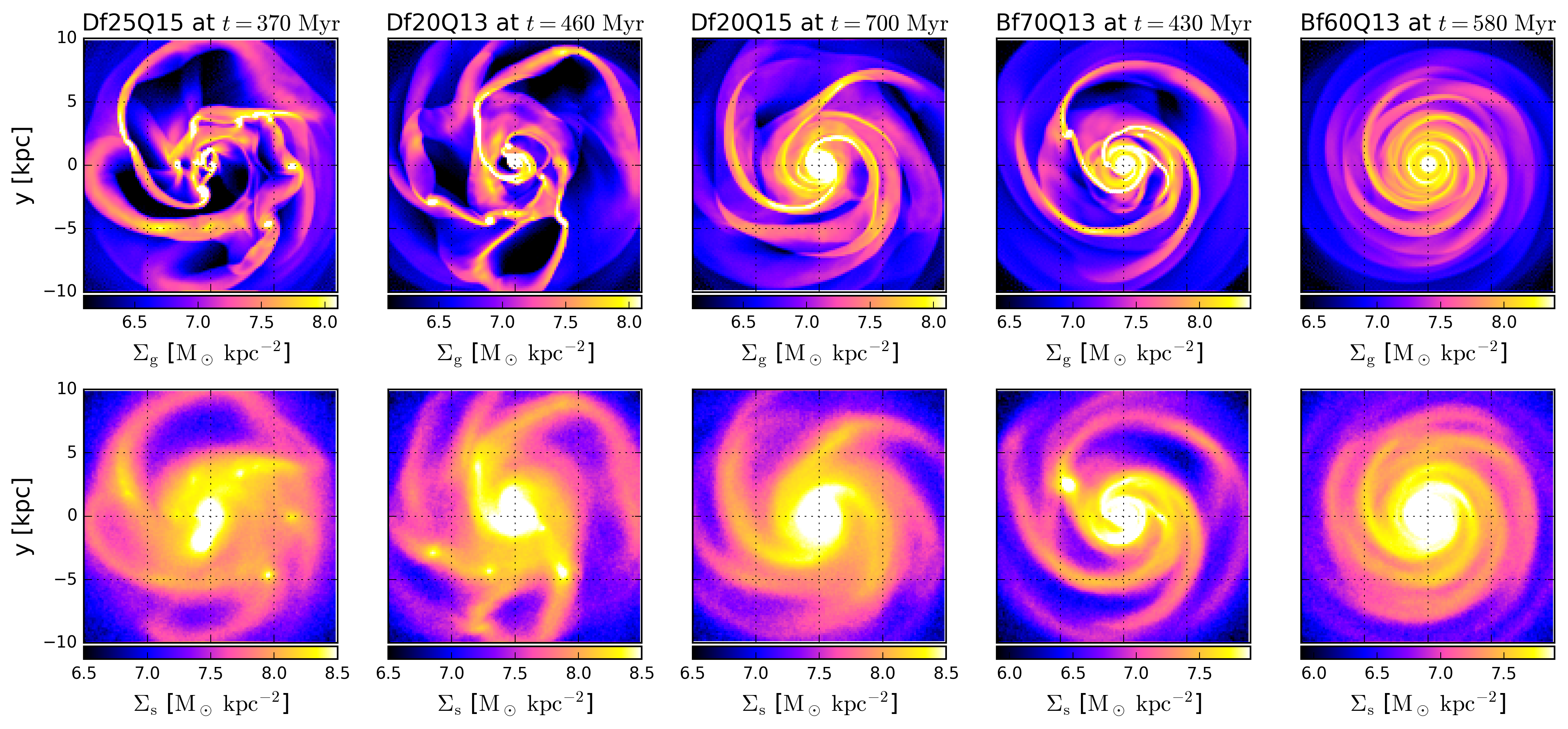}
		\caption{Face-on surface density distributions of gas (top) and stars (bottom) in the runs: Df25Q15 at $t=370~{\rm Myr}$, Df20Q13 at $t=460~{\rm Myr}$, Df20Q15 at $t=700~{\rm Myr}$, Bf70Q13 at $t=430~{\rm Myr}$ and Bf60Q13 at $t=580~{\rm Myr}$, from left to right.}
		\label{faces}
	\end{minipage}
\end{figure*}
\begin{figure}
	\includegraphics[bb=0 0 415 238,width=\hsize]{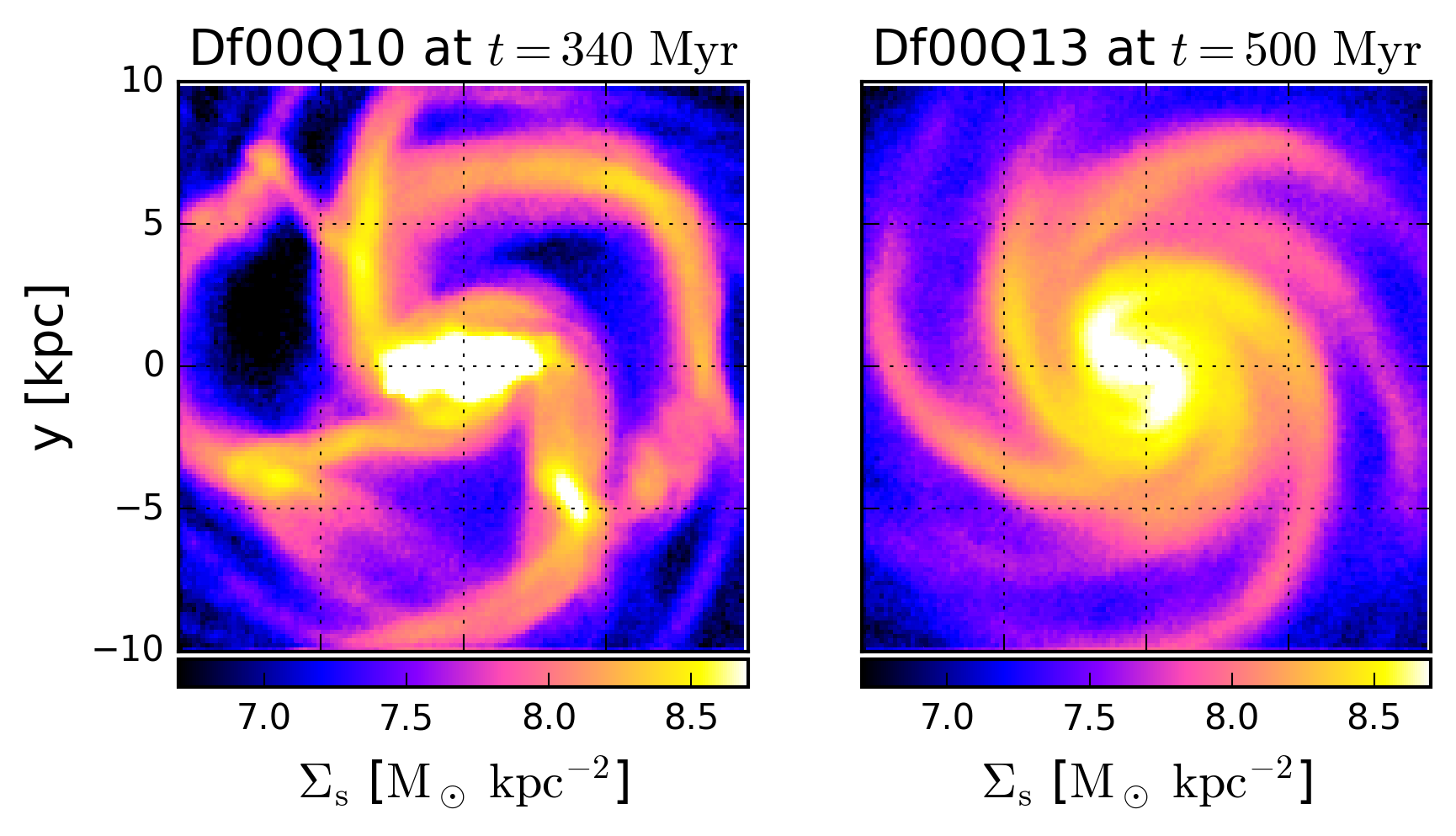}
	\caption{Same as the bottom panels in Fig. \ref{faces} but for the $N$-body runs: Df00Q10 at $t=340~{\rm Myr}$ (left) and Df00Q13 at $t=500~{\rm Myr}$ (right).}
	\label{facesNbody}
\end{figure}
In Figs. \ref{faces} and \ref{facesNbody}, we show surface density distributions of gas and stars in our galaxy models listed in Table \ref{modellist} after development of spiral arms. We find that some models are subject to SAI involving formation of clumpy structures (Df25Q15, Df20Q13, Df00Q10 and Bf70Q13), whereas the other model form stable and long-lived spiral arms without fragmenting into clumps (Df20Q15, Df00Q13 and Bf60Q13). In these unstable runs, some spiral arms break up into clumpy structures in their discs at early times $t\lsim300~{\rm Myr}$. In Df20Q15, although we see formation of only a few clumps at very late times $t\gsim900~{\rm Myr}$ (see Section \ref{nonlinear}), we consider that this model is stable against SAI until $t\simeq900~{\rm Myr}$. The two-component disc-dominant models, Df25Q15, Df20Q13 and Df00Q10, have similar initial conditions but slightly different in $Q_{\rm min}$ and $f_{\rm g}$ from one another, and they are used for our fiducial test of our theory (see Section \ref{unstable}, \ref{stable} and \ref{nonlinear}). The gas-less disc-dominant models Df00Q10 and Df00Q13 are aimed to examine the validity of the fluid approximation applied to the stellar dispersion relation (equation \ref{DRstar_prev}, see Section \ref{Nbodyruns}). The bulge-dominant models Df60Q13 and Df70Q13 have mass distributions largely different from the disc-dominant ones. Since their initial conditions have high $Q$ values in their inner disc regions, spiral-arm fragmentation only occurs in highly gas-rich runs. In these runs, we aim to prove general robustness of our analysis (see Section \ref{bulge}).

\subsection{Data analysis}
\label{ana}
To test our linear perturbation theory, we perform map-based two-dimensional analysis by utilizing snapshots of our simulations. First, we need to obtain $\Sigma$, $\sigma$, $\Omega$ and $W$ for each of gas and stellar component. In computing $\Sigma$, $\sigma$ and $\Omega$, we apply two-dimensional Gaussian kernels of the full width at half maximum $0.5~{\rm kpc}$ to gas and stellar distributions in the snapshots. Then, we vertically integrate the physical quantities weighted by mass and compute the mean and a dispersion of $v_{\rm \phi}$. As mentioned in Section \ref{basiceq}, $\sigma_{\rm g}^2\equiv c^2+\sigma_{\phi,{\rm g}}^2$ and $\sigma_{\rm s}^2\equiv\sigma_{\phi,{\rm s}}^2$, where $\sigma_{\rm \phi}$ is dispersion of $v_{\rm \phi}$. Angular velocity is computed as $\Omega=\overline{v_{\rm \phi}}/R$ for each of gas and stars. We make polar plots of $\Sigma$, $\sigma$ and $\Omega$ as functions of $(R,\phi)$.

It is not obvious or straightforward how to detect a spiral arm in the simulations and measure a half width $W$ of the spiral arm. For the sake of automatic detection of spiral arms, we perform one-dimensional Gaussian fitting along the radial direction at a given $\phi$ in the polar maps of $\Sigma_{\rm g}$ and $\Sigma_{\rm s}$. The Gaussian function is given as 
$\tilde{\Sigma}(R,\xi,\phi)=\Sigma(R,\phi)\exp[-\xi^2/2w^2]$, where $\xi$ represents radial offset from $R$, and $\Sigma(R,\phi)$ is an actual surface density in the simulation. The goodness-of-fit at $(R,\phi)$ is evaluated as
\begin{equation}
\chi^2(R,\phi) \equiv \frac{1}{2W}\int^W_{-W}\left[\frac{\tilde{\Sigma}(R, \xi,\phi) - \Sigma(R+\xi,\phi)}{\tilde{\Sigma}(R,\xi,\phi)}\right]^2~\textrm{d}\xi,
\label{Xi2}
\end{equation}
and we look for $W$ that gives the minimum value of $\chi^2(R,\phi)$; this $W$ is defined to be the half width of the spiral arm. If there is a ridge of a spiral arm at $R$ and the density distribution is nearly Gaussian, the fitting procedures return $\chi^2(R,\phi)$ significantly lower than unity. Thus, we compute $W_{\rm g}$, $W_{\rm s}$ and $\chi^2_{\rm g+s}\equiv\chi^2_{\rm g}+\chi^2_{\rm s}$ in all regions of the polar coordinates and define spiral arms to be regions where $\log\chi^2_{\rm g+s}<-0.25$. Although the threshold of $\chi^2_{\rm g+s}$ is an arbitrary value, the computations of quantities such as $S_{\rm 2}$, $\lambda_{\rm MU}$ and $t_{\rm g}$ are independent of the threshold. This definition of spiral arms requires both gas and stars to have their density peaks at similar radii. This means that if only either gas or stars has a spiral arm, our scheme regards that there is not an arm.\footnote{SAI could, however, occur even in such cases. We do not take into account this kind of instability caused solely by a gas or a stellar arm since we do not find this instability in our simulations.} Our measurement of $W$ assumes a pitch angle $\theta=0$ for an arm, therefore our scheme overestimates the true width by a factor of $1/\cos\theta$. If a pitch angle largely deviates from $\theta=0$, the tight-winding approximation assumed in our analysis may not hold.

\section{Results}
\label{result}
\subsection{Unstable cases}
\label{unstable}
As seen in the leftmost panels of Fig. \ref{faces}, the disc-dominant model Df25Q15 forms spiral arms unstable against fragmentation after a few dynamical time-scales from the beginning of the simulation. Although this model has the same $Q_{\rm min}$ as the stable disc-dominant model Df20Q15 in their initial states, the gas fraction of Df25Q15 is higher than that of the stable model by 0.05. Therefore, it appears that more gas-rich galaxies tend to be unstable against spiral-arm fragmentation, and the clump formation in Df25Q15 could be attributed to the higher gas fraction.

\begin{figure*}
	\begin{minipage}{\hsize}
		\includegraphics[bb=0 0 3842 3925,width=\hsize]{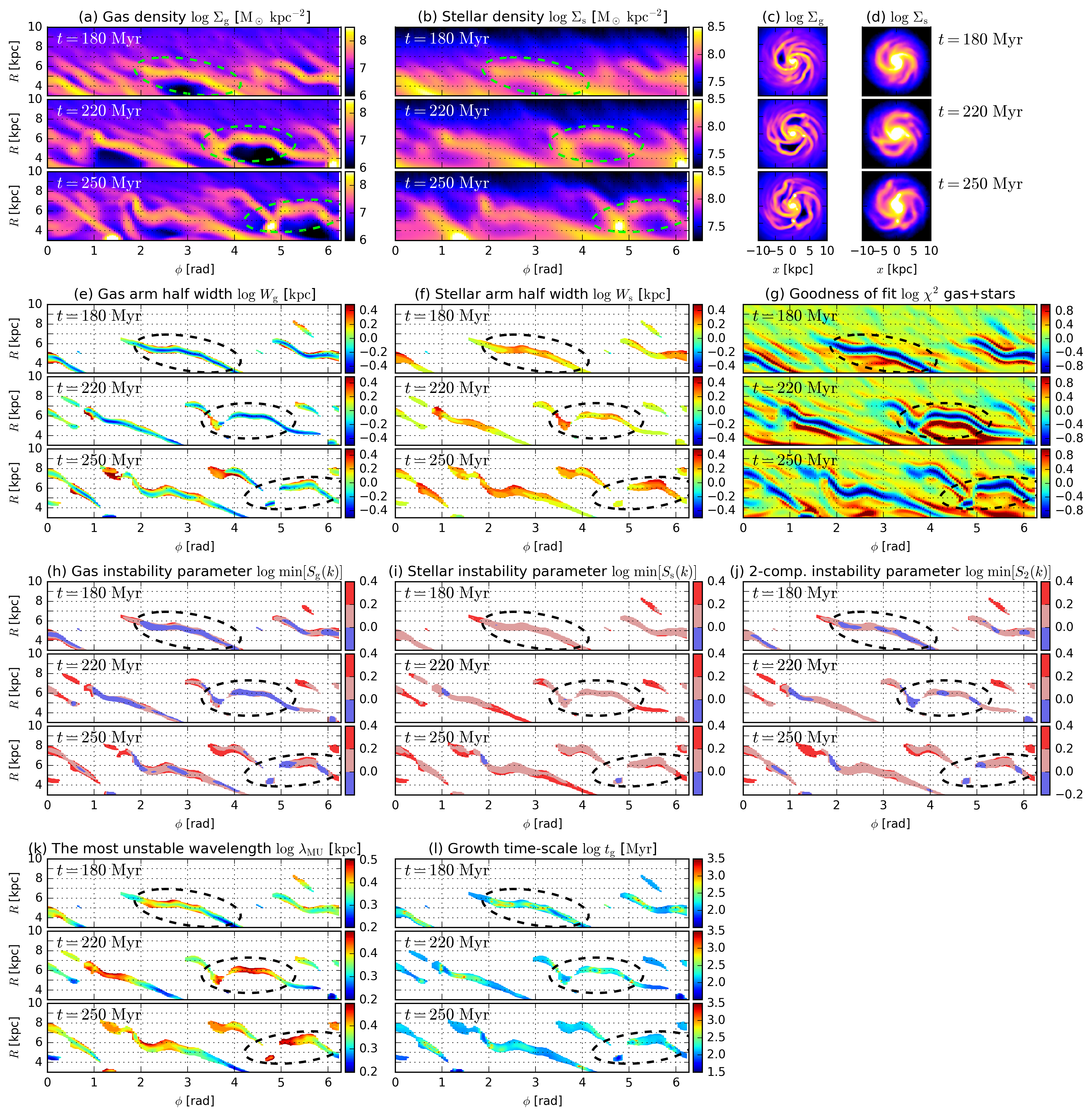}
		\caption{Polar-map analysis of the model Df25Q15 at $t=180$, $220$ and $250~{\rm Myr}$. \textit{Panels a and b}: surface density distributions of gas and stars in the polar coordinates. \textit{Panels c and d}: surface density distributions of gas and stars in the face-on Cartesian coordinates. \textit{Panels e and f}: half widths of gas and stellar spiral arms. \textit{Panel g}: goodness-of-fit $\chi^2_{\rm g+s}$ computed with equation (\ref{Xi2}) for the density distributions. \textit{Panels h and i}: the instability parameters $\min[S(k)]$ computed from the single-component analysis for each of the gas and stellar component, where unstable values $<1$ are indicated with blue colour. \textit{Panel j}: the two-component instability parameters $\min[S_{\rm 2}(k)]$ for the spiral arms. \textit{Panels k and l}: the wavelengths and the growth time-scales of the most unstable perturbations in the two-component analysis. In Panels e, f, h, i, j, k and l, the inter-arm regions where $\log\chi^2_{\rm g+s}>-0.25$ are uncoloured, according to our definition of spiral arms. The spiral-arm regions detected by the threshold of $\log\chi^2_{\rm g+s}<-0.25$ are consistent with high-density filamentary structures in Panels a and b. In Panels k and l, the values of $\lambda_{\rm MU}$ and $t_{\rm g}$ in stable regions with $\min[S_{\rm 2}(k)]>1$ correspond to wavelengths and oscillation time-scales of the perturbations with the lowest frequencies. The dashed ellipses in each panel trace the unstable spiral arm we focus on. The fragmenting arm marked with the ellipses have three segments indicating $\min[S_{\rm 2}(k)]<1$ whose sizes are larger than $\lambda_{\rm MU}$ at $t=180~{\rm Myr}$, and then the segments collapse into three clumps at $t=250~{\rm Myr}$. The estimated $t_{\rm g}$ in the segments is longer than the actual clump-formation time-scale by a factor of $2$--$4$.}
		\label{FullMaps}
	\end{minipage}
\end{figure*}
Fig. \ref{FullMaps} shows our polar-map analysis for the snapshots of the model Df25Q15 at $t=180$, $220$ and $250~{\rm Myr}$. In these snapshots, a spiral arm fragments into massive clumps, and the dashed ellipses in the figure chase the arm fragmenting. Panels a and b show gas and stellar surface densities in the polar coordinates. Although the arm has not fragmented yet at $t=180~{\rm Myr}$, the spiral arm breaks up into three clumps at $t=250~{\rm Myr}$. Since Panel g indicates low values of $\chi^2_{\rm g+s}$ along the spiral arm seen in the surface density distributions, our scheme appears to detect a spiral arm correctly. Panels e and f show half widths, $W_{\rm g}$ and $W_{\rm s}$, measured for the detected spiral arm. 

Panels h and i in Fig. \ref{FullMaps} show the single-component instability parameters, $\min[S_{\rm g}(k)]$ and $\min[S_{\rm s}(k)]$, computed using equation (\ref{Crit1C}) for each of the gas and stellar component. For the encircled spiral arm, the single-component analysis indicates that the gas is expected to be unstable since $\min[S_{\rm g}(k)]<1$ (blue colour) within the arm, whereas the stars in the arm are stable since $\min[S_{\rm s}(k)]>1$ (red colour). This means that the fragmenting instability of the spiral arm is dominated by gas. Panel j shows the two-component instability parameters $\min[S_{\rm 2}(k)]$ and indicates that the fragmenting spiral arm has three segments where $\min[S_{\rm 2}(k)]<1$ at $t=180~{\rm Myr}$. Panel k shows the most unstable wavelengths $\lambda_{\rm MU}$ estimated from the two-component analysis, which are $\lambda_{\rm MU}\simeq 2~{\rm kpc}$ and shorter than or comparable to the sizes of the segments indicating $\min[S_{\rm 2}(k)]<1$ within the unstable segments at $t=180~{\rm Myr}$, therefore these segments are expected to collapse. Panel l shows the growth time-scales of the most unstable perturbations, which are $t_{\rm g}\simeq 100$--$300~{\rm Myr}$ before the fragmentation (at $t=180~{\rm Myr}$). As seen in Panels a and b, the unstable spiral arm actually takes $\simeq70~{\rm Myr}$ to form the clumps, which is about $2$--$4$ times shorter than the predicted growth time-scale $t_{\rm g}$. However, note that the clump formation time-scale in the simulation is a crude estimate from our visual inspection, which is defined as a time elapsed since a region with $\min[S_{\rm 2}(k)]<1$ appears to when a clump forms clearly in the unstable arm.

The linear perturbation analysis described in Section 2 is cased on several assumptions. In Appendix \ref{assumptionvalidity}, we discuss the validity of the assumptions of linearity of the initial density perturbations and the rigid rotations within spiral arms.

Another unstable model Df20Q13 also forms some massive clumps via SAI (see Fig. \ref{faces}). This model has the same gas fraction and mass distributions as the stable model Df20Q15 but a lower $Q_{\rm min}$. Hence, the fragmentation in Df20Q13 is attributed to the lower $Q_{\rm min}$ in the initial condition.

\begin{figure*}
	\begin{minipage}{\hsize}
		\includegraphics[bb=0 0 3807 2475,width=\hsize]{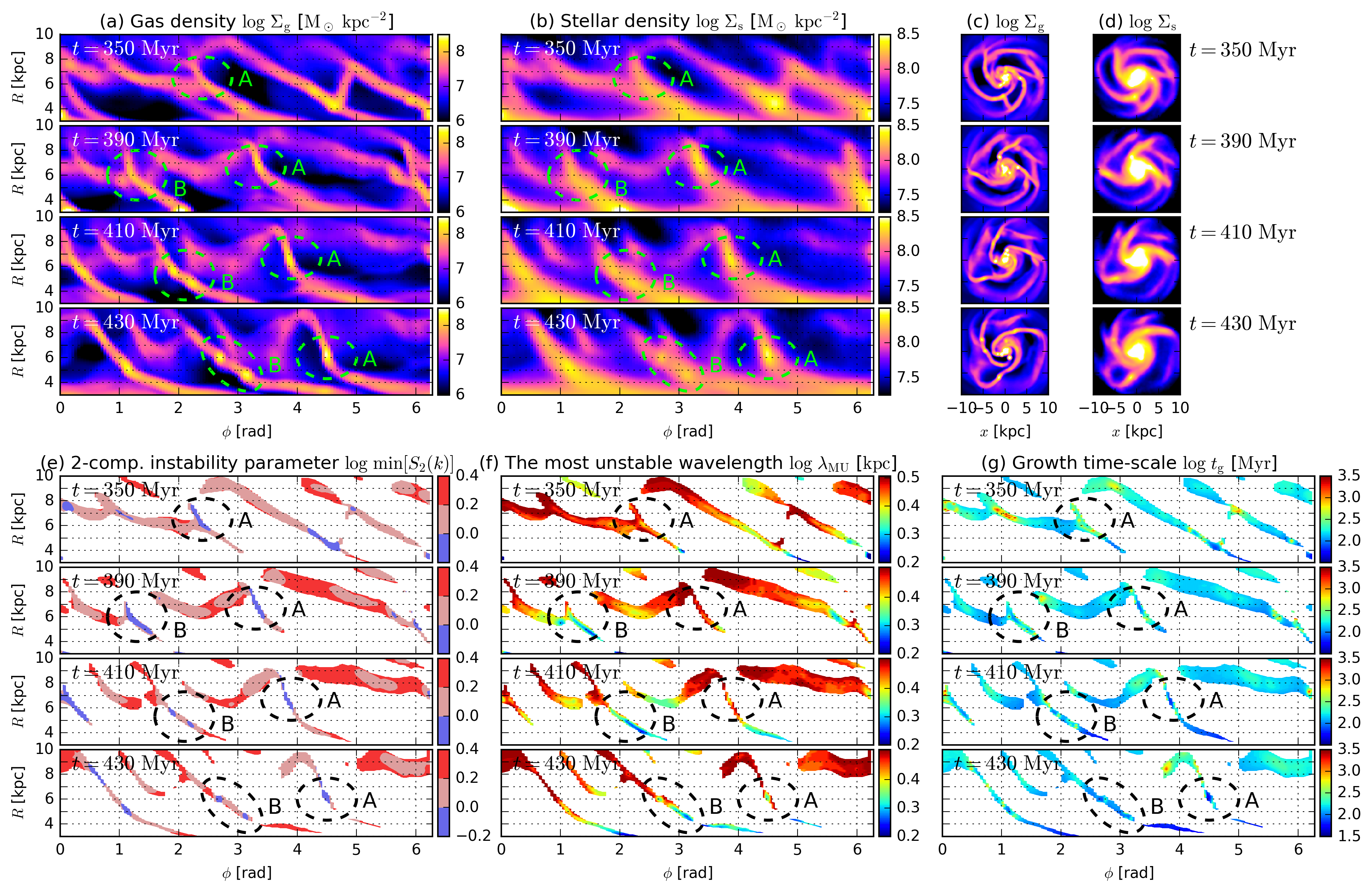}
		\caption{Same as Fig. \ref{FullMaps} but for the run Df20Q13 at $t=350$, $390$ $410$ and $430~{\rm Myr}$. Results of $W_{\rm g}$, $W_{\rm s}$ and $\chi^2_{\rm g+s}$ are not shown in this figure. In the bottom panels, the inter-arm regions where $\log\chi^2_{\rm g+s}>-0.25$ are uncoloured. The dashed ellipses labelled as `A' and `B' in each panel trace two fragmenting structures. The spiral arms in Regions A and B start indicating $\min[S_{\rm 2}(k)]<1$ at $t=350$ and $390~{\rm Myr}$, and they fragment into one and two clumps, respectively, at $t=430~{\rm Myr}$. The estimated $t_{\rm g}\simeq 200$ and $100~{\rm Myr}$ within the segments in Regions A and B before their fragmentation, are longer than actual clump formation time-scales by a factor of $2$--$3$.}
		\label{ReduceMap1}
	\end{minipage}
\end{figure*}
Fig. \ref{ReduceMap1} shows the results of our polar-map analysis for Df20Q13 at $t=350$, $390$, $410$ and $430~{\rm Myr}$. In these snapshots, we find two examples of spiral-arm fragmentation, which are labelled as Region A and B in Fig. \ref{ReduceMap1} (dashed ellipses). The spiral arm in Region A starts indicating $\min[S_{\rm 2}(k)]<1$ (blue colour in Panel e) at $t=350~{\rm Myr}$, but it has not fragmented yet. It appears that the low values of $\min[S_{\rm 2}(k)]$ in Region A are driven by interaction with a bifurcated spiral arm and/or by bending of the arm. Eventually, the spiral arm in Region A collapses into a single massive clump about $80~{\rm Myr}$ later (at $t=430~{\rm Myr}$). In Panels f and g at $t=350~{\rm Myr}$, the most unstable wavelengths $\lambda_{\rm MU}\simeq3~{\rm kpc}$ in the fragmenting arm are comparable to the size of the segment with $\min[S_{\rm 2}(k)]<1$ in Region A. The predicted growth time-scale $t_{\rm g}\simeq200~{\rm Myr}$ is a few times longer than the actual formation time-scale of the clump in Region A  (about $80~{\rm Myr}$). Meanwhile, Region B begins to indicate $\min[S_{\rm 2}(k)]<1$ at $t=390~{\rm Myr}$, where the low $\min[S_{\rm 2}(k)]$ seems to be induced by interactions with neighbouring arms. The unstable segment in Region B forms two clumps about $40~{\rm Myr}$ later (at $t=430~{\rm Myr}$). From the two-component analysis, the most unstable wavelengths and the growth time-scales are estimated to be $\lambda_{\rm MU}\simeq1$--$2~{\rm kpc}$ and $t_{\rm g}\simeq100~{\rm Myr}$ from our analysis. The wavelengths $\lambda_{\rm MU}$ are shorter than the size of the segment with $\min[S_{\rm 2}(k)]<1$ in Region B before the fragmentation, whereas the predicted growth time-scales $t_{\rm g}$ are longer than the actual collapse time-scale only by a factor of $2$--$3$.

As we showed above, our linear perturbation analysis is able to predict well spiral-arm fragmentation in our simulations of disc-dominant galaxy models shown above although it appears that the growth time-scales are overestimated approximately by a factor of $2$--$4$. Especially, the instability condition of $\min[S_{\rm 2}(k)]<1$ appears to be remarkably robust in our simulations. Note that we find a few cases of non-linear fragmentation with $\min[S_{\rm 2}(k)]>1$ in Df20Q15 (see Section \ref{nonlinear}) although such cases are quite rare in our simulations. In Toomre's instability analysis, although the instability condition against radial perturbations in a disc is analytically derived to be $Q<1$ for a razor-thin disc, previous studies using $N$-body simulations of disc galaxies have demonstrated that the discs are actually somewhat unstable and can form spiral arms until $Q\lsim1.7$ due to non-linear and/or global instability \citep[e.g.][]{t:64,bt:08,fbs:11,mk:14}\footnote{\citet{hs:15} proposed that their thin discs can be unstable until $Q<1.3$ using $N$-body simulations with resolutions higher than the other studies.}. If the formation of spiral arms is attributed to Toomre instability, the range of $1<Q\lsim1.7$ can be considered as a non-linear regime of the disc instability, in which discs can be unstable without satisfying the Toomre's instability condition. However, in our SAI analysis, our instability condition of $\min[S_{\rm 2}(k)]<1$ appears to be quite accurate to predict spiral-arm fragmentation, and there seems to be little non-linear regime of SAI, at least in our simulated models shown above.

\subsubsection{Prediction of clump masses}
\label{clumpmass}
\begin{figure}
	\includegraphics[bb=0 0 1359 1925,width=\hsize]{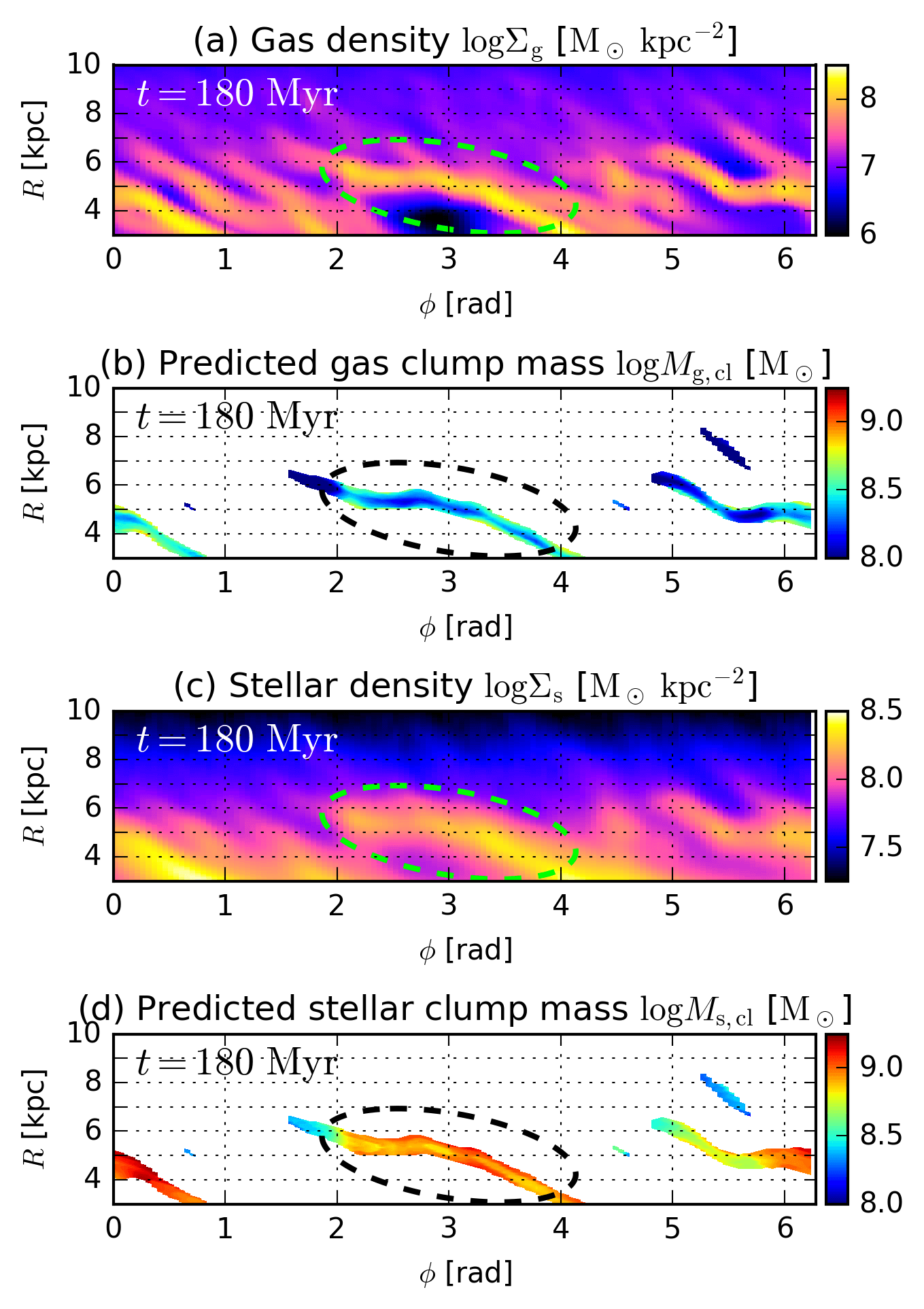}
	\caption{Surface density distributions and predicted clump masses in the snapshot of the model Df25Q15 at $t=180~{\rm Myr}$. \textit{Panel a}: gas density. \textit{Panel b}: predicted clump masses of gas obtained from the polar-map analysis in Fig. \ref{FullMaps} as $M_{\rm g,cl}\sim1.4W_{\rm g}\Sigma_{\rm g}\lambda_{\rm MU}$. \textit{Panel c}: stellar density. \textit{Panel d}: predicted clump masses of stars estimated as $M_{\rm s,cl}\sim1.4W_{\rm s}\Sigma_{\rm s}\lambda_{\rm MU}$. In the fragmenting arm marked with the ellipses, our analysis predicts $M_{\rm g,cl}\simeq10^{8}$--$10^{8.5}~{\rm M_\odot}$ and $M_{\rm s,cl}\simeq10^{8.7}$--$10^{9}~{\rm M_\odot}$.}
	\label{ClumpMassMap}
\end{figure}
Our analysis can also predict a mass of a clump forming via SAI. For the fragmenting arm in the run of Df25Q15 (Fig. \ref{FullMaps}), the most unstable wavelength $\lambda_{\rm MU}\simeq 2~{\rm kpc}$ is longer than and comparable to the widths of gas and stellar arms, $2W_{\rm g}\simeq 1~{\rm kpc}$ and $2W_{\rm s}\simeq 2~{\rm kpc}$, before the fragmentation (at $t=180~{\rm Myr}$; see Panels e and f in Fig. \ref{FullMaps}). Therefore, the unstable mode can be expected to collapse along the spiral arm: i.e. one-dimensional collapse.\footnote{Since $2W_{\rm s}\simeq\lambda_{\rm MU}$, the mode of collapse can become two-dimensional, rather than one-dimensional, for stars. When $2W\simeq\lambda_{\rm MU}$, however, the predicted clump masses are nearly the same between the two modes (see Fig. \ref{AnaObs} and Section \ref{scalingrelation})} If we assume that gas and stars included within $\lambda_{\rm MU}$ collapse into a single clump in an unstable arm and that line-mass $\Upsilon$ is approximately constant along the arm within $\lambda_{\rm MU}$, then the gas and stellar masses within the collapsing clump are estimated to be $M_{\rm g,cl}\sim\Upsilon_{\rm g}\lambda_{\rm MU}\simeq1.4W_{\rm g}\Sigma_{\rm g}\lambda_{\rm MU}$ and $M_{\rm s,cl}\sim1.4W_{\rm s}\Sigma_{\rm s}\lambda_{\rm MU}$, respectively. In Fig. \ref{ClumpMassMap}, Panels b and d show $M_{\rm g,cl}$ and $M_{\rm s,cl}$ predicted from our polar-map analysis at $t=180~{\rm Myr}$ in Df25Q15; $M_{\rm g,cl}\simeq10^{8}$--$10^{8.5}~{\rm M_\odot}$ and $M_{\rm s,cl}\simeq10^{8.7}$--$10^{9}~{\rm M_\odot}$ in the unstable arm. As we showed in Fig. \ref{FullMaps}, the unstable spiral arm has the three segments with $\min[S_{\rm 2}(k)]<1$ and indeed fragments into the three clumps. Panels b and d in Fig, \ref{ClumpMassMap} indicate that the predicted clump masses do not vary largely along the arm. So, all of the resultant clumps are expected to have similar masses.

\begin{figure}
	\includegraphics[bb=0 0 2192 1982,width=\hsize]{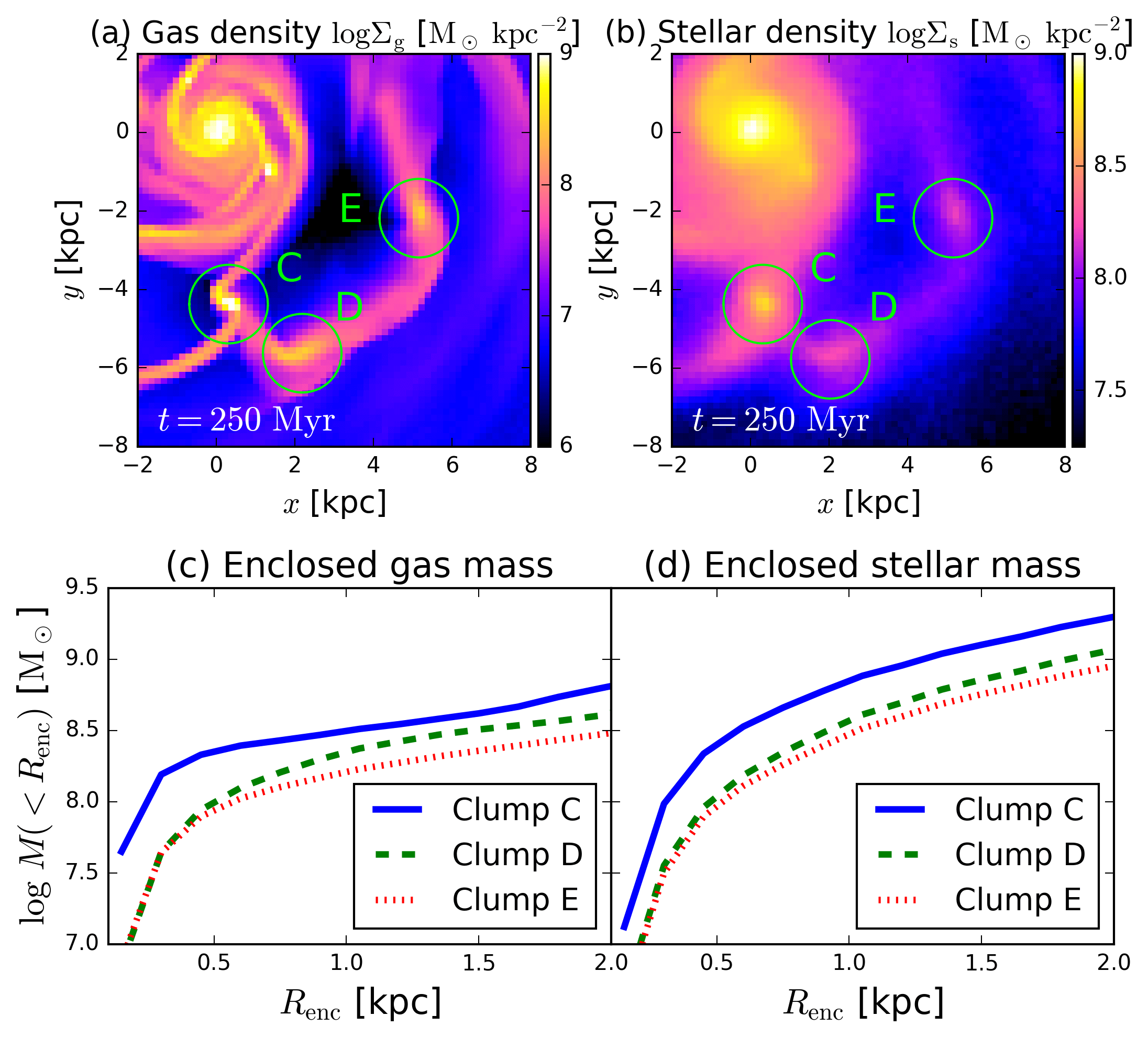}
	\caption{\textit{Panels a and b}: face-on surface density distributions of gas and stars in the model Df25Q15 after clump formation ($t=250~{\rm Myr}$). The three clumps are marked with the green circles. \textit{Panels c and d}: gas and stellar masses cylindrically enclosed within a distance from the clump centre for each clump. The clump centre is defined to be the position of the highest surface density of each clump. As a reference, the sizes of the green circles in the top panels correspond to $R_{\rm enc}=1~{\rm kpc}$ from the clump centres. The three clumps have gas and stellar masses of $M_{\rm g}\simeq M_{\rm s}\simeq10^{8}$--$10^{8.5}~{\rm M_\odot}$ within $R_{\rm enc}=0.5~{\rm kpc}$.}
	\label{EncMass}
\end{figure}
In Fig. \ref{EncMass}, Panels a and b show face-on surface densities of gas and stars around the clumps after their formation (at $t=250~{\rm Myr}$). The green circles in the panels illustrate the positions of the three clumps; they are labelled as Clump C, D and E. Panels c and d show gas and stellar masses enclosed within distance $R_{\rm enc}$ from the centre of each clump. Although Clump C has the largest masses in gas and stars, the differences from the other clumps are within a factor of $\simeq3$. Their enclosed masses of gas are nearly constant in $R_{\rm enc}\gsim0.5~{\rm kpc}$, whereas their stellar masses still increase gently outside $R_{\rm enc}=0.5~{\rm kpc}$. If we define the clump radii to be $R_{\rm cl}=0.5~{\rm kpc}$, we can estimate that the simulated clumps have their gas and stellar masses $M_{\rm g}(<0.5~{\rm kpc})\simeq M_{\rm s}(<0.5~{\rm kpc})\simeq10^{8}$--$10^{8.5}~{\rm M_\odot}$. If $R_{\rm cl}=1~{\rm kpc}$, $M_{\rm s}(<1~{\rm kpc})\simeq10^{8.5}$--$10^{9}~{\rm M_\odot}$. Hence, the masses of these simulated clumps are approximately consistent with the predicted values from our analysis shown in Panels b and d of Fig. \ref{ClumpMassMap}. We further discuss the analytical prediction of properties and scaling relations of clumps forming via SAI in Section \ref{scalingrelation}.

\subsection{A stable case}
\label{stable}
To demonstrate further the robustness of our instability condition of  $\min[S_{\rm 2}(k)]<1$, we examine the stable model Df20Q15 in which neither spiral-arm fragmentation nor clump formation occurs (see Fig. \ref{faces}), except in a very late phase of the run. From the stable states of arms in this run, the spiral arms are expected to have $S_{\rm 2}>1$ for all $k$. 

\begin{figure}
	\includegraphics[bb=0 0 1675 3925,width=\hsize]{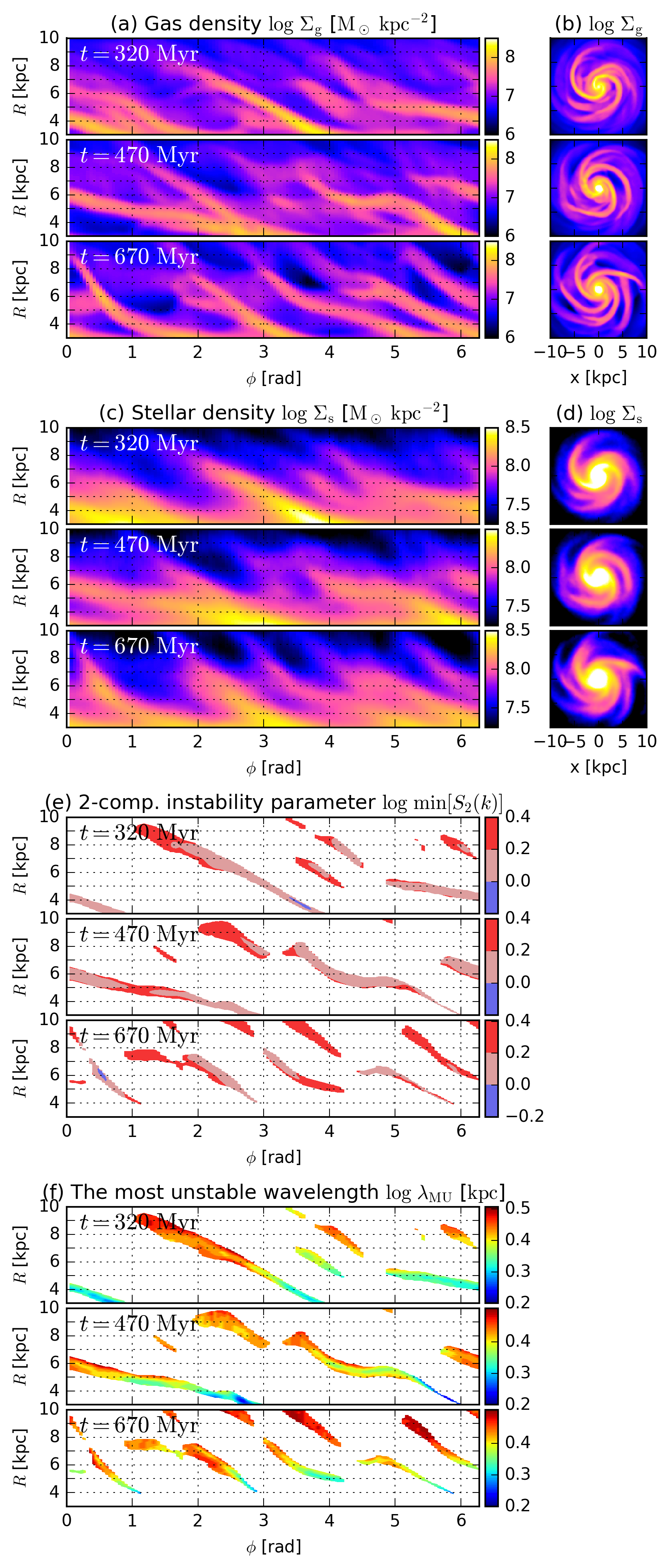}
	\caption{Polar-map analysis for the stable model Df20Q15 at $t=320$, $470$ and $670~{\rm Myr}$. In this run, no spiral arms fragment to form clumps until $t\simeq900~{\rm Myr}$, and Panel e indicates $\min[S_{\rm 2}(k)]>1$. There are small segments with $\min[S_{\rm 2}(k)]<1$ at $(R,\phi)=(4~{\rm kpc}, 3.5~{\rm rad})$ at $t=320~{\rm Myr}$ and $(6~{\rm kpc}, 0.5~{\rm rad})$ at $t=670~{\rm Myr}$; however these segments with $\min[S_{\rm 2}(k)]<1$ do not fragment in this run since $\lambda_{\rm MU}$ is larger than the sizes of these segments.}
	\label{StableMap1}
\end{figure}
Fig. \ref{StableMap1} shows our polar-map analysis for the model Df20Q15 at $t=320$, $470$ and $670~{\rm Myr}$. Because the time-intervals between the snapshots in this figure are longer than the dynamical time-scale of the galaxy, these snapshots are inconsecutive and can be considered to show independent dynamical states of spiral arms. In all snapshots, the surface density maps show that no spiral arms fragment, and no clumps form (Panels a and b). Consistently with the stable appearance of the spiral arms, the instability parameters indicate high values of $\min[S_{\rm 2}(k)]>1$ in the arms (Panel e). Although low values of $\min[S_{\rm 2}(k)]<1$ can actually be seen in very small segments at $(R,\phi)=(4~{\rm kpc}, 3.5~{\rm rad})$ at $t=320~{\rm Myr}$ and $(6~{\rm kpc}, 0.5~{\rm rad})$ at $t=670~{\rm Myr}$, we find that the sizes of the segments indicating $\min[S_{\rm 2}(k)]<1$ are significantly smaller than their most unstable wavelengths (Panel f). Because the segments cannot settle the perturbations whose wavelengths are larger than their own sizes, the unstable modes of $\lambda_{\rm MU}$ cannot exist in these segments with $\min[S_{\rm 2}(k)]<1$.\footnote{Even if the most unstable wavelength is larger than the size of segment indicating $\min[S_{\rm 2}(k)]<1$, it is still possible that the shortest unstable wavelength is smaller than the segment and can collapses.} Indeed, these segments do not fragment and disappear within $<30~{\rm Myr}$. Thus, the spiral arms in the run Df20Q15 are marginally stable, which can be explained by our linear perturbation analysis indicating high values of $\min[S_{\rm 2}(k)]>1$ or too long unstable wavelengths. However, we find a few cases of non-linear fragmentation with $\min[S_{\rm 2}(k)]>1$ at very late times in this run (see below).

\subsection{Non-linear fragmentation induced by interactions}
\label{nonlinear}
Our analysis is able to characterise well spiral-arm fragmentation and clump formation following. It should be recalled, however, that our theory is based on the linearised local perturbation equations (\ref{linearlized1}, \ref{linearlized2} and \ref{phimom}). Gravitational instability of a spiral arm may actually be induced by non-linear and/or global effects. In addition, although our linear perturbation analysis assumes an equilibrium state for a spiral arm, spiral arms in our simulations are actually not static. In reality, spiral arms can snake, merge with one another and be torn by sharing rotation, especially in gas-rich and/or kinematically cold discs. In such cases, the linear approximation in our analysis might not be accurate enough to describe the evolution of perturbations; moreover the tight-winding approximation might no longer be valid. If it is the case, our instability parameter $S_{\rm 2}$ may fail to predict spiral-arm fragmentation and clump formation.

\begin{figure}
	\includegraphics[bb=0 50 1648 3650,width=\hsize]{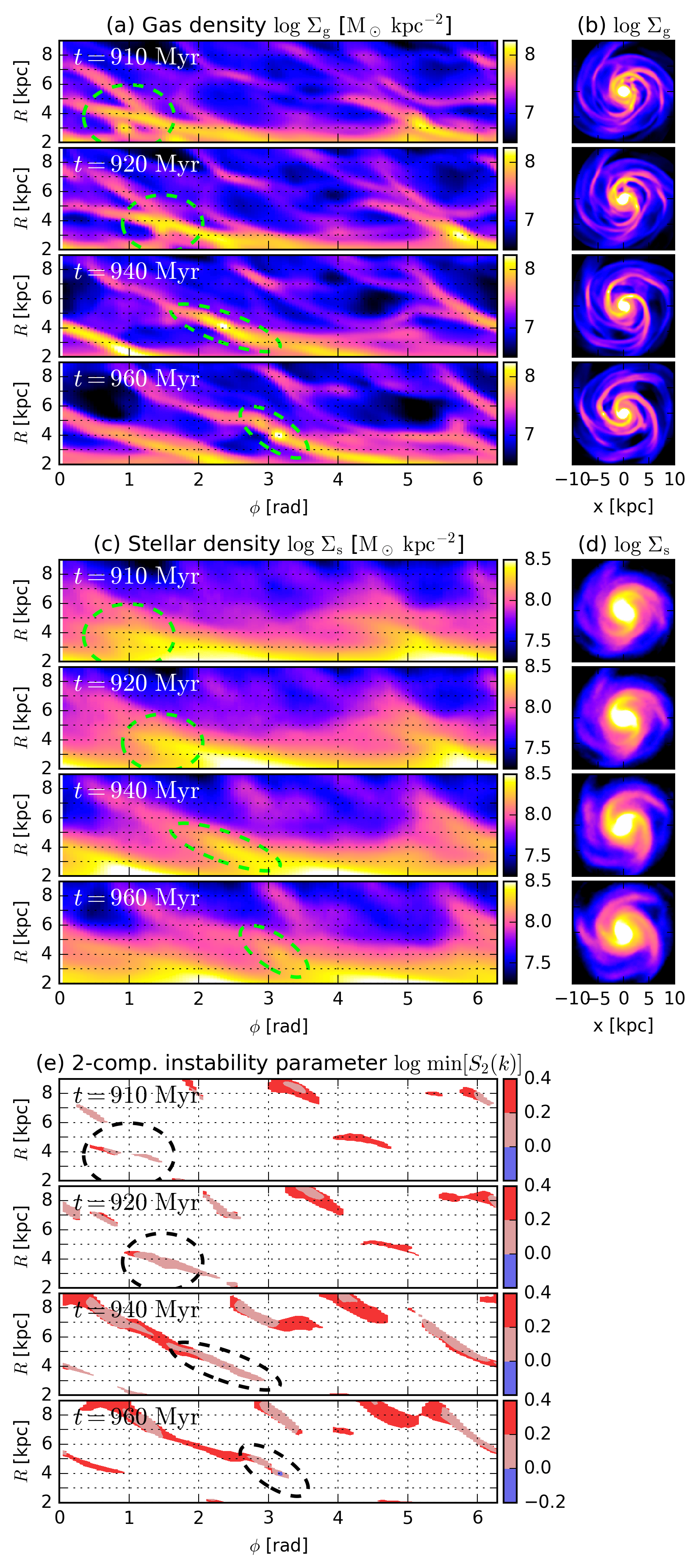}
	\caption{Polar-map analysis for the stable model Df20Q15 at $t=910$, $920$, $940$ and $960~{\rm Myr}$. The dashed ellipses encircle a clump forming in the snapshots, where two arms and a gas cloud merge into a single arm from $t=910$ to $940~{\rm Myr}$, and then the clump forms at $t=960~{\rm Myr}$ in spite of $\min[S_{\rm 2}(k)]>1$ in the fragmenting arm.}
	\label{NonlinearMap}
\end{figure}
As we showed in Section \ref{stable}, the model Df20Q15 is stable for spiral-arm fragmentation until $t\simeq900~{\rm Myr}$, and the spiral arms generally indicate $\min[S_{\rm 2}(k)]>1$. However, we find a few cases of clump formation with $\min[S_{\rm 2}(k)]$ higher than unity at late times $t\gsim900~{\rm Myr}$ in this simulation. Fig. \ref{NonlinearMap} shows polar maps of $\Sigma_{\rm g}$, $\Sigma_{\rm s}$ and $\min[S_{\rm 2}(k)]$ at $t=910$, $920$, $940$ and $960~{\rm Myr}$ in the run Df20Q15. The non-linear clump formation is highlighted with dashed ellipses in the figure, and the clump can be seen clearer in the gas density maps. In the gas density maps (Panel a) at $t=910$ and $920~{\rm Myr}$, the clump appears to start forming with interaction between two arms and a gas cloud; the dashed ellipses encircle these structures. Then, at $t=940~{\rm Myr}$, these structures merge into a single spiral arm that still have smooth density distribution. Eventually, the merged spiral arm forms a clumpy structure within the arm at $t=960~{\rm Myr}$. Panel e shows, however, that $\min[S_{\rm 2}(k)]$ remains higher than unity during the interaction and the clump formation. Hence, in this case, it appears that our analysis fails to predict the actual fragmentation shown in the figure. We find that such a non-linear fragmentation of a spiral arm only occurs in late times, $t\gsim900~{\rm Myr}$, in the run Df20Q15. 

As seen in the gas density maps at $t=910$ and $920~{\rm Myr}$ of Fig. \ref{NonlinearMap}, the non-linear fragmentation may be induced by mergers between the spiral arms and/or the gas cloud. However, even though interactions and mergers between neighbouring arms occur frequently in the unstable runs such as Df25Q15 and Df20Q13, our analysis using $S_{\rm 2}$ can predict fragmentation quite well in those runs. Besides the merger-induced fragmentation, since the spiral arms are relatively long-lived in this stable run, perturbations that are presumed to be stable in the linear analysis might grow slowly in non-linear regime. 

\subsection{Fragmentation in the $N$-body disc models}
\label{Nbodyruns}
In equations (\ref{DRgas_prev}) and (\ref{DRstar_prev}), our analysis assumes the same form of dispersion relation for gas and stars: the fluid approximation for stars. However, the stellar component does not necessarily follow the same dispersion relation because of the collisionless nature. Stellar orbits in a disc potential generally have epicyclic motions which enable stars to periodically leave and return to a spiral arm. This effect may become important for the perturbations whose wavelengths are smaller than the typical amplitude of epicyclic motions of stars \citep[e.g.][]{t:64,r:01}.\footnote{The influence by epicyclic motions is estimated to be $\lsim10$ per cent in calculating Toomre's instability parameter $Q$ if a Schwartzchild distribution function is assumed for a stellar disc \citep{t:64,bt:08,e:11}.} Hence, it is important to examine whether our analysis can characterise spiral-arm fragmentation in $N$-body models. In addition, comparison between models with and without gas may give us clues to understand why gas-rich and -poor galaxies tend to form giant clumps and spiral arms.

The gas-less disc models are generally more stable against SAI even though the initial mass distributions of the galactic structures are the same. As seen in Fig. \ref{faces}, our disc models with $f_{\rm g}=0.2$ become unstable for spiral-arm fragmentation when $Q_{\rm min}\lsim1.3$, whereas the model with $f_{\rm g}=0$ is still stable even though $Q_{\rm min}=1.3$ in the initial state (Df00Q13). However, stellar spiral arms in the $N$-body disc can fragment if $Q_{\rm min}\lsim1.0$ (Df00Q10), which form relatively extended stellar structures resembling clumps.

\begin{figure}
	\includegraphics[bb=0 50 1675 3925,width=\hsize]{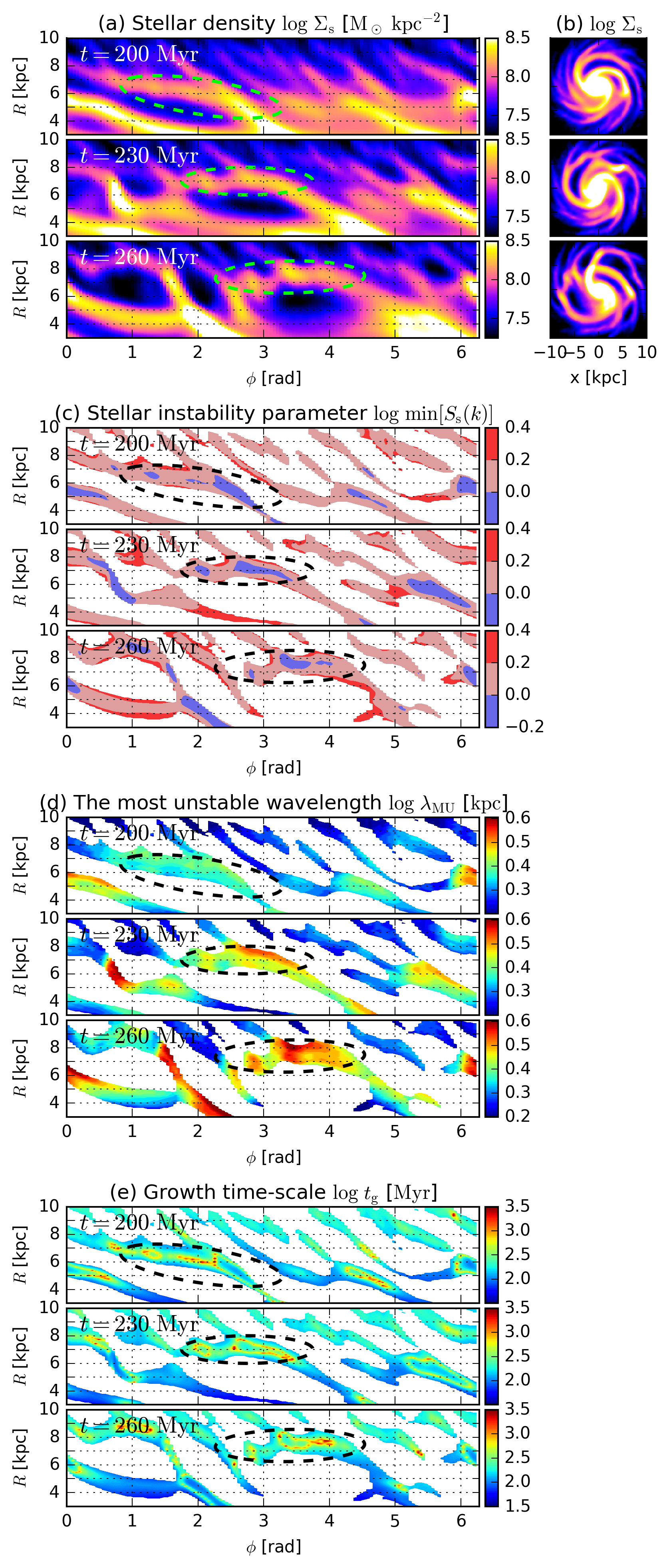}
	\caption{Polar-map analysis for the model Df00Q10 at $t=200$, $230$ and $260~{\rm Myr}$. The dashed ellipses encircle a fragmenting arm, within which some segments indicate $\min[S_{\rm s}(k)]<1$.}
	\label{UnstableNbodyMaps}
\end{figure}
\begin{figure}
	\includegraphics[bb=0 0 1675 1975,width=\hsize]{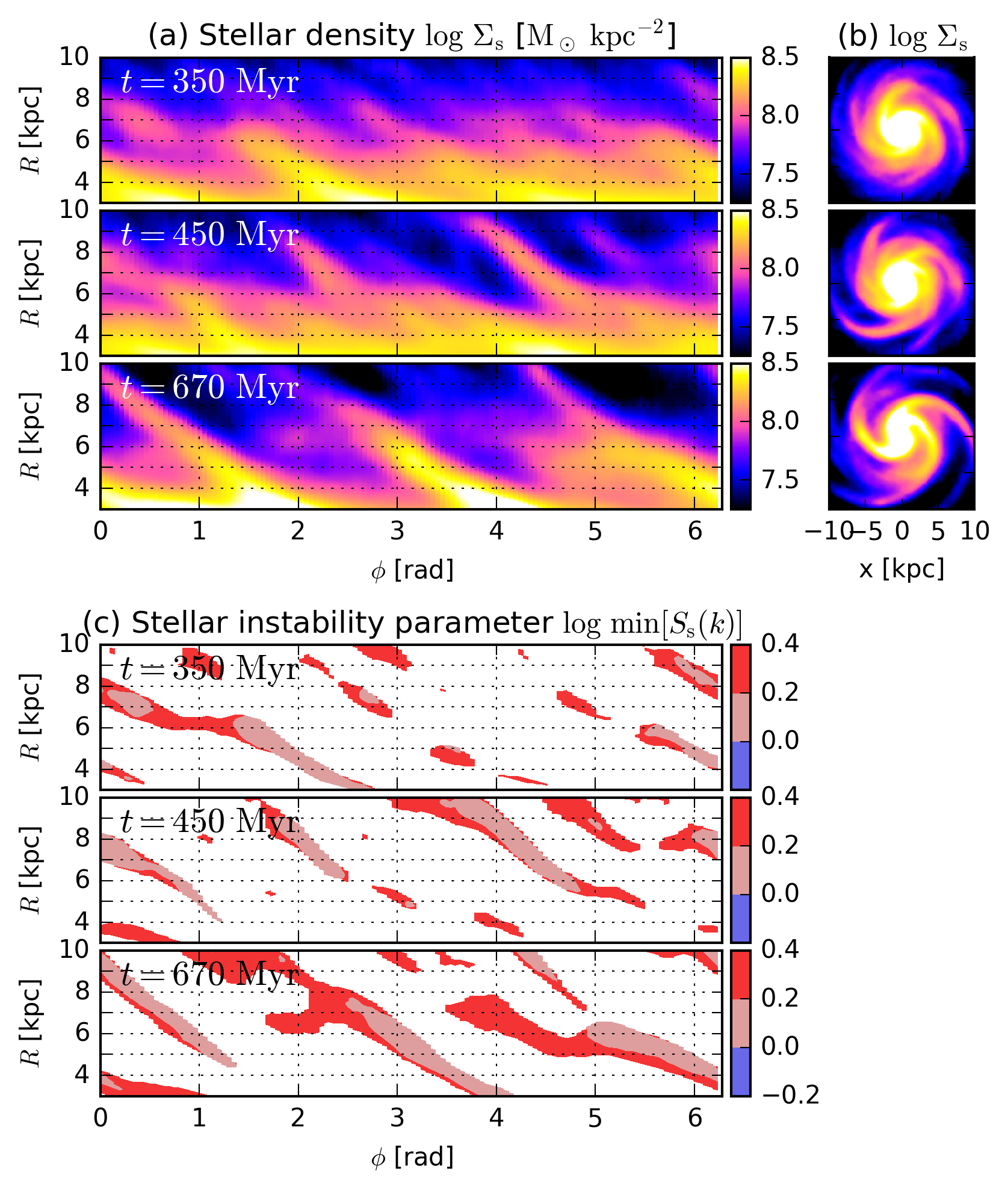}
	\caption{Polar-map analysis for the model Df00Q13 at $t=350$, $450$ and $670~{\rm Myr}$. No spiral arms fragment in this run, and Panel c indicates $\min[S_{\rm s}(k)]>1$ in the arms.}
	\label{StableNbodyMaps}
\end{figure}
Now that we focus on purely stellar discs, we use the single-component analysis (equations \ref{DL1}, \ref{DL11} and \ref{Crit1C}) applied to a stellar component to compute $S_{\rm s}$, $\lambda_{\rm MU}$ and $t_{\rm g}$. Stellar spiral arms are defined to be regions where $\log\chi^2_{\rm s}<-0.25$ with the same Gaussian fitting scheme. Fig. \ref{UnstableNbodyMaps} shows snapshots where spiral-arm fragmentation occurs in the model Df00Q10. In Panel a, the spiral arm marked with the dashed ellipse at $t=200~{\rm Myr}$ fragment into two (or three) clumpy structures at $t=260~{\rm Myr}$. In Panel c, the encircled spiral arm indicates $\min[S_{\rm s}(k)]<1$ within a few segments at $t=200~{\rm Myr}$. In Panel d, $\lambda_{\rm MU}\simeq2~{\rm kpc}$ in the arm before the fragmentation ($t=200~{\rm Myr}$). Panel e shows $t_{\rm g}\simeq100$--$200~{\rm Myr}$ in the unstable segments at $t=200~{\rm Myr}$. The estimated growth time-scales $t_{\rm g}$ are longer than the actual clump-formation time-scale in this simulation by a factor of $2$--$4$. Fig. \ref{StableNbodyMaps} shows the stable model Df00Q13 where no stellar arms fragment (Panels a and b). Panel c indicates $\min[S_{\rm s}(k)]>1$ in all spiral arms in the snapshots. The high values of $\min[S_{\rm s}(k)]$ are consistent with the stability of the stellar arms in this $N$-body run.

As shown above, our analysis appears to predict well the fragmentation of stellar arms in the $N$-body runs, as well as in the two-component models. This result means that adopting the fluid approximation to the stellar dispersion relation does not deteriorate the accuracy of our analysis too much at least in our models. It is also inferred that a stellar spiral arm would follow a dispersion relation similar to that of gas. This could be because $\lambda_{\rm MU}$ is typically longer than amplitude of epicyclic motion of a star in a spiral arm.

The $N$-body model Df00Q13 has the same mass-distribution model and $Q_{\rm ini}$ as the two-component model Df20Q13 in their initial conditions; however spiral arms in Df00Q13 are stable whereas those in Df20Q13 fragment and form clumps. This results is indicative that the gas component is the driver of the SAI in the model Df20Q13. Thus, as we mentioned in Section \ref{unstable}, a gas fraction in a spiral arm is an important factor of fragmentation and clump formation following. This is consistent with the result that the instability is dominated by the gas in the fragmenting spiral arm in the run Df25Q15, in which $\min[S_{\rm g}(k)]<1$ and $\min[S_{\rm s}(k)]>1$ (see Fig. \ref{FullMaps}). However, presence of gas is not a necessary condition of SAI since the $N$-body run Df00Q10 shows fragmentation of spiral arms.

\subsection{Fragmentation in the bulge-dominant models}
\label{bulge}
The galaxy models shown above are generated in the same initial mass distributions modelling after a low-redshift disc galaxy like the Milky Way. However, disc galaxies, especially massive and quiescent ones, often host bulges more massive than their discs. To see whether our instability analysis can characterise SAI even in a significantly different mass distribution, we run simulations with the bulge-dominant models (see Table \ref{modellist}) and adopt our analysis to them. 

\begin{figure*}
	\begin{minipage}{\hsize}
		\includegraphics[bb=0 0 3780 2475,width=\hsize]{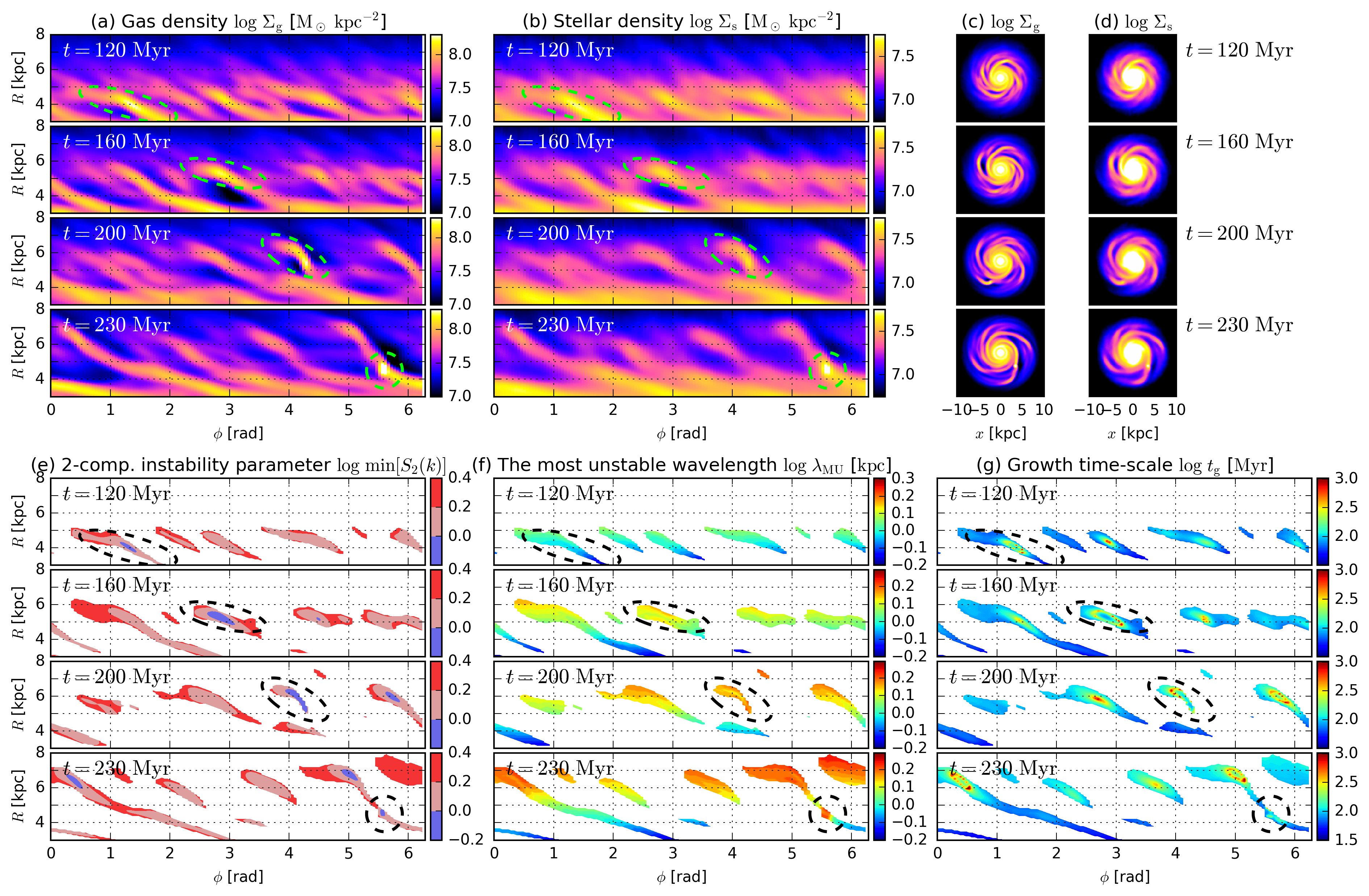}
		\caption{Polar-map analysis for the model Bf70Q13 at $t=120$, $160$ $200$ and $230~{\rm Myr}$. The dashed ellipses encircle a fragmenting arm, within which there is a segment indicating $\min[S_{\rm 2}(k)]<1$. The predicted growth time-scales $t_{\rm g}\simeq100$--$200~{\rm Myr}$ are longer than the actual time-scale of the clump formation by a factor of $\lsim2$.}
		\label{ReduceMapsBulge}
	\end{minipage}
\end{figure*}
Because of the massive bulges, inner disc regions of the bulge-dominant models have high $Q$ in the initial conditions (see the bottom panel of Fig. \ref{VcircQ}). In spite of the high gas fraction $f_{\rm g}=0.7$ in Bf70Q13, this model is only marginally unstable for spiral-arm fragmentation; only a few clumps form during the run. Fig. \ref{ReduceMapsBulge} shows snapshots and our polar-map analysis where the first clump forms via spiral-arm fragmentation in this run. The ellipse in each panel traces a segment with $\min[S_{\rm 2}(k)]<1$ collapsing into the clump. Besides the collapsing segment, there are other two segments with $\min[S_{\rm 2}(k)]<1$ --- at $(R,\phi)=(6~{\rm kpc}, 5.7~{\rm rad})$ at $t=200~{\rm Myr}$\footnote{This segment corresponds to the one at $(R,\phi)=(6.5~{\rm kpc}, 0.3~{\rm rad})$ at $t=230~{\rm Myr}$.} and $(7~{\rm kpc}, 5.0~{\rm rad})$ at $t=230~{\rm Myr}$ in Panel e; however these segments do not collapse and disappear within $100~{\rm Myr}$ since their $\lambda_{\rm MU}$ are longer than the sizes of the segments (Panel f). Panel g indicates that the growth time-scales within the collapsing segment are predicted to be $t_{\rm g}\simeq100$--$200~{\rm Myr}$ from our analysis, which are longer than the actual time-scale of the clump formation by a factor of $\lsim2$.

\begin{figure}
	\includegraphics[bb=0 0 1675 2950,width=\hsize]{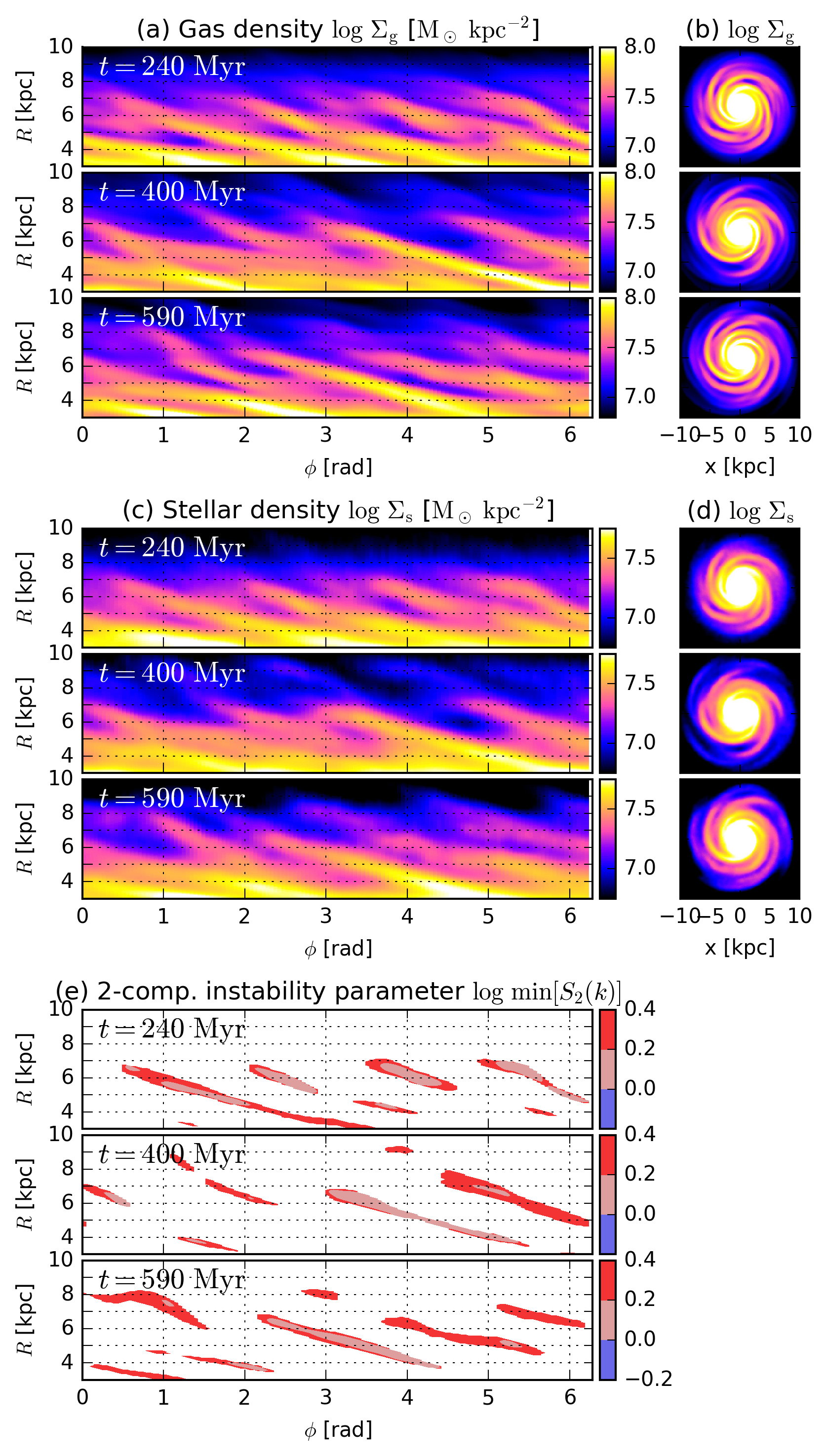}
	\caption{Polar-map analysis for snapshots in the model Bf60Q13 at $t=240$, $400$ and $590~{\rm Myr}$. No spiral arms fragment in this run, and Panel c shows $\min[S_{\rm 2}(k)]>1$ in the arms.}
	\label{StablePlotBulge}
\end{figure}
Fig. \ref{StablePlotBulge} shows polar plots of gas and stellar surface densities and the instability parameters in the model Bf60Q13. This model has a gas fraction $f_{\rm g}=0.6$ slightly lower than Bf70Q13. As seen in Panels from a to d, spiral arms in this run do not fragment, and mergers between the arms rarely occur. Hence, this model is considered to be stable against SAI. The high values of $\min[S_{\rm 2}(k)]$ shown in Panel e are consistent with the absence of spiral-arm fragmentation and clump formation.

Thus far, our linear perturbation analysis can characterise well fragmenting instability of spiral arms even in the bulge-dominant models with the high gas fractions. The applicability of our analysis is not be limited to a specific galaxy model, and we could expect our instability analysis to be applicable to various types of disc galaxies although the tests should be repeated further with other mass distributions and/or different methodologies of hydrodynamic simulations. 

To summarise our results shown above, the tests using our simulations demonstrate the robustness of our linear perturbation analysis. Generally, our instability parameter $S$ can characterise fragmentation of spiral arms, and the mass of a clump forming via SAI can be estimated as $M_{\rm cl}\sim\Upsilon\lambda_{\rm MU}$ when $\lambda_{\rm MU}$ is longer than the spiral arm width (Section \ref{clumpmass}). Although our analysis assumes the fluid approximation for the stellar dispersion relation, we confirm that our analysis is also applicable to stellar spiral arms in the gas-less simulations (Section \ref{Nbodyruns}). The applicability of our analysis is independent from external potentials of bulges and haloes (Section \ref{bulge}). However, we find a few cases of non-linear fragmentation in which spiral arms can fragment with $\min[S(k)]>1$ although such cases are rare in our simulations (Section \ref{nonlinear}).

\subsection{Comparison with other instability parameters}
\subsubsection{Toomre's instability parameter $Q$}

As we mentioned in Section \ref{Intro}, TTI have proposed that Toomre's instability parameter $Q$ measured on a spiral arm can be used as a fragmentation parameter of SAI in their Keplarian gas discs. Although we propose the more general fragmentation condition, $\min[S_{\rm 2}(k)]<1$, for SAI in multi-component systems such as disc galaxies, computing $S$ is more expensive than measuring $Q$ since it requires to know an arm width $W$ for each component. Hence, if the TTI condition $Q\lsim0.6$ is still valid for multi-component galactic discs, it would be easier to use $Q$ to predict fragmentation by SAI.

Conditions of Toomre instability in a multi-component disc have been discussed by a number of studies \citep[e.g.][]{m:81,r:85,ws:94,j:96,r:01,rw:11}. \citet{js:84b,js:84} invented their formulation of a multi-component $Q$ parameter, in which their dispersion relation assumes a fluid-fluid disc system. Meanwhile, \citet{r:01} proposed a different multi-component model which does not assume the fluid approximation for a stellar component but uses the reduction factor \citep{t:64,bt:08}.

In what follows, we use the two-component parameter $Q_{\rm 2}$ proposed by \citet[][see their equation A1]{idm:16},
\begin{equation}
  \frac{1}{Q_{\rm 2}} = \frac{2\left[1-\exp\left(-p_{\rm s}^2\right)I_0\left(p_{\rm s}^2\right)\right]}{Q_{\rm s}p_{\rm s}} + \frac{2p_{\rm g}}{Q_{\rm g}\left(1+p_{\rm g}^2\right)},
\label{RafikovQ}
\end{equation}
where $Q_{\rm g}\equiv\sigma_{\rm T}\kappa_{\rm g}/(\pi G\Sigma_{\rm g})$, $Q_{\rm s}\equiv\sigma_{R,{\rm s}}\kappa_{\rm s}/(\pi G\Sigma_{\rm s})$, $p_{\rm g}\equiv k_{\rm R}\sigma_{\rm T}/\kappa_{\rm g}$, $p_{\rm s}\equiv k_{\rm R}\sigma_{R,{\rm s}}/\kappa_{\rm s}$ and $\sigma_{\rm T}^2\equiv c^2+\sigma_{R,{\rm g}}^2$, and $k_{\rm R}$ is wavenumber of radial perturbation. In this formulation, in order to capture highly turbulent states of clumpy discs, $\kappa_{\rm g}$ and $\kappa_{\rm s}$ are locally measured from the actual rotation velocities for gas and stellar components individually:
\begin{equation}
  \kappa^2=2\frac{\overline{v_{\phi}}}{R}\left(\frac{\mathrm{d}\overline{v_{\phi}}}{\mathrm{d}R} + \frac{\overline{v_{\phi}}}{R}\right).
  \label{kappa}
\end{equation} 
The above formulation of $Q_{\rm 2}$ is based on that of \citet{r:01}; however in his formulation $\kappa$ is measured from the circular velocities and shared between gas and stars. Because the above $Q_{\rm 2}$ is a function of $k_{\rm R}$, we look for the most unstable perturbation that gives $\min[Q_{\rm 2}(k_{\rm R})]$.

\begin{figure}
	\includegraphics[bb=0 0 1359 1445,width=\hsize]{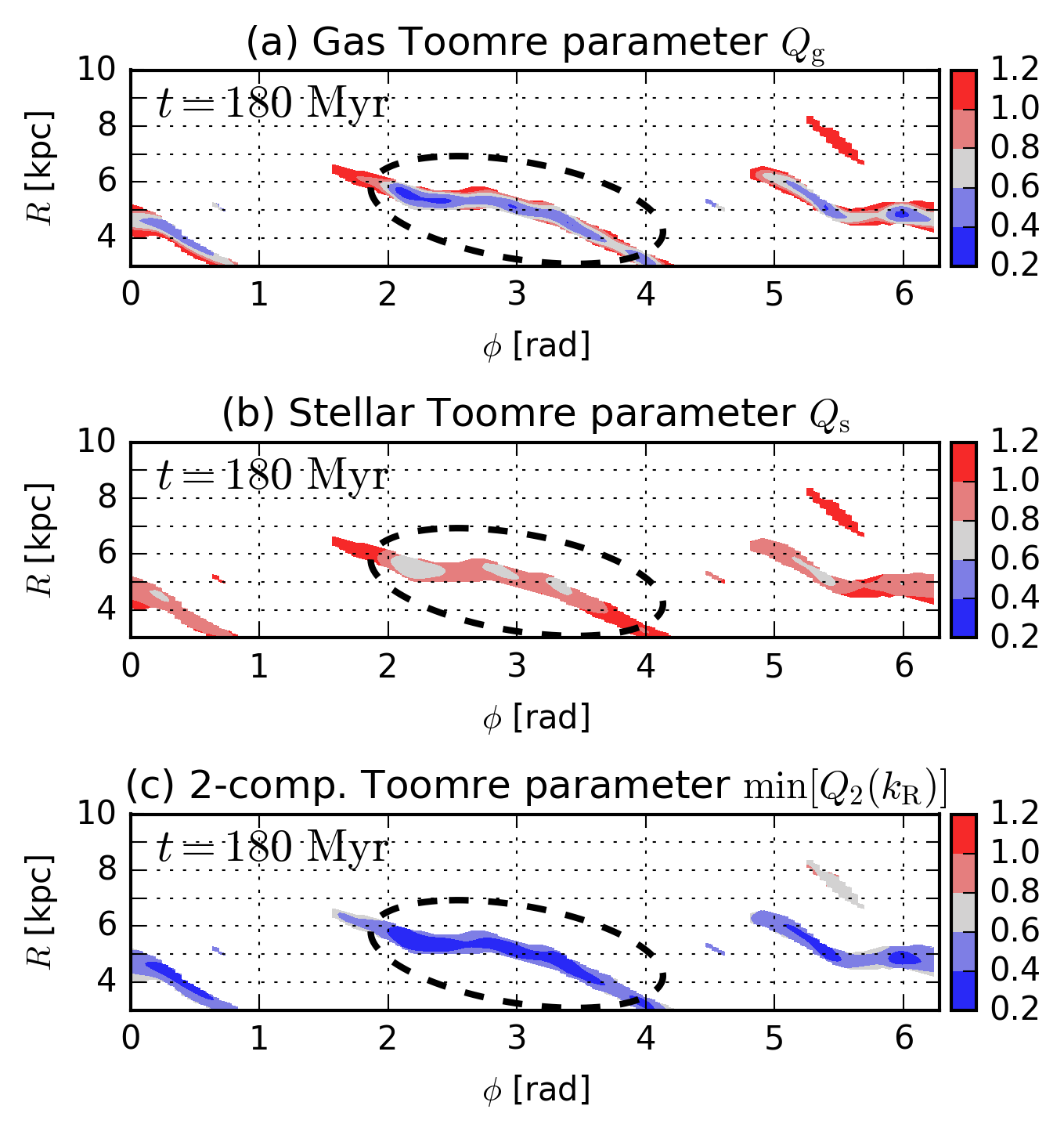}
	\caption{The Toomre analysis for the model Df25Q15 at $t=180~{\rm Myr}$. \textit{Panels a and b}: the single-component Toomre parameters for gas and stars. \textit{Panel c}: the two-component Toomre instability parameter $\min[Q_{\rm 2}(k_{\rm R})]$. The fragmenting arm encircled with the dashed ellipses satisfies the TTI condition $\min[Q_{\rm 2}(k_{\rm R})]\lsim0.6$.}
	\label{ToomreUnstable}
\end{figure}
\begin{figure}
	\includegraphics[bb=0 0 1359 1445,width=\hsize]{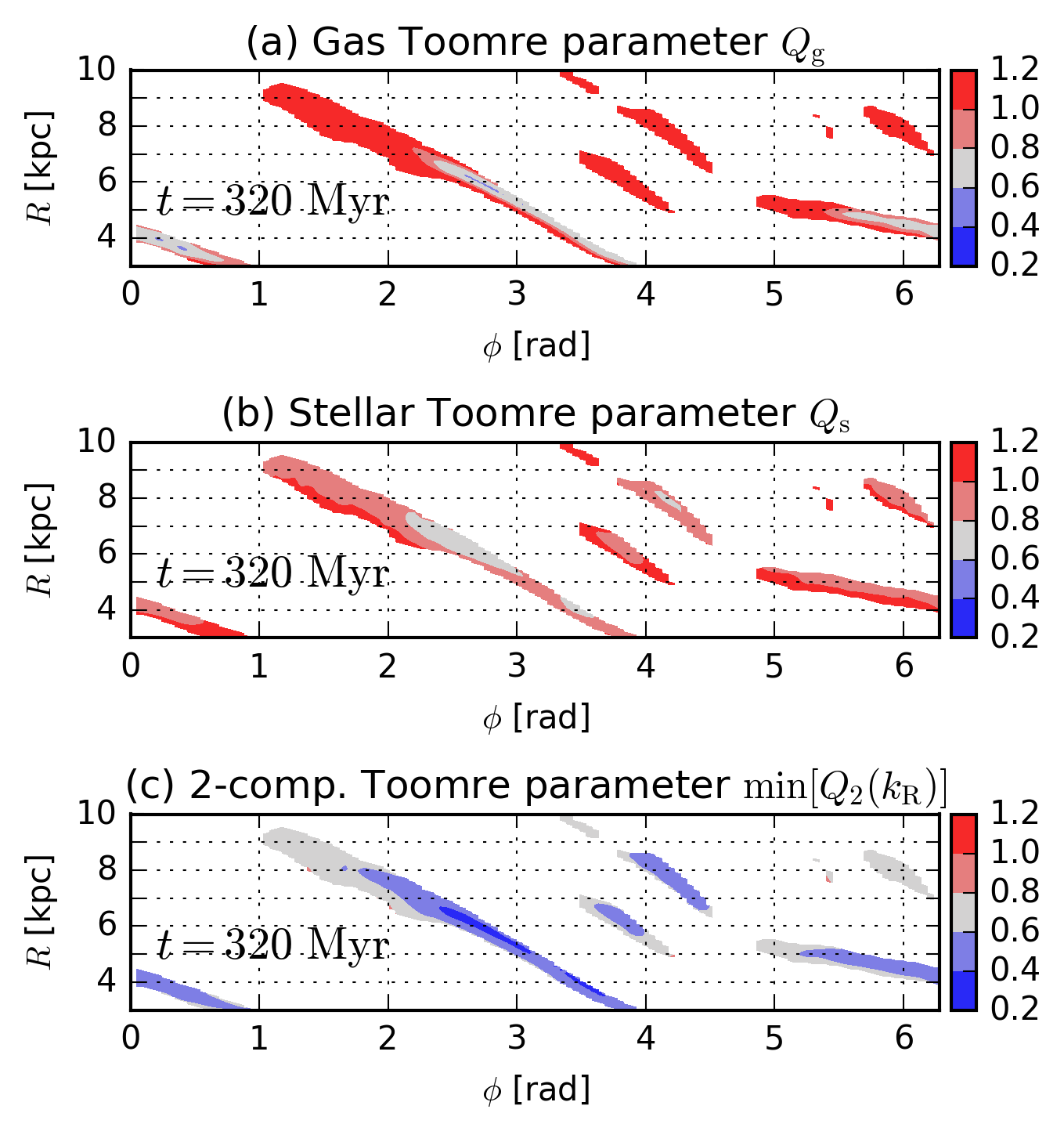}
	\caption{Same as Fig. \ref{ToomreUnstable} but for the non-fragmenting model Df20Q15 at $t=320~{\rm Myr}$. No spiral arms fragment in this simulation; however the arms indicate $\min[Q_{\rm 2}(k_{\rm R})]<0.6$ in this snapshots.}
	\label{ToomreStable}
\end{figure}
Fig. \ref{ToomreUnstable} shows the results of our Toomre analysis for the snapshot of the fragmenting run Df25Q15 at $t=180~{\rm Myr}$. In Panel c, the fragmenting arm marked with the dashed ellipses indicates $\min[Q_{\rm 2}(k_{\rm R})]<0.6$, which is consistent with the instability analysis of TTI. Fig. \ref{ToomreStable} indicates the same results but for the stable model Df20Q15 at $t=320~{\rm Myr}$. In this run, no spiral arms fragment, therefore the arms are expected to indicate $\min[Q_{\rm 2}(k_{\rm R})]\gsim0.6$. However, most of arms even in this run satisfies the TTI condition: $\min[Q_{\rm 2}(k_{\rm R})]\lsim0.6$ (Panel c). Especially, the most prominent arm stretching between $\phi\sim2$--$4~{\rm rad}$ partially indicates $\min[Q_{\rm 2}(k_{\rm R})]<0.4$ significantly lower than the TTI fragmenting criterion, regardless of the stability against SAI of this arm. Thus, the TTI condition $\min[Q_{\rm 2}(k_{\rm R})]\lsim0.6$ appears to over-predict fragmentation of spiral arms in two-component galactic discs; in other words, spiral arms do not necessarily fragment even if $\min[Q_{\rm 2}(k_{\rm R})]\lsim0.6$ although arms would be stable if $\min[Q_{\rm 2}(k_{\rm R})]\gsim0.6$. In two-component galactic discs, our instability condition $\min[S_{\rm 2}(k)]<1$ appears to be more accurate to characterise spiral-arm fragmentation. Our analysis considers not only gas but also stars in the disc, and thus is able to describe the fragmentation more accurately than the fragmenting condition of TTI. Fragmentation condition using $Q_{\rm 2}$ would need calibration for the criterion in multi-component systems.

\subsubsection{The instability condition of isothermal filaments}
Spiral arms may be approximated to be filaments. \citet{o:64} has derived the self-gravitating equilibrium density distribution of an isolated infinite filament of isothermal gas,
\begin{equation}
  \rho_{\rm gf}(r) = \rho_{\rm gf,0}\left[1+\left(\frac{r}{H_0}\right)^2\right]^{-2},
  \label{filament}
\end{equation}
where $H_0$ is the scale radius of the filament, and $\rho_{\rm gf,0}$ corresponds to the gas density on the filament axis at $r=0$. The line-mass of the gas filament is given as
\begin{equation}
	\Upsilon_{\rm gf}=2\pi\int^\infty_0\rho_{\rm gf}(r)r\ \textrm{d}r=\pi\rho_{\rm gf,0}H_0^2,
	\label{linemassIM}
\end{equation}
which is determined solely by sound velocity of the isothermal gas: $\Upsilon_{\rm gf,c}=2c^2/G$ \citep[see][]{o:64}. \citet{im:92,im:97} demonstrated that the isothermal filament is unstable against axisymmetric `sausage-type' perturbations along the filament when the actual line-mass $\Upsilon_{\rm gf}\simeq\Upsilon_{\rm gf,c}$. Hence, if the spiral-arm fragmentation seen in our simulations is caused by the filament instability discussed by \citet{im:92,im:97}, the fragmenting spiral arms are expected to have line-masses close to $\Upsilon_{\rm gf,c}$, and stable arms would have $\Upsilon_{\rm gf}\ll\Upsilon_{\rm gf,c}$ in our simulations.

The surface density profile of equation (\ref{filament}), projected perpendicularly to the filament axis, is
 \begin{equation}
	\Sigma_{\rm gf}(R) = \frac{\pi\rho_{\rm gf,0}H_0^4}{2\left(R^2+H_0^2\right)^{3/2}}=\Sigma_{\rm gf,0}\left[1+\left(\frac{R}{H_0}\right)^2\right]^{-\frac{3}{2}},
	\label{projectedSD}
\end{equation}
where $\Sigma_{\rm gf,0}\equiv\Sigma_{\rm gf}(0)=\pi\rho_{\rm gf,0}H_0/2$. The line-mass of the filament is written as $\Upsilon_{\rm gf}=2\Sigma_{\rm gf,0}H_0$. The surface density profile given by equation (\ref{projectedSD}) decreases to $\Sigma_{\rm gf}(R)=0.3\Sigma_{\rm gf,0}$ at $R=1.15H_0$. Based on the procedures described in Section \ref{ana}, we perform similar analysis to fits the polar maps of $\Sigma_{\rm g}(R,\phi)$ in our simulations with equation (\ref{projectedSD}), while redefining $\tilde{\Sigma}_{\rm g}(R,\xi,\phi)\equiv\Sigma_{\rm g}(R,\phi)[1+(\xi/H_0)^2]^{-3/2}$ and $W=1.15H_0$ in equation (\ref{Xi2}). Then, we compute $H_0$ that gives the lowest $\chi^2$ and determine actual line-masses $\Upsilon_{\rm gf}(R,\phi)\equiv2\Sigma_{\rm g}H_0$ in the polar maps of our simulations.

\begin{figure}
	\includegraphics[bb=0 0 1394 970,width=\hsize]{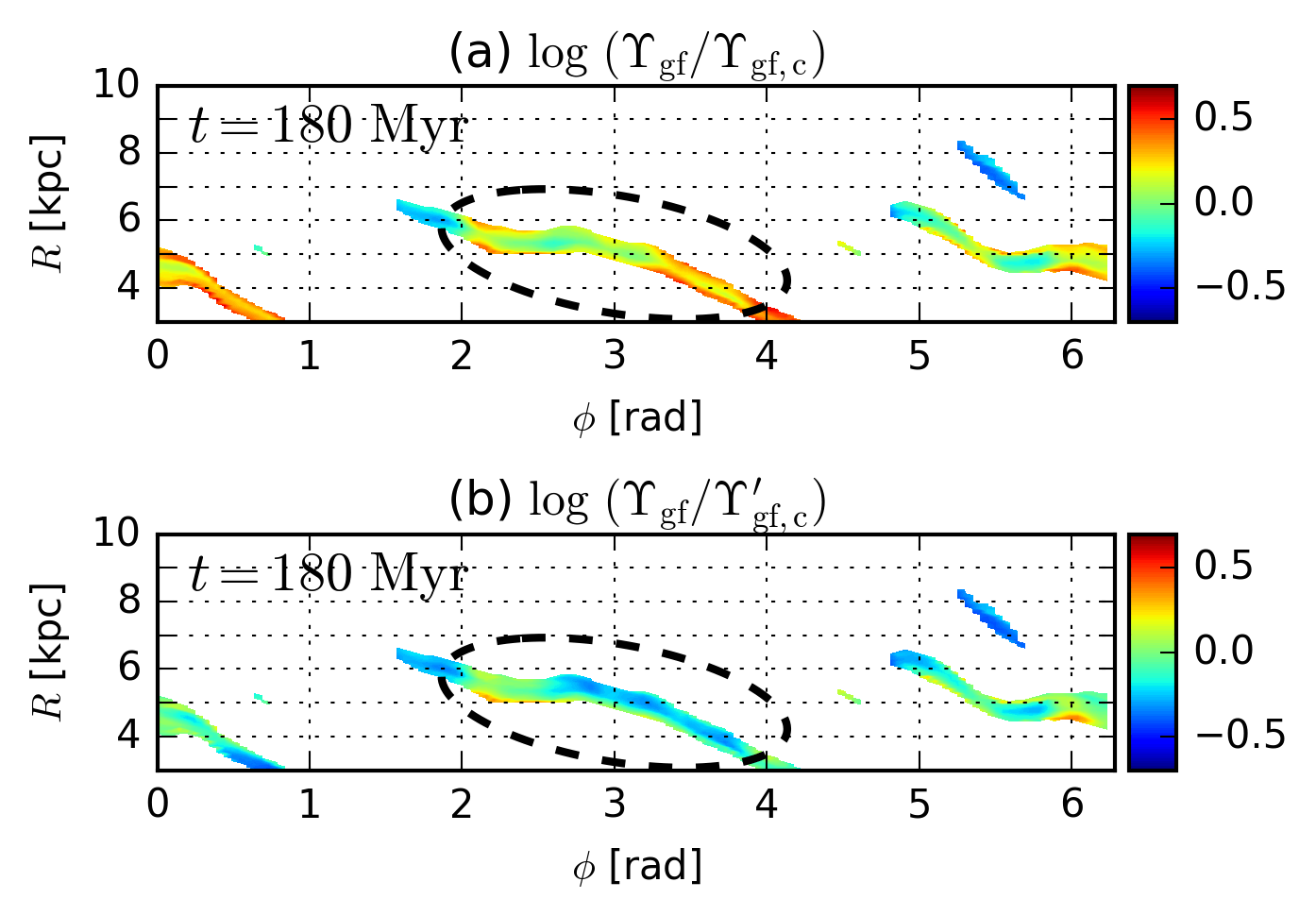}
	\caption{The filament instability analysis for the model Df25Q15 at $t=180~{\rm Myr}$. \textit{Panels a}: $\Upsilon_{\rm gf}/\Upsilon_{\rm gf,c}$, in which $\Upsilon_{\rm gf,c}=2c_{\rm s}^2/G$. \textit{Panels b}: $\Upsilon_{\rm gf}/\Upsilon_{\rm gf,c}'$, in which $\Upsilon_{\rm gf,c}'\equiv2(c^2+\sigma_{{\rm g},R}^2)/G$. The line-masses range $\Upsilon_{\rm gf}/\Upsilon_{\rm gf,c}\simeq1$--$2$ in the fragmenting arm in Panel a, whereas $\Upsilon_{\rm gf}/\Upsilon_{\rm gf,c}'\simeq0.5$--$1$ in Panel b. So, the actual line-masses are approximately close to the equilibrium values.}
	\label{IM_Unstable}
\end{figure}
\begin{figure}
	\includegraphics[bb=0 0 1394 970,width=\hsize]{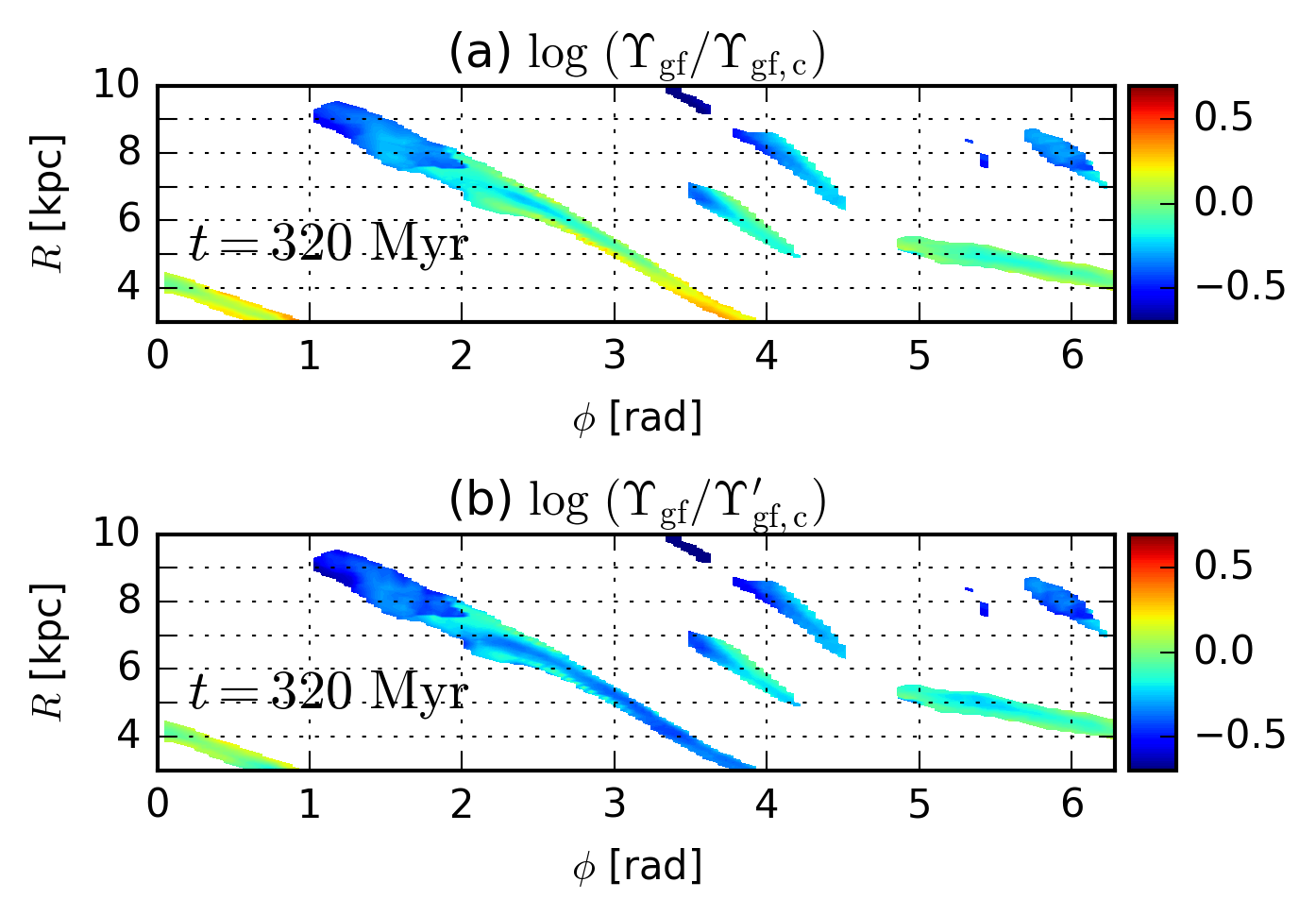}
	\caption{Same as Fig. \ref{IM_Unstable} but for the non-fragmenting model Df20Q15 at $t=320~{\rm Myr}$. In the longest arm stretching between $\phi=1$--$4~{\rm rad}$, the line-masses range $\Upsilon_{\rm gf}/\Upsilon_{\rm gf,c}\simeq0.5$--$1$ in Panel a and $\Upsilon_{\rm gf}/\Upsilon_{\rm gf,c}'\simeq0.4$--$0.6$ in Panel b. In the second longest arm between $\phi=5$--$1~{\rm rad}$, however, $\Upsilon_{\rm gf}$ is close to $\Upsilon_{\rm gf,c}$ and $\Upsilon_{\rm gf,c}'$. Thus, this arm is expected to fragment but actually does not.}
	\label{IM_Stable}
\end{figure}
Panels a in Fig. \ref{IM_Unstable} and \ref{IM_Stable} show the ratios of $\Upsilon_{\rm gf}/\Upsilon_{\rm gf,c}$ for the models Df25Q15 at $t=180~{\rm Myr}$ and Df25Q15 at $t=180~{\rm Myr}$. In Panels b, we take into account the contribution by turbulent pressure to the equilibrium line-mass as $\Upsilon_{\rm gf,c}'\equiv2(c^2+\sigma_{{\rm g},R}^2)/G$. In Fig. \ref{IM_Unstable}, the fragmenting arm encircled with the dashed ellipses has line-masses similar to the equilibrium values: $\Upsilon_{\rm gf}/\Upsilon_{\rm gf,c}\simeq1$--$2$ and $\Upsilon_{\rm gf}/\Upsilon_{\rm gf,c}'\simeq0.5$--$1$. Therefore, the fragmentation of this arm is consistent with the filament instability analysis of \citet{im:92,im:97}. In Fig. \ref{IM_Stable}, the longest arm stretching between $\phi=1$--$4~{\rm rad}$ indicates $\Upsilon_{\rm gf}/\Upsilon_{\rm gf,c}\simeq0.5$--$1$ in Panel a and $\Upsilon_{\rm gf}/\Upsilon_{\rm gf,c}'\simeq0.4$--$0.6$ in Panel b. These low values are consistent with the stability of the arm in this simulation. However, the second longest arm stretching between $\phi=5$--$1~{\rm rad}$ indicates $\Upsilon_{\rm gf}/\Upsilon_{\rm gf,c}\simeq\Upsilon_{\rm gf}/\Upsilon_{\rm gf,c}'\simeq1$. Therefore this arm is expected to fragment. This result appears to be inconsistent with the fact that no spiral arms fragment in the run Df25Q15. Because spiral-arm fragmentation in rotating multi-component discs is intricate more than the filament instability assumed in \citet{o:64} and \citet{im:92,im:97}, the criteria of $\Upsilon_{\rm gf,c}$ and $\Upsilon_{\rm gf,c}'$ does not necessarily appear to be accurate. Therefore, our instability analysis for SAI using the parameter $S$ is needed for characterising spiral-arm fragmentation.

The spiral arms in our simulations are quite different from the situation assumed in \citet{o:64} and \citet{im:92,im:97}. Especially in inner disc regions, Coriolis force by disc rotation can prevent large-scale perturbations within the arms from fragmenting. Although the theory of \citet{o:64} and \citet{im:92,im:97} assumes self-gravitating gas filament in isolation, our simulations have stellar components. The gas fractions within spiral arms in the runs Df25Q15 and Df20Q15 are $f_{\rm g}\sim0.5$, therefore gas in the spiral arms could not be completely self-gravitating. Under the stellar potential, the density profile and the line-mass of an equilibrium gas filament could differ from those proposed by \citet{o:64}. In the simulations and real galaxies, a gas filament can continuously accrete surrounding gas. \citet{cwh:16,cwd:17} perform hydrodynamic simulations for isolated non-equilibrium filaments accreting gas and argue that their filament fragmentation can be related to various properties of gas accretion and have a multi-modal dispersion relation depending on turbulent states such as compressive/solenoidal energy distribution and super/subsonic turbulent velocity.

\section{Discussion}
\label{discussion}
\subsection{Scaling relations}
\label{scalingrelation}
Using our analysis presented in Section \ref{singleanalysis}, we can obtain scaling relations between some physical properties of clumps forming via SAI, which could be useful to compare with observations for giant clumps in disc galaxies. Here, we assume that gravitational fragmenting instability is dominated by gas, and a stellar component is solely stable: $S_{\rm g}\lsim1$ and $S_{\rm s}\gg1$. In this case, since $S_{\rm 2}\simeq S_{\rm g}\ll S_{\rm s}$, one can ignore the stellar component in the perturbation analysis. Hereafter, all variables, such as $\Upsilon$, $\Sigma$, $\sigma$, $\Omega$ and $W$, represent physical properties of gas in a spiral arm unless otherwise stated. 

The function of $\omega^2(k)$ given by the dispersion relation (equation \ref{DL1}) is downward convex (see Fig. \ref{DRrev}), therefore $\omega^2$ becomes the minimum at the most unstable mode $k_{\rm MU}$ and simultaneously $\mathrm{d}\omega^2/\mathrm{d}k=0$. If $\min[S(k)]=1$, the dispersion relation has the only solution for $\omega^2=0$ at $k_{\rm MU}$.

The function $f(kW)$ in the dispersion relation varies as shown in Fig. \ref{fkw}. Here, we approximate the function as $f(kW)\simeq F_{\rm 0}\times(kW)^{-\alpha}$, where $F_{\rm 0}\simeq1$, and $\alpha$ varies with $kW$. With this approximation and equation (\ref{linemass}), the dispersion relation (equation \ref{DL1}) is rewritten as 
\begin{equation}
\omega^2 \simeq \sigma^2k^2 - \pi GF_{\rm 0}A\Sigma W^{1-\alpha}k^{2-\alpha} + 4\Omega^2.
\label{DLdiscuss}
\end{equation}
Using this approximated dispersion relation, the condition of $\mathrm{d}\omega^2/\mathrm{d}k=0$ at $k_{\rm MU}$ gives,\footnote{Although the derivative also becomes $\mathrm{d}\omega^2/\mathrm{d}k=0$ at $k=0$ when $\alpha\neq1$, this solution is not physically meaningful.} 
\begin{equation}
k_{\rm MU} = \left(\frac{\pi\left(2-\alpha\right)GF_{\rm 0}A\Sigma W^{1-\alpha}}{2\sigma^2} \right)^{\frac{1}{\alpha}}.
\label{generalkmu}
\end{equation}
Considering the boundary case of $S(k_{\rm MU})=1$, the condition of $\omega^2=0$ at $k_{\rm MU}$ gives the unstable wavelength as
\begin{equation}
\lambda_{\rm MU} = 2\pi\left(\frac{\pi\alpha GF_{\rm 0}A\Sigma W^{1-\alpha}}{8\Omega^2}\right)^{\frac{1}{2-\alpha}}.
\label{generallmu}
\end{equation}
As in Toomre analysis \citep[e.g.][]{bt:08}, the perturbation that grows first can be generally characterised by $\lambda\simeq\lambda_{\rm MU}$ in marginally unstable cases of $\min[S(k)]\lsim1$. 

If a gas disc is described with the averaged density $\Sigma_{\rm g,d}$ and typical radius $R_{\rm d}$, the total gas mass in the disc is 
\begin{equation}
M_{\rm g,d} \sim \pi\Sigma_{\rm g,d}R_{\rm d}^2 = \pi\frac{\Sigma}{\beta}R_{\rm d}^2 = \eta M_{\rm tot},
\label{totmass}
\end{equation}
where $\beta$ is density contrast of gas between the spiral arm and the disc: $\beta\equiv\Sigma/\Sigma_{\rm g,d}$, and $\eta$ is the mass fraction of the gas disc to the total mass (baryon and dark matter) within $R_{\rm d}$. Circular velocity $V$ is related as $V^2=GM_{\rm tot}/R_{\rm d}$ and $V\simeq \overline{v_{\rm \phi}}=R_{\rm d}\Omega$. Then, equation (\ref{generallmu}) is approximated as
\begin{equation}
\lambda_{\rm MU} \sim 2\pi\left(\frac{1}{8}\alpha F_{\rm 0}A\eta \beta W^{1-\alpha}R_{\rm d}\right)^{\frac{1}{2-\alpha}}.
\label{lmu}
\end{equation}
When the mode $\lambda_{\rm MU}$ collapses into a single clump, we assume that the half wavelength contracts to a clump radius\footnote{Some previous studies assume that $\lambda_{\rm MU}/4$ contracts.}, 
\begin{equation}
R_{\rm cl}=\varepsilon\frac{\lambda_{\rm MU}}{2},
\label{clumpsize}
\end{equation}
where $0<\varepsilon\lsim1$. 

In what follows, we discuss two cases of $k_{\rm MU}W\ll1$ and $\gg1$ as large- and small-scale SAI models. We expect modes of collapse to differ between the two cases: one-dimensional collapse for $k_{\rm MU}W\ll1$ and two-dimensional collapse for $\gg1$ (see below). The latter case behaves very similar to Toomre instability (Section \ref{SSana}). We compare scaling relations obtained from these analytic models with observations of low-redshift clumpy galaxies in Section \ref{ana_obs}.

\subsubsection{One-dimensional collapse mode}
\label{LSana}
First, we discuss the case where $k_{\rm MU}W\ll1$, i.e. $\lambda_{\rm MU}\gg 2W$. Because the unstable wavelength is longer than the width of the spiral arm, the unstable perturbation is expected to collapse along the arm, as in Section \ref{clumpmass}. We consider that gas and stars within the collapsing region form a single clump. Therefore, the total baryon mass of the single clump is estimated as 
\begin{equation}
M_{\rm cl} \sim \frac{\Upsilon}{f_{\rm g}}\lambda_{\rm MU} = A\frac{\Sigma}{f_{\rm g}} W\lambda_{\rm MU},
\label{Mcl_longscale}
\end{equation}
where $f_{\rm g}$ is the gas fraction within the spiral arm: $f_{\rm g}\equiv\Sigma/(\Sigma+\Sigma_{\rm s})$, and $W$ is assumed to be the same between gas and stars in the unstable arm.\footnote{As we showed in Fig. \ref{FullMaps}, generally a stellar arm appears to be wider than a gas arm, therefore equation (\ref{Mcl_longscale}) can underestimate a stellar mass in a clump.}  Using equations (\ref{totmass}), (\ref{clumpsize}) and $V^2=GM_{\rm tot}/R_{\rm d}$ and assuming velocity distribution within the clump to be isotropic, the one-dimensional velocity dispersion can be estimated to be 
\begin{equation}
\sigma_{\rm cl}^2\simeq\frac{1}{3}\frac{GM_{\rm cl}}{R_{\rm cl}} = \frac{2}{3}G\frac{\Sigma}{\varepsilon f_{\rm g}} WA\simeq\frac{2}{3\pi}\eta\beta A\left(\varepsilon f_{\rm g}\right)^{-1}\frac{W}{R_{\rm d}}V^2.
\end{equation}
Using equations (\ref{lmu}) and (\ref{clumpsize}) to eliminate $\eta$,
\begin{equation}
\sigma_{\rm cl}^2\simeq\frac{16}{3}\left(\pi\epsilon\right)^{\alpha-3}\left(\alpha F_{\rm 0}f_{\rm g}\right)^{-1}\left(\frac{W^{\alpha/2}R_{\rm cl}^{1-\alpha/2}}{R_{\rm d}}V\right)^2.
\end{equation}
Then, we obtain the scaling relation for clump sizes,
\begin{equation}
\frac{W^{\alpha/2}R_{\rm cl}^{1-\alpha/2}}{R_{\rm d}} \simeq \sqrt{\frac{3}{16}\left(\pi\epsilon\right)^{3-\alpha}\alpha F_{\rm 0}f_{\rm g}}\frac{\sigma_{\rm cl}}{V}.
\label{largescalerelation}
\end{equation}
Because the exponent $\alpha\lsim0.5$ for $kW\lsim0.5$ (see Fig. \ref{fkw}), equation (\ref{largescalerelation}) means that the dependence of $R_{\rm cl}$ on ${R_{\rm d}\sigma_{\rm cl}}/V$ is at the most $R_{\rm cl}\propto({R_{\rm d}\sigma_{\rm cl}}/V)^{1.3}$, by assuming $\alpha=0.5$, when $W$ and $f_{\rm g}$ are constant. 

Another important scaling relation is mass fraction between a single clump and the host disc: clump mass fraction. From the definition of $\eta$ and by assuming the host disc to have the same gas fraction as the spiral arm, i.e.  $f_{\rm g}\simeq M_{\rm g,d}/(M_{\rm g,d}+M_{\rm s,d})$, the total mass of baryon in the disc is $M_{\rm d}\equiv M_{\rm g,d}+M_{\rm s,d}=\eta M_{\rm tot}/f_{\rm g}$. From equations (\ref{totmass}), (\ref{lmu}) and (\ref{Mcl_longscale}), we obtain 
\begin{equation}
\frac{M_{\rm cl}}{M_{\rm d}} \simeq 2\left[\frac{1}{8}\alpha F_0\left( A\beta\right)^{3-\alpha}\eta\left(\frac{W}{R_{\rm d}}\right)^{3-2\alpha}\right]^{\frac{1}{2-\alpha}},
\label{massrel_long}
\end{equation}
If $\alpha=0.5$, $M_{\rm cl}/M_{\rm d}\propto\beta^{1.7}\eta^{0.7} (W/R_{\rm d})^{1.3}$, meaning that clump mass fractions decrease with $R_{\rm d}$ in this model.

\subsubsection{Two-dimensional collapse mode}
\label{SSana}
\subparagraph{Small-scale SAI:}
When $k_{\rm MU}W\gg1$, i.e. $\lambda_{\rm MU}\ll 2W$, the exponent $\alpha\simeq1$ (see Fig. \ref{fkw}). In this case, the approximated dispersion relation (equation \ref{DLdiscuss}) becomes a quadratic function of $k$,
\begin{equation}
\omega^2 \simeq \sigma^2k^2-\pi GF_{\rm 0}A\Sigma k + 4\Omega^2,
\label{DR_small}
\end{equation}
as in Toomre analysis. However, our SAI models treat the dispersion relation of azimuthal perturbations $k\equiv k_\phi$, and the pressure term also considers the azimuthal component of turbulent velocities (i.e. $\sigma^2\equiv c^2+\sigma_\phi^2$), which are physically different from the Toomre analysis. When $\alpha=1$, equation (\ref{lmu}) becomes
\begin{equation}
\lambda_{\rm MU} \sim\frac{\pi}{4}F_{\rm 0}A\eta \beta R_{\rm d}.
\label{lmu_small}
\end{equation}
Now that we consider the case of $\lambda_{\rm MU}\ll 2W$, the unstable perturbation is deeply embedded within the spiral arm. Therefore, we consider that a round region with a radius of $\lambda_{\rm MU}/2$ collapses (two-dimensional collapse). The mass and the velocity dispersion of the clump are estimated to be 
\begin{equation}
M_{\rm cl} \sim \pi\frac{\Sigma}{f_{\rm g}}\left(\frac{\lambda_{\rm MU}}{2}\right)^2 = \frac{\pi^2}{64} F_{\rm 0}^2A^2\eta ^3\beta^3f_{\rm g}^{-1}M_{\rm tot},
\label{mcl_mtot}
\end{equation}
and
\begin{equation}
\sigma_{\rm cl}^2\simeq\frac{1}{3}\frac{GM_{\rm cl}}{R_{\rm cl}} = \frac{\pi}{24}\left(\varepsilon f_{\rm g}\right)^{-1}F_{\rm 0}A\eta ^2\beta^2V^2.
\end{equation}
Using equations (\ref{clumpsize}) and (\ref{lmu_small}) to eliminate $\eta$,
\begin{equation}
\frac{R_{\rm cl}}{R_{\rm d}} \simeq \sqrt{\frac{3\pi}{8}F_{\rm 0}Af_{\rm g}\varepsilon^3}\frac{\sigma_{\rm cl}}{V}.
\label{smallscalerelation}
\end{equation}
In this model, the above scaling relation does not depend on $W$ since both dispersion relation (i.e. $\lambda_{\rm MU}$) and $M_{\rm cl}$ are independent of $W$. This implies that the small perturbations $k_{\rm MU}W\gg1$ do not react to the spiral arm but just to the local density.

For the clump mass fraction in this case, using $M_{\rm d}=\eta M_{\rm tot}/f_{\rm g}$, equation (\ref{mcl_mtot}) becomes
\begin{equation}
\frac{M_{\rm cl}}{M_{\rm d}} \simeq \frac{\pi^2}{64} F_0^2A^2\eta^2\beta^3.
\label{massrel_short}
\end{equation}
It is noteworthy that the clump mass fraction above is independent from $R_{\rm d}$.

It should be recalled, however, that our analysis assumes an infinitesimal thickness for the arm. Perturbations in such small scales, $\lambda_{\rm MU}\ll 2W$, may actually be shorter than a vertical thickness of the spiral arm. If it is the case, the mode of collapse could become three-dimensional, rather than two-dimensional \citep[e.g.][]{e:15,eh:15}, and the scaling relations for clump properties could deviate from the prediction discussed above.

\subparagraph{Toomre instability:}
We also perform Toomre instability analysis to derive the scaling relations with the same physical parameters. Similar analysis can be found in previous studies \citep[e.g.][]{el:08,dsc:09,fga:17,rck:17}. Again, we assume that Toomre instability is dominated by gas, and a stellar disc is solely Toomre stable: $Q_{\rm g}\lsim1$ and $Q_{\rm s}\gg1$. In this case, since $Q_{\rm 2}\simeq Q_{\rm g}\ll Q_{\rm s}$ \citep[e.g.][]{js:84_2,js:84b,ws:94,j:96,rw:11}, one can ignore the stellar disc in the perturbation analysis. The Toomre analysis considers axisymmetric linear perturbations $k_R$ propagating in the radial direction in a rotating disc \citep{s:60,t:64}, in which the dispersion relation is given as  
\begin{equation}
\omega^2 = \sigma_{\rm T}^2k_R^2-2\pi G\Sigma_{\rm g,d}k_R + \kappa^2,
\end{equation}
where $\kappa$ is epicyclic frequency which can be described as $\kappa=aV/R_{\rm d}$ with $1<a<2$ \citep[e.g.][]{bt:08}. The first term in the right-hand side represents the radial force to stabilise radial perturbations $k_R$ by thermal and turbulent pressure although it is azimuthal in our SAI models.

The conditions of $\mathrm{d}\omega^2/\mathrm{d}k_R=0$ at $k_R=k_{R,{\rm MU}}$ and  $Q\equiv\sigma_{\rm T}\kappa/(\pi G\Sigma_{\rm g,d})=1$, one can obtain the characteristic wavelength $\lambda_{R,{\rm MU}}$ in a disc with $Q\lsim1$ as
\begin{equation}
\lambda_{R,{\rm MU}} \simeq \frac{2\pi^2 G\Sigma_{\rm g,d}R_{\rm d}^2}{a^2V^2} = \frac{2\pi \eta GM_{\rm tot}}{a^2V^2} = \frac{2\pi\eta R_{\rm d}}{a^2}.
\label{lmuT}
\end{equation}
As the previous studies do, we assume that a clump forms via the two-dimensional collapse within $\lambda_{R,{\rm MU}}/2$. The mass and the velocity dispersion of the clump are estimated to be 
\begin{equation}
M_{\rm cl} \sim \pi\frac{\Sigma_{\rm g,d}}{f_{\rm g}}\left(\frac{\lambda_{R,{\rm MU}}}{2}\right)^2 = \pi^2a^{-4} \eta ^3f_{\rm g}^{-1}M_{\rm tot},
\label{Mcl_Toomre}
\end{equation}
and 
\begin{equation}
\sigma_{\rm cl}^2\simeq\frac{1}{3}\frac{GM_{\rm cl}}{R_{\rm cl}} = \frac{\pi}{3}\left(\varepsilon f_{\rm g}a^2\right)^{-1}\eta ^2V^2,
\end{equation}
where we assume again that the gas fraction $f_{\rm g}$ is the same between a spiral arm and the disc. Using equations (\ref{clumpsize}) and (\ref{lmuT}) to eliminate $\eta$,
\begin{equation}
\frac{R_{\rm cl}}{R_{\rm d}} \simeq\sqrt{\frac{3\pi}{a^2}\varepsilon^3 f_{\rm g}}\frac{\sigma_{\rm cl}}{V}.
\label{Tana}
\end{equation}
Thus, the Toomre instability analysis also predicts the proportionality between the ratios $R_{\rm cl}/{R_{\rm d}\propto\sigma_{\rm cl}}/V$, which is the same as our small-scale SAI model (equation \ref{smallscalerelation}). 

From equation (\ref{Mcl_Toomre}) and $M_{\rm d}=\eta M_{\rm tot}/f_{\rm g}$, the clump mass fractions in this case are obtained as 
\begin{equation}
\frac{M_{\rm cl}}{M_{\rm d}} \simeq \pi^2a^{-4} \eta^2.
\label{massrel_Toomre}
\end{equation}
The parameter dependence of the above relation is similar to that in our small-scale SAI model (equation \ref{massrel_short}), i.e. $\propto\eta^2$ and independent of $R_{\rm d}$.

\subsubsection{Comparison of the analytic models with observations}
\label{ana_obs}
\begin{figure}
	\includegraphics[bb=0 0 1667 1170,width=\hsize]{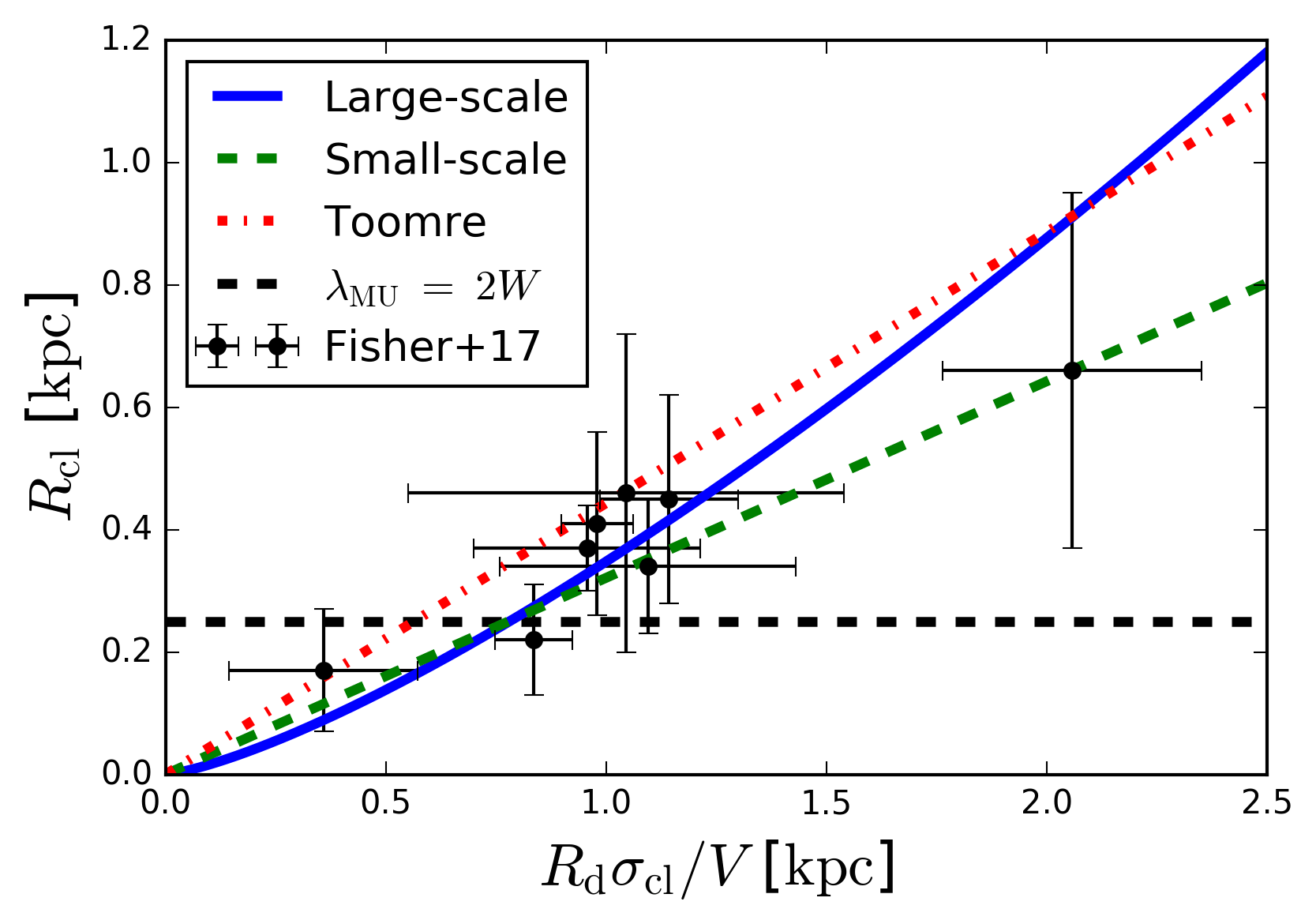}
	\caption{Comparison of scaling relations of clump radius between the analytic models and observations. The blue solid, the green dashed and the red dash-dotted lines delineate the scaling-relations of equations (\ref{smallscalerelation}), (\ref{largescalerelation}) and (\ref{Tana}). The horizontal black dashed line indicates $R_{\rm cl}$ given by equation (\ref{clumpsize}) when $\lambda_{\rm MU}=2W$. The black filled circles with the error bars indicate the observations of \citet{fga:17}, in which $R_{\rm d}$ is measured as twice half-light radius for H$\alpha$ emission.}
	\label{AnaObs}
\end{figure}
The analytic scaling relations predicted from our SAI and the Toomre instability models can be compared with properties of giant clumps observed in disc galaxies. In Fig. \ref{AnaObs}, we compare the scaling relations of clump sizes between the analytic models (equations \ref{largescalerelation}, \ref{smallscalerelation} and \ref{Tana}) and the observations of \citet[][reproduced from their table 1, excluding objects classified as merging galaxies]{fga:17} for clumpy galaxies at redshifts $z=0.07$--$0.14$. Although their observed sample is limited to the low redshifts, \citet{fgb:14} have argued that their low-redshift clumpy galaxies have high gas fractions similar to those of the high-redshift counterparts \citep[see also][]{ggm:14,wfm:17}. For the analytic models in Fig. \ref{AnaObs}, we set the parameters to $\varepsilon=0.5$, $f_{\rm g}=0.5$, $F_0=1.0$, $A=1.4$, $\alpha=0.5$ (for the large-scale SAI model), $W=0.5~{\rm kpc}$ and $a=\sqrt{3}$. The horizontal dashed line in the figure indicates $R_{\rm cl}$ of the clump that forms from the perturbation of $\lambda_{\rm MU}=2W=1~{\rm kpc}$, above and below which the large- and the small-scale instability modes are expected to be realistic for our SAI models. In the figure, all of the analytic models using the above parameter set are almost consistent with the observations within the error ranges. The non-linearity of the one-dimensional collapse model is not significant enough to largely deviate from the linear relations predicted by the two-dimensional collapse models. Thus, our SAI models are not inconsistent with the observations of the low-redshift giant clumps. 

It should be noted, however, that the analytic models depend on some parameters that can change the coefficients of the scaling relations. Especially, values of $\varepsilon$ and $f_{\rm g}$ are highly uncertain and may differ between individual giant clumps. What we emphasize is that the large-scale mode of our SAI model also predicts the scaling relation similar to that of the Toomre analysis, and the current observations still cannot distinguish these models. Therefore, the SAI models that our study discusses could be possible mechanisms to form giant clumps in gas-rich disc galaxies.

The scaling relations of clump mass fractions given by equations (\ref{massrel_long}), (\ref{massrel_short}) and (\ref{massrel_Toomre}) have the appreciably different parameter-dependence and are therefore expected to be distinguishable in comparison with observations. We use the H$\alpha$ observations of \citet{fga:17} to estimate clump mass fractions in their sample.\footnote{They kindly provided us their observational data for $L_{\rm cl}/L_{\rm d}$.} If $f_{\rm g}$ is constant between a clump and its host galaxy, $M_{\rm cl}/M_{\rm d} \simeq M_{\rm g,cl}/M_{\rm g,d}$. To estimate $M_{\rm g,cl}/M_{\rm g,d}$ from the H$\alpha$ observations, we use the relation between H${\rm \alpha}$ luminosity and star-formation rate, i.e. $L_{\rm H\alpha}\propto SFR$. Surface densities of star-formation rate and gas are related as $\Sigma_{\rm SFR}\propto\Sigma_{\rm g}^{1.5}$ \citep{s:59,k:89}, therefore $L_{\rm H\alpha}/R^2\propto M_{\rm g}^{1.5}/R^3$. Then, we approximate
\begin{equation}
\frac{M_{\rm g,cl}}{M_{\rm g,d}} \propto \left(\frac{L_{\rm cl}R_{\rm cl}}{L_{\rm d}R_{\rm d}}\right)^{\frac{1}{1.5}},
\label{LumiToMass}
\end{equation}
where $L_{\rm cl}$ and $L_{\rm d}$ are the H$\alpha$ luminosities within a clump and its host galaxy. Here, we use the median values of luminosities and radii among clumps in each galaxy for $L_{\rm cl}$ and $R_{\rm cl}$. If the total mass inside $R_{\rm d}$ is dominated by baryon in the disc, we can assume $\eta\simeq f_{\rm g}$. From these assumptions, the scaling relation (equation \ref{massrel_long}) for the one-dimensional collapse model with $\alpha=0.5$ can be approximated as
\begin{equation}
\frac{M_{\rm g,cl}}{M_{\rm g,d}} \propto f_{\rm g}^{0.7}R_{\rm d}^{-1.3},
\label{massrel_1D}
\end{equation}
and the relations (equation \ref{massrel_short} and \ref{massrel_Toomre}) for the two-dimensional collapse models are  
\begin{equation}
\frac{M_{\rm g,cl}}{M_{\rm g,d}} \propto f_{\rm g}^2,
\label{massrel_2D}
\end{equation}
where we ignore the dependence on $\beta$, $W$ and $a$.

\begin{figure}
	\includegraphics[bb=0 0 1710 1188,width=\hsize]{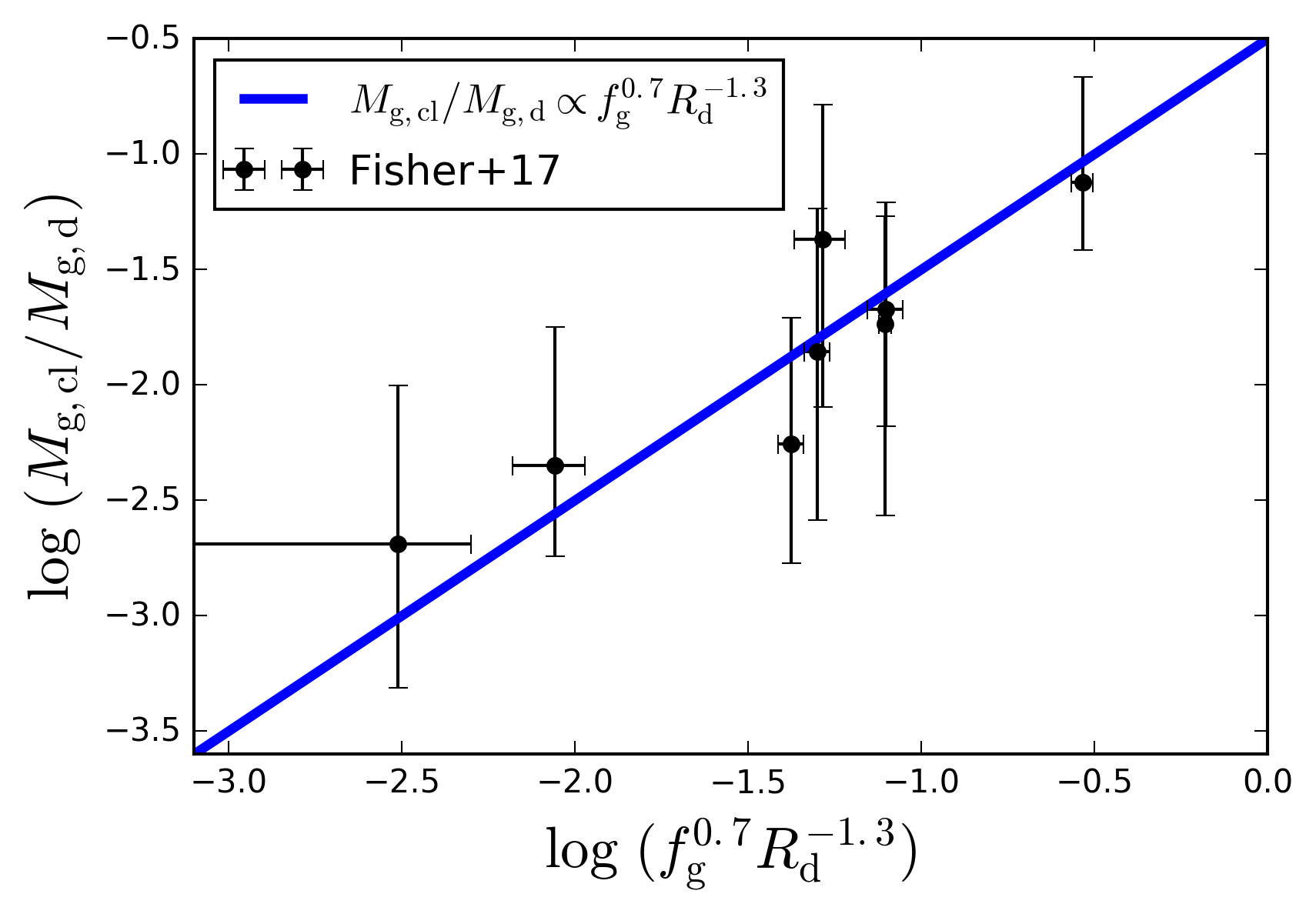}
	\includegraphics[bb=0 0 1710 1188,width=\hsize]{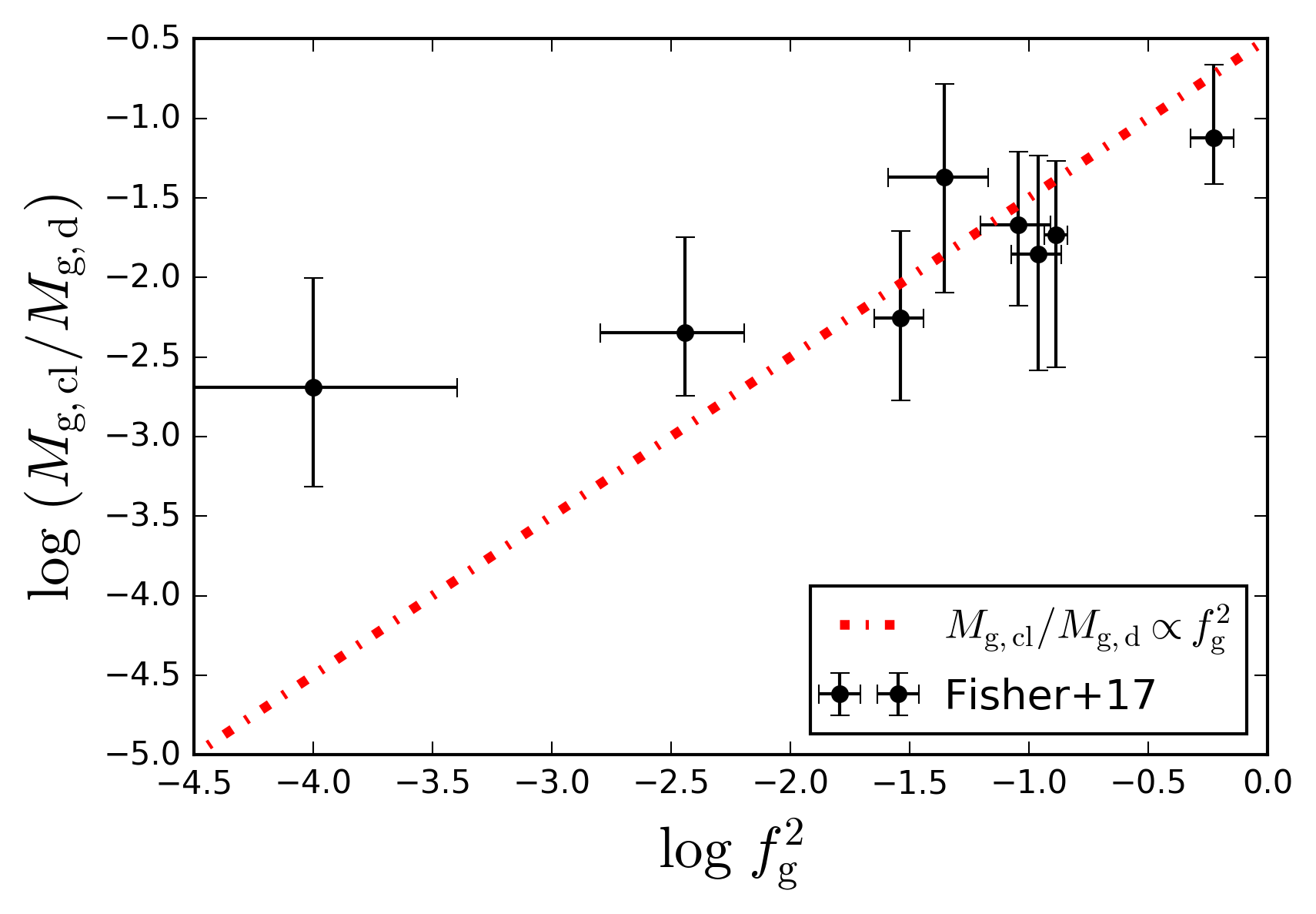}
	\caption{Comparison of the scaling relations of clump mass fractions between the analytic models and the observations. \textit{top panel:} our one-dimensional collapse model with $\alpha=0.5$. \textit{bottom panel:} our two-dimensional collapse models. The filled circles and the error bars indicate the median values among all clumps in each galaxy and variations between the clumps, including observational uncertainties. In the top panel, $R_{\rm d}$ is in the unit of kpc. The model predictions are elevated to fit the best the data points.}
\label{ClumpMassFraction}
\end{figure}
In Fig. \ref{ClumpMassFraction}, we compare the scaling relations of the analytic models (equations \ref{massrel_1D} and \ref{massrel_2D}) with the observations. The diagonal lines correspond to the model predictions, and the data points align diagonally if the observations are consistent with the analytic models. The observations appear to be more consistent with the one-dimensional collapse model (the top panel) than the two-dimensional ones (the bottom panel); the observations may show significant deviation from the prediction in the bottom panel, especially for low $f_{\rm g}^2$. From this result, it may be suggested that the one-dimensional collapse mode based on our SAI model can describe better the observed properties of giant clumps. Our result in Fig. \ref{ClumpMassFraction}, however, could not conclusively reject the two-dimensional collapse models such as Toomre instability. First of all, the observed sample in \citet{fga:17} has only a handful of clumpy galaxies. In addition, the most gas-poor galaxy in their sample has $f_{\rm g}=0.01$, which is atypical for clumpy galaxies. If we exclude this galaxy as an outlier, the consistency with the models becomes comparable between the two panels. Moreover, the analytic scaling relations do not take into account the parameters such as $\beta$ and $W$ that may correlate with $f_{\rm g}$ and/or $R_{\rm d}$. It is also possible that the one- and two-dimensional collapse modes operate in a single galaxy, and/or the instability modes may be different between individual galaxies.

\subsection{SAI in more realistic situations and limitations of our models}
\label{real}
We test our idealized analytic model with the simple $N$-body/hydrodynamic simulations that include the minimal physics. As mentioned in Section \ref{basiceq}, our analysis is based on the tight-winding approximation for a spiral arm in self-gravitating equilibrium. As seen in our simulations, however, a spiral arm can deform, interact actively with other structures and have a large pitch angle. In Section \ref{result}, we demonstrated that our analysis using the instability parameter $S$ is quite robust in our simulations, and non-linear fragmentation with $\min[S(k)]>1$ are rarely seen. However, it is unclear how frequently the non-linear fragmentation can occur in real galaxies. 

The pressure term in equation (\ref{phimom}) is derived by assuming barotropic fluid \citep{t:64,bt:08} which is applicable to isothermal gas. Generally, however, galactic gas does not have such a simple equation of state and can cool by the dissipative nature. For Toomre instability, \citet{e:11} analytically argues that galactic discs can be unstable up to $Q\simeq2$--$3$ if gas dissipation is as rapid as a local dynamical time-scale. Galactic spiral arms can also be destabilised by gas cooling, and our instability criterion, $\min[S(k)]<S_{\rm crit}=1$, can become higher if dissipation is effective. \citet{l:17} argues that kinematic dissipation by collisions between molecular clouds can play an important role in disc instability and may explain the morphological difference between spiral and clumpy galaxies. Additionally, our analysis assumes that density distributions of spiral arms can be fitted with Gaussian functions, as TTI did; however, actual density distributions in gas and/or stellar arms may not be Gaussian.

Our simulations are designed to dedicate to test our linear perturbation analysis and therefore lacking some important physics. First of all, our simulations do not have star formation or stellar feedback processes such as supernovae and radiation from massive stars. In real galaxies, formation of their spiral arms and giant clumps generally involves active star formation which consumes gas to create stars and can weaken cooling effect by lowering gas density. This implies that star formation may prevent a spiral arm from fragmenting. Stellar feedback processes can dramatically affect thermal and kinematic states inside a spiral arm. They can be strong heating sources for gas and may stabilise the arm. Meanwhile, the feedback can be violent enough to destroy the arm and make it non-equilibrium. Although we show that our analytic prediction for clump masses is in agreement with our simulation results in Section \ref{clumpmass}, clumps can considerably decrease their masses due to stellar feedback while causing strong outflows \citep[e.g.][]{n:96,nsg:12,g:12,bpr:14,tms:14,mdc:17,rck:17}. In this sense, our analytic prediction could overestimate actual clump masses; the scaling relations discussed in Section \ref{scalingrelation} do not take such feedback effects into account either.

Our simulation models have `well-tailored' structures in isolation. The gas and stellar discs are assumed to be as thin as $z_{\rm d}=50~{\rm pc}$ at all radii in the initial conditions, in order to marginalize thickness effect on the instability; $z_{\rm d}$ is the same as the minimum softening length in our simulations. For example, however, the Milky Way has $z_{\rm d}\simeq50$ and $300~{\rm pc}$ for the gas and the stellar (thin) discs \citep[e.g.][]{jib:08,ig:13,hg:15,brs:16}, and high-redshift clumpy discs are observed to be even thicker \citep[e.g.][]{ee:06,bem:09,gsu:17}. Generally, disc thickness can stabilise a disc and a spiral arm by weakening their self-gravity on the plane. A number of studies have proposed their parameterised models and analytical treatments to take into account the thickness effects in Toomre instability parameters or its criterion \citep[e.g.][]{gl:65,r:92,e:11,rw:11,bbs:14}; however our current analysis for SAI is based on the assumption of razor-thin arms. In addition, asymmetric structures can cause non-linear effects on the instability. Galactic discs can be warped, and non-axisymmetric bulges, bars and haloes may affect SAI by inducing non-circular motions and/or resonance in discs \citep[e.g.][]{amr:13,hs:15}. In cosmological context, minor mergers and external gas accretion can significantly impact on disc dynamics and clump formation \citep[e.g.][]{bkw:12,keg:16}.

To further examine our SAI analysis and study physical mechanisms of clump formation in more realistic galaxy models, we are planning to utilize the cosmological simulations of the Auriga project \citep{ggm:17} in our future work.  

\subsection{SAI as a possible mechanism of  giant clump formation}
Previous studies have proposed Toomre instability to be a possible mechanism of giant clump formation in highly gas-rich discs in their early formation stages \citep[e.g.][]{SN:93,n:96,n:98,n:99,g:08,gnj:11,dsc:09,fkb:12,fkr:13,fga:17} although \citet{idm:16} claimed that giant clumps do not necessarily start to form with $Q<1$ in their cosmological simulations. In this study, we propose that SAI could be an alternative mechanism to form giant clumps and possibly explain the morphological difference between spiral and clumpy galaxies. As TTI discussed for proto-planet formation,\footnote{TTI have discussed as follows; a proto-planetary disc can form spiral arms via Toomre instability if $Q\lsim1$ in the disc, and then the spiral arms can fragment into proto-planets via the spiral-arm (TTI) instability if $Q\lsim0.6$ in the spiral arms.} we suppose that giant clump could form via two-stage instability: disc instability to form spiral arms followed by SAI to form giant clumps. The first stage is the phase of spiral-arm formation which could be triggered by marginally unstable state of the disc with $Q\lsim2$, possibly via swing amplification mechanism \citep{t:81}. We infer that spiral galaxies such as the Milky Way in the low-redshift Universe would correspond to this stage; they are unstable enough to form their spiral arms but stable enough to maintain their arms. Then, some galaxies forming spiral arms can experience the second stage where SAI operates if the instability parameter $S\lsim1$ in their arms. Their unstable arms can fragment into giant clumps in their disc regions, and they are observed as clumpy galaxies.

As we show in Section \ref{result}, our simulated galaxies with high gas fractions and/or cold kinematic states with low $Q_{\rm min}$ tend to be unstable for spiral-arm fragmentation provided their mass distributions are the same. Clumpy galaxies have observed to have gas fractions $f_{\rm g}\sim0.2$--$0.4$ typically higher than spiral galaxies, independent from redshifts \citep{fgb:14}. In addition, observations of high-redshift clumpy galaxies have estimated Toomre's instability parameters $Q$ to be quite low in their discs \citep{g:08,gnj:11,fga:17}. Thus, in terms of gas fraction and $Q$, our simulations indicating clump formation preferentially in gas-rich discs with low $Q_{\rm min}$ are not contradictory to these observations. \citet{lss:12} have found a grand-design spiral galaxy at the redshift $z=2.18$. Interestingly, this high-redshift galaxy has a bright clump in the spiral arm. They mentioned that this galaxy is the only object displaying regular spiral morphology in their sample of 306 galaxies at this redshift. The rarity of grand-design spiral galaxies in their sample implies that high-redshift disc galaxies may not able to form regular spiral structures, or spiral arms in such galaxies may be short-lived and dissolved soon after its emergence. Alternatively, it is also possible that most of current observations have missed spiral arms in high-redshift discs because of their faintness. \citet{ee:14} argued, from their visual inspection for observed galaxy images, that the onset of spiral structures in galaxies appears to occur at a redshift $z\sim1.4$--$1.8$ and that clumpy galaxies further than this redshift do not have prominent spiral arms. Our clump formation scenario based on SAI assumes that the emergence of spiral arms predates the formation of giant clumps; therefore our SAI model may be feasible for clumpy galaxies at $z\lsim2$.

The absence of giant clumps in most of low-redshift spiral galaxies may imply that SAI does not operate in these disc galaxies. If our SAI model discussed in this study is the mechanism of clump formation, it is implied that spiral arms of low-redshift disc galaxies indicate $\min[S(k)]>1$. This means that observations for such a nearby spiral galaxy could also give us important clues to understand mechanisms of clump formation. Although it is still challenging to observationally estimate values of $S$ in high-redshift galaxies, it may be possible in low-redshift spiral galaxies. 

\section{Conclusions and summary}
\label{conclusions}
We extend the linear perturbation theory, previously presented by TTI, for a spiral arm to a multi-component model and analytically obtain an instability parameter and its criterion. If the instability condition is satisfied, the spiral arm is gravitationally unstable and expected to fragment into clumpy structures. Calculating the instability parameter requires to know properties of a spiral arm: surface density $\Sigma$, angular rotation velocity $\Omega$, azimuthal component of velocity dispersion (and sound velocity for gas) $\sigma$ and width of the spiral arm $W$ for each component.

Next, we perform $N$-body/hydrodynamics simulations using our disc galaxy models in isolation with an isothermal equation of motion, examine our instability analysis by utilizing the simulation data. We find our linear perturbation analysis to be able to characterise remarkably well fragmentation by SAI in our simulations; a spiral arm breaks out into clumps only if the instability condition is satisfied. However, our analysis generally overestimates growth time-scales of the unstable perturbations by a factor of $2$--$4$. We also find a few cases of non-linear fragmentation which forms clumps without satisfying the instability condition, although such cases are rare in our simulations. In our SAI model, a clump is expected to form via filamentary (one-dimensional) collapse of an arm if the most unstable wavelength is longer than the arm width. Therefore, the clump mass can be estimated to be the product between a line-mass of the arm and the most unstable wavelength. Our analytical prediction is in agreement with actual clump masses in our simulations.

We discuss whether SAI can be a physical mechanism of giant clump formation in observed disc galaxies. Using our linear perturbation analysis, we obtain expected scaling relations of physical properties of giant clumps. In the case of the one-dimensional collapse of our SAI model, the scaling relation of clump sizes only weakly deviates from a linear relation predicted from the Toomre's instability analysis. For the scaling relation of clump mass fractions, the observations of low-redshift clumpy galaxies may prefer the one-dimensional collapse mode of our SAI model, rather than two-dimensional modes such as Toomre instability; however the observations still cannot distinguish these models conclusively. Neither scaling relations are inconsistent with the current observations of giant clumps, therefore we expect that our SAI model could be a possible mechanism of giant clump formation in disc galaxies.

\section*{Acknowledgments}
We are grateful for the helpful comments and careful reading of an anonymous referee. This work was begun with inspiring discussion with Andreas Burkert while SI was visiting the University Observatory Munich, thanks to a grant from the Hayakawa Sachio Fund awarded by the Astronomical Society of Japan. The observational data used in Section \ref{ana_obs} were kindly provided in part by David B. Fisher. We thank Sanemichi Takahashi, Yusuke Tsukamoto and Shugo Michikoshi for their useful and valuable comments, and Ken-ichi Tadaki and Takatoshi Shibuya for their fascinating comments on our observational comparison in Section \ref{discussion}, and Volker Springel for kindly providing the simulation code {\sc Arepo}. This study was supported by World Premier International Research Center Initiative (WPI), MEXT, Japan and by SPPEXA through JST CREST JPMHCR1414. SI receives the funding from KAKENHI Grant-in-Aid for Young Scientists (B), No. 17K17677. The numerical computations presented in this paper were carried out on Cray XC30 at Center for Computational Astrophysics, National Astronomical Observatory of Japan.


\begin{thebibliography}{}

\bibitem[\protect\citeauthoryear{{Athanassoula}, {Machado} \&
  {Rodionov}}{{Athanassoula} et~al.}{2013}]{amr:13}
{Athanassoula} E.,  {Machado} R.~E.~G.,    {Rodionov} S.~A.,  2013, \mnras,
  429, 1949

\bibitem[\protect\citeauthoryear{{Baba}, {Morokuma-Matsui}, {Miyamoto}, {Egusa}
  \& {Kuno}}{{Baba} et~al.}{2016}]{bmm:16}
{Baba} J.,  {Morokuma-Matsui} K.,  {Miyamoto} Y.,  {Egusa} F.,    {Kuno} N.,
  2016, \mnras, 460, 2472

\bibitem[\protect\citeauthoryear{{Bassett} et~al.,}{{Bassett}
  et~al.}{2014}]{bgf:14}
{Bassett} R.,  et~al., 2014, \mnras, 442, 3206

\bibitem[\protect\citeauthoryear{{Behrendt}, {Burkert} \&
  {Schartmann}}{{Behrendt} et~al.}{2015}]{bbs:14}
{Behrendt} M.,  {Burkert} A.,    {Schartmann} M.,  2015, \mnras, 448, 1007

\bibitem[\protect\citeauthoryear{{Binggeli}, {Sandage} \& {Tammann}}{{Binggeli}
  et~al.}{1988}]{bst:88}
{Binggeli} B.,  {Sandage} A.,    {Tammann} G.~A.,  1988, \araa, 26, 509

\bibitem[\protect\citeauthoryear{{Binney} \& {Tremaine}}{{Binney} \&
  {Tremaine}}{2008}]{bt:08}
{Binney} J.,  {Tremaine} S.,  2008, Galactic Dynamics Second Edition.
Princeton Univ. Press, Princeton

\bibitem[\protect\citeauthoryear{{Bird}, {Kazantzidis} \& {Weinberg}}{{Bird}
  et~al.}{2012}]{bkw:12}
{Bird} J.~C.,  {Kazantzidis} S.,    {Weinberg} D.~H.,  2012, \mnras, 420, 913

\bibitem[\protect\citeauthoryear{{Bournaud}, {Elmegreen} \&
  {Martig}}{{Bournaud} et~al.}{2009}]{bem:09}
{Bournaud} F.,  {Elmegreen} B.~G.,    {Martig} M.,  2009, ApJL, 707, L1

\bibitem[\protect\citeauthoryear{{Bournaud} et~al.,}{{Bournaud}
  et~al.}{2014}]{bpr:14}
{Bournaud} F.,  et~al., 2014, \apj, 780, 57

\bibitem[\protect\citeauthoryear{{Bovy}, {Rix}, {Schlafly}, {Nidever},
  {Holtzman}, {Shetrone} \& {Beers}}{{Bovy} et~al.}{2016}]{brs:16}
{Bovy} J.,  {Rix} H.-W.,  {Schlafly} E.~F.,  {Nidever} D.~L.,  {Holtzman}
  J.~A.,  {Shetrone} M.,    {Beers} T.~C.,  2016, \apj, 823, 30

\bibitem[\protect\citeauthoryear{{Buck}, {Macci{\`o}}, {Obreja}, {Dutton},
  {Dom{\'{\i}}nguez-Tenreiro} \& {Granato}}{{Buck} et~al.}{2017}]{bmo:17}
{Buck} T.,  {Macci{\`o}} A.~V.,  {Obreja} A.,  {Dutton} A.~A.,
  {Dom{\'{\i}}nguez-Tenreiro} R.,    {Granato} G.~L.,  2017, \mnras, 468, 3628

\bibitem[\protect\citeauthoryear{{Ceverino}, {Dekel}, {Tweed} \&
  {Primack}}{{Ceverino} et~al.}{2015}]{cdt:14}
{Ceverino} D.,  {Dekel} A.,  {Tweed} D.,    {Primack} J.,  2015, \mnras, 447,
  3291

\bibitem[\protect\citeauthoryear{{Clarke}, {Whitworth}, {Duarte-Cabral} \&
  {Hubber}}{{Clarke} et~al.}{2017}]{cwd:17}
{Clarke} S.~D.,  {Whitworth} A.~P.,  {Duarte-Cabral} A.,    {Hubber} D.~A.,
  2017, \mnras, 468, 2489

\bibitem[\protect\citeauthoryear{{Clarke}, {Whitworth} \& {Hubber}}{{Clarke}
  et~al.}{2016}]{cwh:16}
{Clarke} S.~D.,  {Whitworth} A.~P.,    {Hubber} D.~A.,  2016, \mnras, 458, 319

\bibitem[\protect\citeauthoryear{{Dekel}, {Sari} \& {Ceverino}}{{Dekel}
  et~al.}{2009}]{dsc:09}
{Dekel} A.,  {Sari} R.,    {Ceverino} D.,  2009, \apj, 703, 785

\bibitem[\protect\citeauthoryear{{Elmegreen}}{{Elmegreen}}{2011}]{e:11}
{Elmegreen} B.~G.,  2011, \apj, 737, 10

\bibitem[\protect\citeauthoryear{{Elmegreen}}{{Elmegreen}}{2015}]{e:15}
{Elmegreen} B.~G.,  2015, \apjl, 814, L30

\bibitem[\protect\citeauthoryear{{Elmegreen} \& {Elmegreen}}{{Elmegreen} \&
  {Elmegreen}}{2006}]{ee:06}
{Elmegreen} B.~G.,  {Elmegreen} D.~M.,  2006, \apj, 650, 644

\bibitem[\protect\citeauthoryear{{Elmegreen}, {Elmegreen}, {S{\'a}nchez
  Almeida}, {Mu{\~n}oz-Tu{\~n}{\'o}n}, {Dewberry}, {Putko}, {Teich} \&
  {Popinchalk}}{{Elmegreen} et~al.}{2013}]{ees:13}
{Elmegreen} B.~G.,  {Elmegreen} D.~M.,  {S{\'a}nchez Almeida} J.,
  {Mu{\~n}oz-Tu{\~n}{\'o}n} C.,  {Dewberry} J.,  {Putko} J.,  {Teich} Y.,
  {Popinchalk} M.,  2013, \apj, 774, 86

\bibitem[\protect\citeauthoryear{{Elmegreen} \& {Hunter}}{{Elmegreen} \&
  {Hunter}}{2015}]{eh:15}
{Elmegreen} B.~G.,  {Hunter} D.~A.,  2015, \apj, 805, 145

\bibitem[\protect\citeauthoryear{{Elmegreen} \& {Elmegreen}}{{Elmegreen} \&
  {Elmegreen}}{2014}]{ee:14}
{Elmegreen} D.~M.,  {Elmegreen} B.~G.,  2014, \apj, 781, 11

\bibitem[\protect\citeauthoryear{{Escala} \& {Larson}}{{Escala} \&
  {Larson}}{2008}]{el:08}
{Escala} A.,  {Larson} R.~B.,  2008, \apjl, 685, L31

\bibitem[\protect\citeauthoryear{{Feng}, {Lin}, {Wang} \& {Taam}}{{Feng}
  et~al.}{2014}]{flw:14}
{Feng} C.-C.,  {Lin} L.-H.,  {Wang} H.-H.,    {Taam} R.~E.,  2014, \apj, 785,
  103

\bibitem[\protect\citeauthoryear{{Fisher} et~al.,}{{Fisher}
  et~al.}{2014}]{fgb:14}
{Fisher} D.~B.,  et~al., 2014, \apjl, 790, L30

\bibitem[\protect\citeauthoryear{{Fisher} et~al.,}{{Fisher}
  et~al.}{2017a}]{fga:17}
{Fisher} D.~B.,  et~al., 2017a, \apjl, 839, L5

\bibitem[\protect\citeauthoryear{{Fisher} et~al.,}{{Fisher}
  et~al.}{2017b}]{f:17}
{Fisher} D.~B.,  et~al., 2017b, \mnras, 464, 491

\bibitem[\protect\citeauthoryear{{Forbes}, {Krumholz} \& {Burkert}}{{Forbes}
  et~al.}{2012}]{fkb:12}
{Forbes} J.,  {Krumholz} M.,    {Burkert} A.,  2012, \apj, 754, 48

\bibitem[\protect\citeauthoryear{{Forbes}, {Krumholz}, {Burkert} \&
  {Dekel}}{{Forbes} et~al.}{2014}]{fkr:13}
{Forbes} J.~C.,  {Krumholz} M.~R.,  {Burkert} A.,    {Dekel} A.,  2014, \mnras,
  438, 1552

\bibitem[\protect\citeauthoryear{{F{\"o}rster Schreiber} et~al.,}{{F{\"o}rster
  Schreiber}  et~al.}{2009}]{fgb:09}
{F{\"o}rster Schreiber} N.~M.,  et~al., 2009, \apj, 706, 1364

\bibitem[\protect\citeauthoryear{{Fujii}, {Baba}, {Saitoh}, {Makino}, {Kokubo}
  \& {Wada}}{{Fujii} et~al.}{2011}]{fbs:11}
{Fujii} M.~S.,  {Baba} J.,  {Saitoh} T.~R.,  {Makino} J.,  {Kokubo} E.,
  {Wada} K.,  2011, \apj, 730, 109

\bibitem[\protect\citeauthoryear{{Garland}, {Pisano}, {Mac Low}, {Kreckel},
  {Rabidoux} \& {Guzm{\'a}n}}{{Garland} et~al.}{2015}]{gpm:15}
{Garland} C.~A.,  {Pisano} D.~J.,  {Mac Low} M.-M.,  {Kreckel} K.,  {Rabidoux}
  K.,    {Guzm{\'a}n} R.,  2015, \apj, 807, 134

\bibitem[\protect\citeauthoryear{{Genel} et~al.,}{{Genel}  et~al.}{2012}]{g:12}
{Genel} S.,  et~al., 2012, \apj, 745, 11

\bibitem[\protect\citeauthoryear{{Genzel} et~al.,}{{Genzel}
  et~al.}{2008}]{g:08}
{Genzel} R.,  et~al., 2008, \apj, 687, 59

\bibitem[\protect\citeauthoryear{{Genzel} et~al.,}{{Genzel}
  et~al.}{2011}]{gnj:11}
{Genzel} R.,  et~al., 2011, ApJ, 733, 101

\bibitem[\protect\citeauthoryear{{Genzel} et~al.,}{{Genzel}
  et~al.}{2014}]{gfl:14}
{Genzel} R.,  et~al., 2014, \apj, 785, 75

\bibitem[\protect\citeauthoryear{{Genzel} et~al.,}{{Genzel}
  et~al.}{2017}]{gsu:17}
{Genzel} R.,  et~al., 2017, \nat, 543, 397

\bibitem[\protect\citeauthoryear{{Goldreich} \& {Lynden-Bell}}{{Goldreich} \&
  {Lynden-Bell}}{1965}]{gl:65}
{Goldreich} P.,  {Lynden-Bell} D.,  1965, \mnras, 130, 97

\bibitem[\protect\citeauthoryear{{Grand} et~al.,}{{Grand}
  et~al.}{2017}]{ggm:17}
{Grand} R.~J.~J.,  et~al., 2017, \mnras, 467, 179

\bibitem[\protect\citeauthoryear{{Green} et~al.,}{{Green}
  et~al.}{2014}]{ggm:14}
{Green} A.~W.,  et~al., 2014, \mnras, 437, 1070

\bibitem[\protect\citeauthoryear{{Guo} et~al.,}{{Guo}  et~al.}{2015}]{gfb:14}
{Guo} Y.,  et~al., 2015, \apj, 800, 39

\bibitem[\protect\citeauthoryear{{Hattori} \& {Gilmore}}{{Hattori} \&
  {Gilmore}}{2015}]{hg:15}
{Hattori} K.,  {Gilmore} G.,  2015, \mnras, 454, 649

\bibitem[\protect\citeauthoryear{{Hernquist}}{{Hernquist}}{1990}]{h:90}
{Hernquist} L.,  1990, \apj, 356, 359

\bibitem[\protect\citeauthoryear{Hernquist}{Hernquist}{1993}]{h:93}
Hernquist L.,  1993, ApJ, 86, 389

\bibitem[\protect\citeauthoryear{{Hohl}}{{Hohl}}{1971}]{h:71}
{Hohl} F.,  1971, \apj, 168, 343

\bibitem[\protect\citeauthoryear{{Hopkins}, {Quataert} \& {Murray}}{{Hopkins}
  et~al.}{2012}]{hqm:12}
{Hopkins} P.~F.,  {Quataert} E.,    {Murray} N.,  2012, \mnras, 421, 3522

\bibitem[\protect\citeauthoryear{{Hu} \& {Sijacki}}{{Hu} \&
  {Sijacki}}{2016}]{hs:15}
{Hu} S.,  {Sijacki} D.,  2016, \mnras, 461, 2789

\bibitem[\protect\citeauthoryear{{Inoue}, {Dekel}, {Mandelker}, {Ceverino},
  {Bournaud} \& {Primack}}{{Inoue} et~al.}{2016}]{idm:16}
{Inoue} S.,  {Dekel} A.,  {Mandelker} N.,  {Ceverino} D.,  {Bournaud} F.,
  {Primack} J.,  2016, \mnras, 456, 2052

\bibitem[\protect\citeauthoryear{{Inoue} \& {Gouda}}{{Inoue} \&
  {Gouda}}{2013}]{ig:13}
{Inoue} S.,  {Gouda} N.,  2013, \aap, 555, A105

\bibitem[\protect\citeauthoryear{{Inoue} \& {Saitoh}}{{Inoue} \&
  {Saitoh}}{2011}]{is:11}
{Inoue} S.,  {Saitoh} T.~R.,  2011, \mnras, 418, 2527

\bibitem[\protect\citeauthoryear{{Inoue} \& {Saitoh}}{{Inoue} \&
  {Saitoh}}{2012}]{is:12}
{Inoue} S.,  {Saitoh} T.~R.,  2012, \mnras, 422, 1902

\bibitem[\protect\citeauthoryear{{Inutsuka} \& {Miyama}}{{Inutsuka} \&
  {Miyama}}{1992}]{im:92}
{Inutsuka} S.-I.,  {Miyama} S.~M.,  1992, \apj, 388, 392

\bibitem[\protect\citeauthoryear{{Inutsuka} \& {Miyama}}{{Inutsuka} \&
  {Miyama}}{1997}]{im:97}
{Inutsuka} S.-i.,  {Miyama} S.~M.,  1997, \apj, 480, 681

\bibitem[\protect\citeauthoryear{{Jog}}{{Jog}}{1996}]{j:96}
{Jog} C.~J.,  1996, \mnras, 278, 209

\bibitem[\protect\citeauthoryear{{Jog} \& {Solomon}}{{Jog} \&
  {Solomon}}{1984a}]{js:84_2}
{Jog} C.~J.,  {Solomon} P.~M.,  1984a, \apj, 276, 127

\bibitem[\protect\citeauthoryear{{Jog} \& {Solomon}}{{Jog} \&
  {Solomon}}{1984b}]{js:84b}
{Jog} C.~J.,  {Solomon} P.~M.,  1984b, \apj, 276, 127

\bibitem[\protect\citeauthoryear{{Jog} \& {Solomon}}{{Jog} \&
  {Solomon}}{1984c}]{js:84}
{Jog} C.~J.,  {Solomon} P.~M.,  1984c, \apj, 276, 114

\bibitem[\protect\citeauthoryear{{Juri{\'c}} et~al.,}{{Juri{\'c}}
  et~al.}{2008}]{jib:08}
{Juri{\'c}} M.,  et~al., 2008, \apj, 673, 864

\bibitem[\protect\citeauthoryear{{Kennicutt} Jr.}{{Kennicutt}}{1989}]{k:89}
{Kennicutt} Jr. R.~C.,  1989, ApJ, 344, 685

\bibitem[\protect\citeauthoryear{{Kyziropoulos}, {Efthymiopoulos}, {Gravvanis}
  \& {Patsis}}{{Kyziropoulos} et~al.}{2016}]{keg:16}
{Kyziropoulos} P.~E.,  {Efthymiopoulos} C.,  {Gravvanis} G.~A.,    {Patsis}
  P.~A.,  2016, \mnras, 463, 2210

\bibitem[\protect\citeauthoryear{{Lada} \& {Lada}}{{Lada} \&
  {Lada}}{2003}]{ll:03}
{Lada} C.~J.,  {Lada} E.~A.,  2003, \araa, 41, 57

\bibitem[\protect\citeauthoryear{{Law}, {Shapley}, {Steidel}, {Reddy},
  {Christensen} \& {Erb}}{{Law} et~al.}{2012}]{lss:12}
{Law} D.~R.,  {Shapley} A.~E.,  {Steidel} C.~C.,  {Reddy} N.~A.,  {Christensen}
  C.~R.,    {Erb} D.~K.,  2012, \nat, 487, 338

\bibitem[\protect\citeauthoryear{{Leroy}, {Walter}, {Brinks}, {Bigiel}, {de
  Blok}, {Madore} \& {Thornley}}{{Leroy} et~al.}{2008}]{lwb:08}
{Leroy} A.~K.,  {Walter} F.,  {Brinks} E.,  {Bigiel} F.,  {de Blok} W.~J.~G.,
  {Madore} B.,    {Thornley} M.~D.,  2008, \aj, 136, 2782

\bibitem[\protect\citeauthoryear{{Li}}{{Li}}{2017}]{l:17}
{Li} G.-X.,  2017, preprint (arXiv:1703.10613)

\bibitem[\protect\citeauthoryear{{Lin} \& {Shu}}{{Lin} \& {Shu}}{1966}]{ls:66}
{Lin} C.~C.,  {Shu} F.~H.,  1966, Proceedings of the National Academy of
  Science, 55, 229

\bibitem[\protect\citeauthoryear{{Mandelker}, {Dekel}, {Ceverino}, {DeGraf},
  {Guo} \& {Primack}}{{Mandelker} et~al.}{2017}]{mdc:17}
{Mandelker} N.,  {Dekel} A.,  {Ceverino} D.,  {DeGraf} C.,  {Guo} Y.,
  {Primack} J.,  2017, \mnras, 464, 635

\bibitem[\protect\citeauthoryear{{Michikoshi} \& {Kokubo}}{{Michikoshi} \&
  {Kokubo}}{2014}]{mk:14}
{Michikoshi} S.,  {Kokubo} E.,  2014, \apj, 787, 174

\bibitem[\protect\citeauthoryear{{Morozov}}{{Morozov}}{1981}]{m:81}
{Morozov} A.~G.,  1981, Soviet Astronomy Letters, 7, 5

\bibitem[\protect\citeauthoryear{{Murata} et~al.,}{{Murata}
  et~al.}{2014}]{mkt:14}
{Murata} K.~L.,  et~al., 2014, \apj, 786, 15

\bibitem[\protect\citeauthoryear{Navarro, Frenk \& White}{Navarro
  et~al.}{1997}]{nfw:97}
Navarro J.~F.,  Frenk C.~S.,    White S. D.~M.,  1997, ApJ, 490, 493

\bibitem[\protect\citeauthoryear{{Newman} et~al.,}{{Newman}
  et~al.}{2012}]{nsg:12}
{Newman} S.~F.,  et~al., 2012, \apj, 752, 111

\bibitem[\protect\citeauthoryear{{Noguchi}}{{Noguchi}}{1996}]{n:96}
{Noguchi} M.,  1996, ApJ, 469, 605

\bibitem[\protect\citeauthoryear{{Noguchi}}{{Noguchi}}{1998}]{n:98}
{Noguchi} M.,  1998, Nat, 392, 253

\bibitem[\protect\citeauthoryear{{Noguchi}}{{Noguchi}}{1999}]{n:99}
{Noguchi} M.,  1999, ApJ, 514, 77

\bibitem[\protect\citeauthoryear{{Oklop{\v c}i{\'c}}, {Hopkins}, {Feldmann},
  {Kere{\v s}}, {Faucher-Gigu{\`e}re} \& {Murray}}{{Oklop{\v c}i{\'c}}
  et~al.}{2017}]{okh:17}
{Oklop{\v c}i{\'c}} A.,  {Hopkins} P.~F.,  {Feldmann} R.,  {Kere{\v s}} D.,
  {Faucher-Gigu{\`e}re} C.-A.,    {Murray} N.,  2017, \mnras, 465, 952

\bibitem[\protect\citeauthoryear{{Ostriker}}{{Ostriker}}{1964}]{o:64}
{Ostriker} J.,  1964, \apj, 140, 1056

\bibitem[\protect\citeauthoryear{{Puech}}{{Puech}}{2010}]{p:10}
{Puech} M.,  2010, \mnras, 406, 535

\bibitem[\protect\citeauthoryear{{Rafikov}}{{Rafikov}}{2001}]{r:01}
{Rafikov} R.~R.,  2001, \mnras, 323, 445

\bibitem[\protect\citeauthoryear{{Reina-Campos} \& {Kruijssen}}{{Reina-Campos}
  \& {Kruijssen}}{2017}]{rck:17}
{Reina-Campos} M.,  {Kruijssen} J.~M.~D.,  2017, preprint (arXiv:1704.00732)

\bibitem[\protect\citeauthoryear{{Ribeiro} et~al.,}{{Ribeiro}
  et~al.}{2016}]{r:17}
{Ribeiro} B.,  et~al., 2016, preprint (arXiv:1611.05869)

\bibitem[\protect\citeauthoryear{{Romeo}}{{Romeo}}{1985}]{r:85}
{Romeo} A.~B.,  1985, Master's thesis, Tesi di Laurea, University of Pisa and
  Scuola Normale Superiore, Pisa, Italy

\bibitem[\protect\citeauthoryear{{Romeo}}{{Romeo}}{1992}]{r:92}
{Romeo} A.~B.,  1992, \mnras, 256, 307

\bibitem[\protect\citeauthoryear{{Romeo} \& {Wiegert}}{{Romeo} \&
  {Wiegert}}{2011}]{rw:11}
{Romeo} A.~B.,  {Wiegert} J.,  2011, \mnras, 416, 1191

\bibitem[\protect\citeauthoryear{{Safronov}}{{Safronov}}{1960}]{s:60}
{Safronov} V.~S.,  1960, Annales d'Astrophysique, 23, 979

\bibitem[\protect\citeauthoryear{{Schmidt}}{{Schmidt}}{1959}]{s:59}
{Schmidt} M.,  1959, ApJ, 129, 243

\bibitem[\protect\citeauthoryear{{Shibuya}, {Ouchi}, {Kubo} \&
  {Harikane}}{{Shibuya} et~al.}{2016}]{sok:16}
{Shibuya} T.,  {Ouchi} M.,  {Kubo} M.,    {Harikane} Y.,  2016, \apj, 821, 72

\bibitem[\protect\citeauthoryear{{Shlosman} \& {Noguchi}}{{Shlosman} \&
  {Noguchi}}{1993}]{SN:93}
{Shlosman} I.,  {Noguchi} M.,  1993, \apj, 414, 474

\bibitem[\protect\citeauthoryear{{Springel}}{{Springel}}{2010}]{arepo}
{Springel} V.,  2010, \mnras, 401, 791

\bibitem[\protect\citeauthoryear{{Tadaki}, {Kodama}, {Tanaka}, {Hayashi},
  {Koyama} \& {Shimakawa}}{{Tadaki} et~al.}{2014}]{tkt:13II}
{Tadaki} K.-i.,  {Kodama} T.,  {Tanaka} I.,  {Hayashi} M.,  {Koyama} Y.,
  {Shimakawa} R.,  2014, \apj, 780, 77

\bibitem[\protect\citeauthoryear{{Takahashi}, {Tsukamoto} \&
  {Inutsuka}}{{Takahashi} et~al.}{2016}]{tti:16}
{Takahashi} S.~Z.,  {Tsukamoto} Y.,    {Inutsuka} S.,  2016, \mnras, 458, 3597

\bibitem[\protect\citeauthoryear{{Tamburello}, {Mayer}, {Shen} \&
  {Wadsley}}{{Tamburello} et~al.}{2015}]{tms:14}
{Tamburello} V.,  {Mayer} L.,  {Shen} S.,    {Wadsley} J.,  2015, \mnras, 453,
  2490

\bibitem[\protect\citeauthoryear{{Tenjes}, {Tuvikene}, {Tamm}, {Kipper} \&
  {Tempel}}{{Tenjes} et~al.}{2017}]{ttt:17}
{Tenjes} P.,  {Tuvikene} T.,  {Tamm} A.,  {Kipper} R.,    {Tempel} E.,  2017,
  preprint (arXiv:1701.05815)

\bibitem[\protect\citeauthoryear{{Toomre}}{{Toomre}}{1964}]{t:64}
{Toomre} A.,  1964, ApJ, 139, 1217

\bibitem[\protect\citeauthoryear{{Toomre}}{{Toomre}}{1981}]{t:81}
{Toomre} A.,  1981, in {Fall} S.~M.,  {Lynden-Bell} D.,  eds, Structure and
  Evolution of Normal Galaxies {What amplifies the spirals}.
pp 111--136

\bibitem[\protect\citeauthoryear{{Wang} \& {Silk}}{{Wang} \&
  {Silk}}{1994}]{ws:94}
{Wang} B.,  {Silk} J.,  1994, \apj, 427, 759

\bibitem[\protect\citeauthoryear{{Weiner} et~al.,}{{Weiner}
  et~al.}{2006}]{w:06}
{Weiner} B.~J.,  et~al., 2006, \apj, 653, 1027

\bibitem[\protect\citeauthoryear{{Westfall}, {Andersen}, {Bershady},
  {Martinsson}, {Swaters} \& {Verheijen}}{{Westfall} et~al.}{2014}]{wab:14}
{Westfall} K.~B.,  {Andersen} D.~R.,  {Bershady} M.~A.,  {Martinsson} T.~P.~K.,
   {Swaters} R.~A.,    {Verheijen} M.~A.~W.,  2014, \apj, 785, 43

\bibitem[\protect\citeauthoryear{{White} et~al.,}{{White}
  et~al.}{2017}]{wfm:17}
{White} H.~A.,  et~al., 2017, preprint (arXiv:1707.07005)

\end{thebibliography}

\appendix
\section{The TTI criterion revisited}
\label{TTIana}
TTI has empirically obtained their instability criterion for spiral-arm fragmentation in proto-planetary discs from their simulations: $Q\lsim0.6$ on a spiral arm. Here, we show that the TTI criterion is almost consistent with our instability criterion for a single-component model: $\min[S(k)]<1$ (equation \ref{Crit1C}). 

If we assume isotropy for gas and a rigid rotation, i.e. epicyclic frequency $\kappa=2\Omega$, the Toomre's parameter $Q$ is described, with equation (\ref{linemass}), as
\begin{equation}
Q\equiv \frac{\sigma\kappa}{\pi G\Sigma} = \frac{2A\sigma\Omega W}{\pi G\Upsilon},
\end{equation}
Denoting $x\equiv kW$ and $q\equiv2\Omega W/\sigma$, our instability parameter of equation (\ref{Crit1C}) is reduced to 
\begin{equation}
S\equiv \frac{\sigma^2k^2 + 4\Omega^2}{\pi Gf(kW)\Upsilon k^2}=Q\frac{x^2+q^2}{Af(x)x^2q}.
\end{equation}

\begin{figure}
  \includegraphics[bb=0 0 1625 1170,width=\hsize]{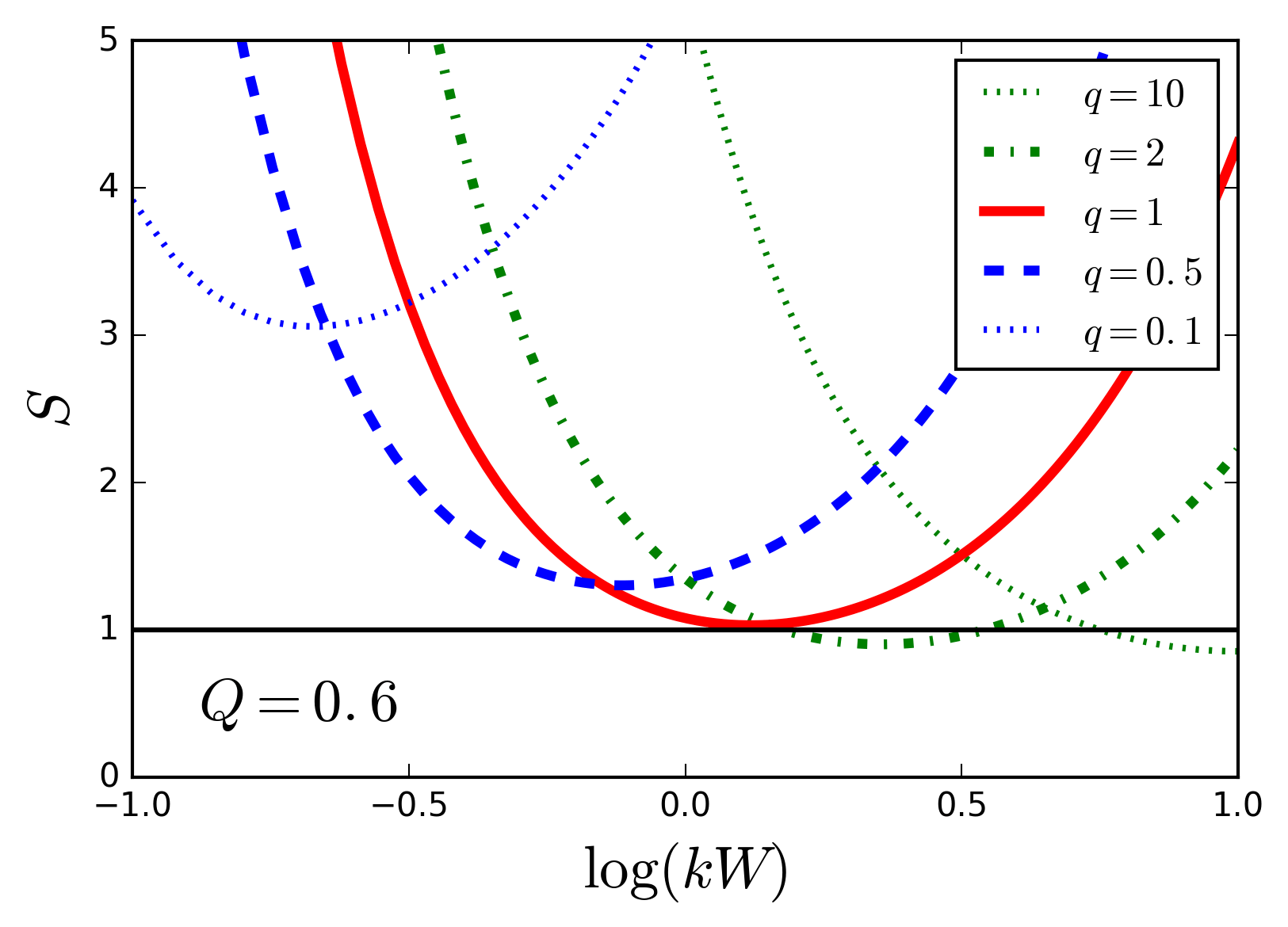}
  \caption{Values of $S$ as functions of $x\equiv kW$ for various $q$ in the case of $Q=0.6$ and $A=1.4$.}
  \label{TTIrev}
\end{figure}
Fig. \ref{TTIrev} shows $S$ as functions of $x$ with $Q=0.6$ and $A=1.4$. It can be seen that $\min[S(x)]\simeq1$ for $q\gsim0.5$ although $\min[S(x)]$ rapidly increases as $q$ decreases for $q\lsim0.5$. This means that our instability condition, $\min[S(x)]\lsim1$, is satisfied in a wide range of $q$ when $Q\lsim0.6$ since $S\propto Q$. Thus, for $q\lsim0.5$, the TTI criterion of $Q\lsim0.6$ can be considered to be accurate and almost consistent with our criterion of $\min[S(x)]<1$. 

\begin{figure}
  \includegraphics[bb=0 0 1699 1170,width=\hsize]{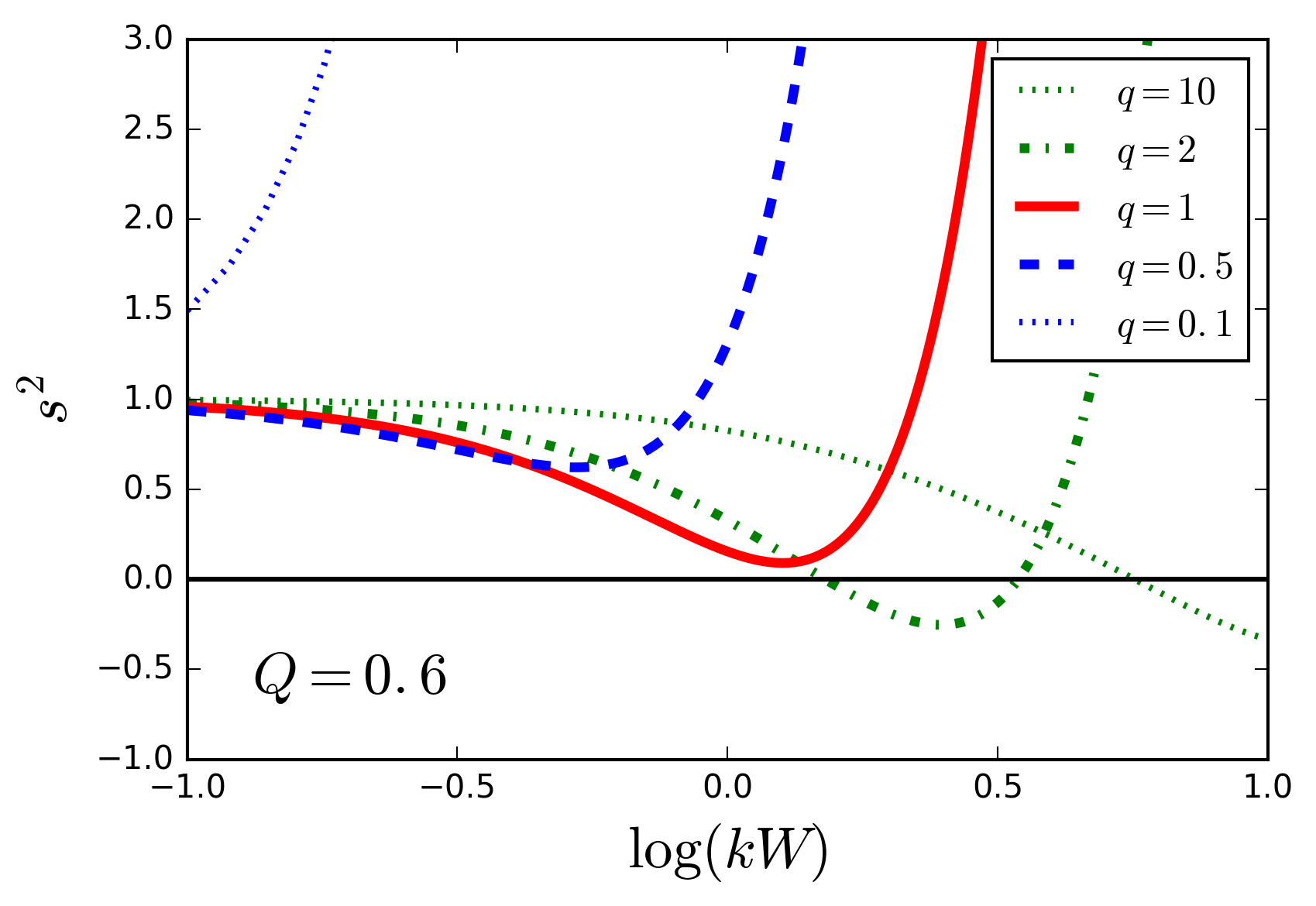}
  \caption{The dispersion relation as a function of $x\equiv kW$ for various $q$ when $Q=0.6$ and $A=1.4$.}
  \label{DRrev}
\end{figure}
The dispersion relation (equation \ref{DL1}) in the single-component model can also be reduced to
\begin{equation}
s^2 = \frac{x^2}{q^2} - \frac{A}{qQ}f(x)x^2 + 1,
\end{equation}
where $s\equiv\omega/(2\Omega)$. Fig. \ref{DRrev} illustrates the above dispersion relations with $Q=0.6$ and $A=1.4$.

\section{The validity of the assumptions}
\label{assumptionvalidity}
\begin{figure}
	\includegraphics[bb=0 0 1805 1156,width=\hsize]{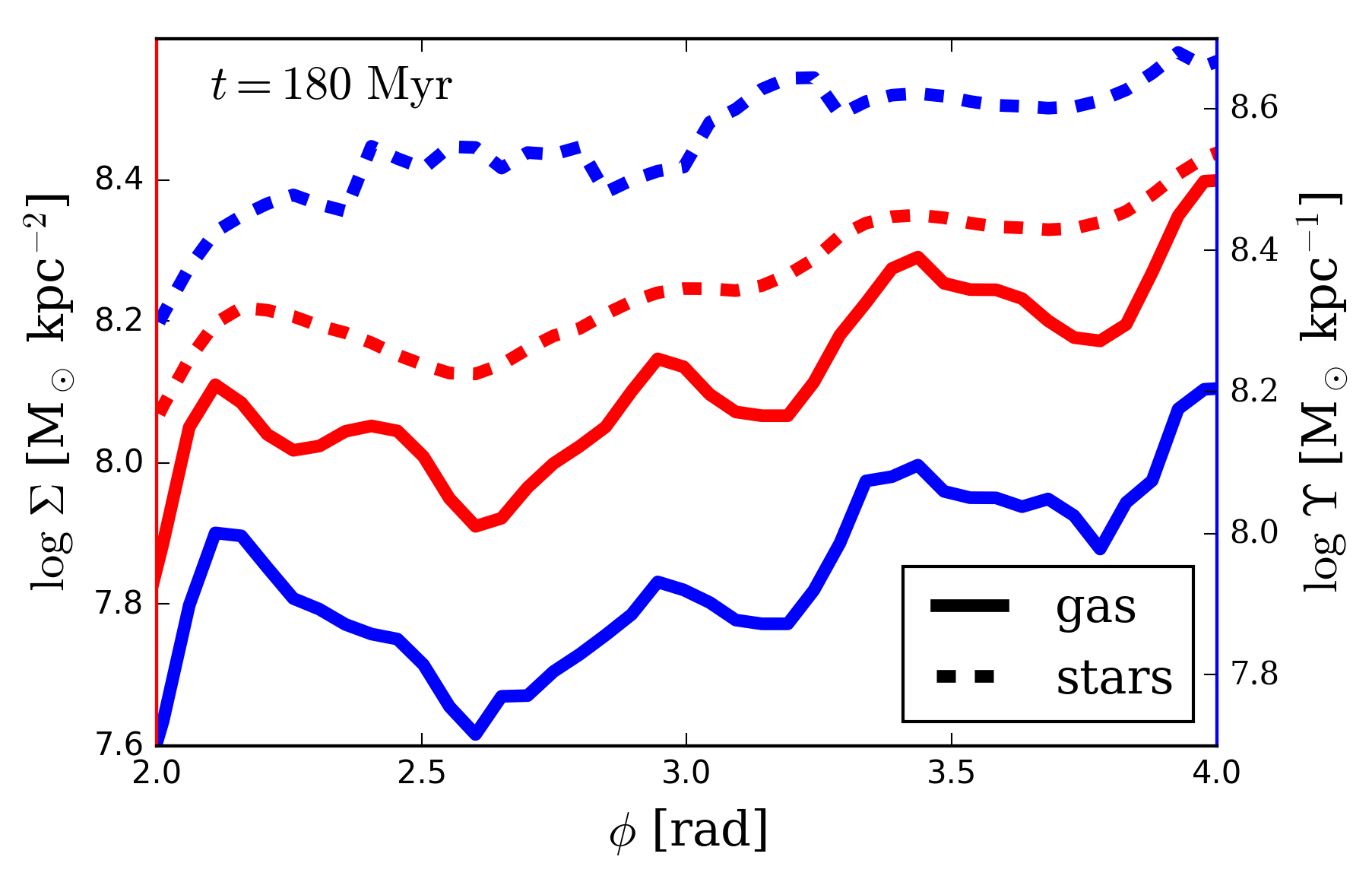}
	\caption{Distributions of $\Sigma$ (the red lines with the left ordinate) and $\Upsilon$ (the blue lines with the right ordinate) for each of gas and stars along the positions where $\chi^2$ becomes the lowest at each $\phi$ in the fragmenting arm in the run Df25Q15 at $t=180~{\rm Myr}$. The spiral arm is marked with the ellipse in the first snapshot of Fig. \ref{FullMaps}.}
	\label{linearity}
\end{figure}
Here we inspect whether our simulations are consistent with the assumptions in the linear perturbation analysis. In the linear analysis, the density fluctuation of the perturbations is presumed to be initially small, i.e. $\delta\Sigma/\Sigma\ll1$ and $\delta\Upsilon/\Upsilon\ll1$. Therefore, a spiral arm should have smooth distributions of $\Sigma$ and $\Upsilon$ before fragmentation. Fig. \ref{linearity} shows distributions of $\Sigma$ and $\Upsilon$ for each of gas and stars along the ridge of the fragmenting spiral arm marked with the ellipse in Fig. \ref{FullMaps} (the model Df25Q15) at $t=180~{\rm Myr}$ (before the fragmentation). Here, the ridge of the arm is defined to be the position at which the Gaussian density fitting gives the lowest $\chi^2$ within the spiral arm at each $\phi$ (see equation \ref{Xi2}). In this arm, the stellar component has the smoother distributions of $\Sigma$ and $\Upsilon$ than the gas component. The fluctuation of $\Sigma_{\rm g}$ and $\Upsilon_{\rm g}$ from their averages appears to be comparable to or lower than the order of unity. Therefore, the density perturbations of both components are approximately linear before the fragmentation in the simulation.

\begin{figure}
	\includegraphics[bb=0 0 1394 1925,width=\hsize]{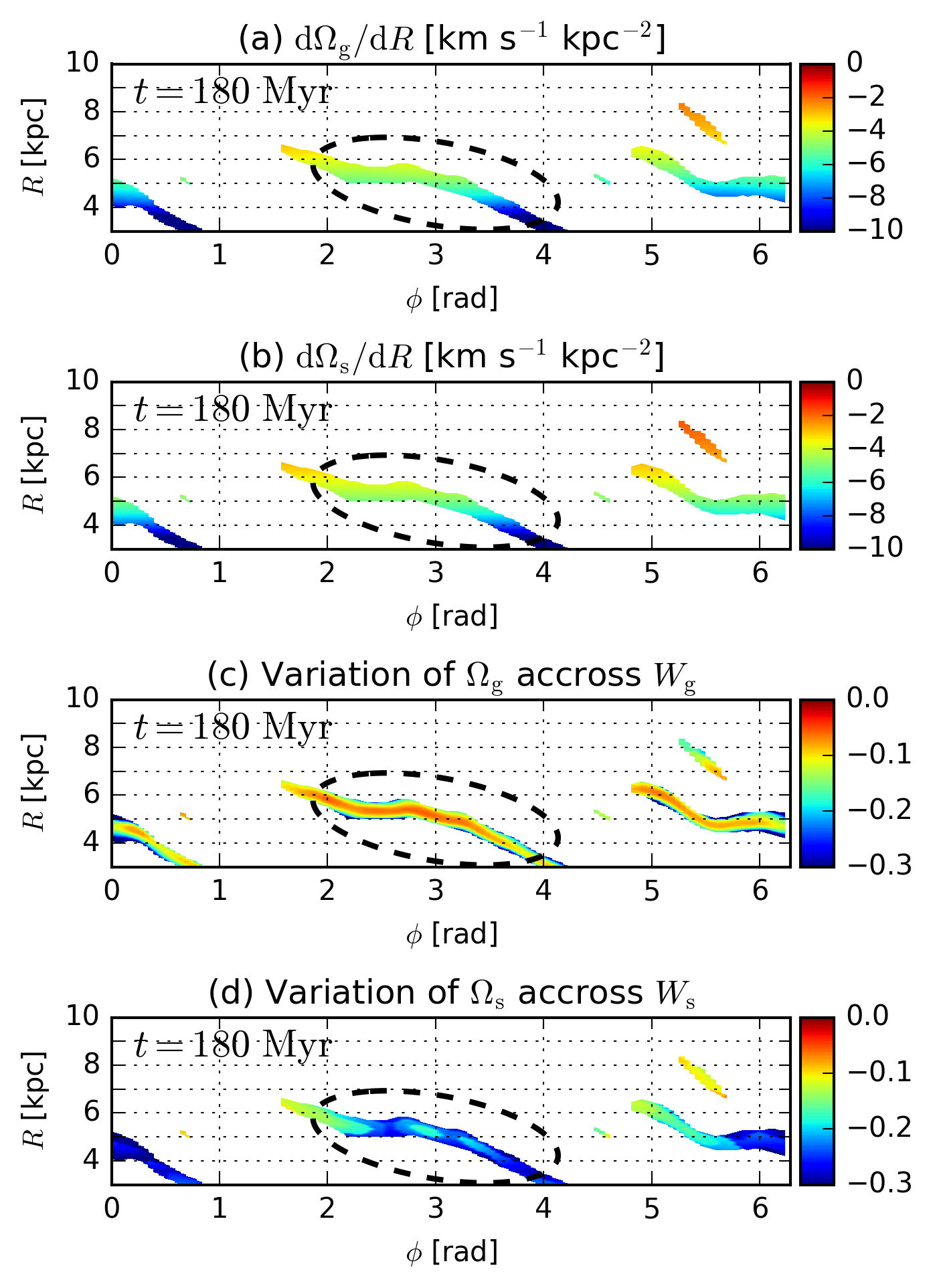}
	\caption{Radial variations of angular velocities in the run Df25Q15 at $t=180~{\rm Myr}$. \textit{Panel a and b:} Radial variations of angular velocities $\mathrm{d}\Omega/\mathrm{d}R$ for the gas and stellar components with in the spiral arms. \textit{Panel c and d:} fractional changes of $\Omega$ for the gas and stars across the half widths $W$ of the arm, i.e. $(W\mathrm{d}\Omega/\mathrm{d}R)/\Omega$.}
	\label{rigidrotation}
\end{figure}
The linear perturbation analysis described in Section \ref{basiceq} also assumes rigid rotation within spiral arms on the grounds that the arms are expected to be self-gravitating. This assumption expects that radial variation of angular velocity $\Omega$ across the arm is small, i.e. $W\mathrm{d}\Omega/\mathrm{d}R\ll\Omega$. In Fig. \ref{rigidrotation}, Panels a and b show polar maps of $\mathrm{d}\Omega/\mathrm{d}R$ in the fragmenting model Df25Q15 at $t=180~{\rm Myr}$. The distributions of $\mathrm{d}\Omega/\mathrm{d}R$ are almost the same between the gas and stars within the spiral arm. Panels c and d show the ratios $(W\mathrm{d}\Omega/\mathrm{d}R)/\Omega$ in the same snapshot; this ratio means a fractional change of $\Omega$ across the half width $W$ of the arm. In Panel c, the fractional changes of $\Omega_{\rm g}$ are quite low in the gas component, $<0.1$. In Panel d, the stellar component indicates the values larger than the gas since $W_{\rm s}>W_{\rm g}$. However, the fractional changes of $\Omega_{\rm s}$ are still $\simeq0.2$. Thus, the rotation velocities in the spiral arm are close to rigid rotations, and the assumption in the linear perturbation analysis appears to hold in the simulation.

\end{document}